\documentclass[aps,prd,12pt,preprint,superscriptaddress,preprintnumbers,nofootinbib,floatfix,longbibliography]{revtex4-2}

\usepackage[T1]{fontenc}
\usepackage[utf8]{inputenc}
\usepackage[english]{babel}
\usepackage{amsmath,amsthm,amssymb,amsfonts,mathrsfs,amsbsy,bm}
\usepackage{tensor}
\usepackage{slashed}
\usepackage{esint}
\usepackage[a4paper, margin=1.6cm]{geometry}
\usepackage{cancel}
\usepackage{graphicx}
\usepackage{multirow}
\usepackage{array}
\usepackage{booktabs}
\usepackage{makecell}
\usepackage{xcolor}
\colorlet{BLUE}{blue}
\usepackage{tikz}
\usetikzlibrary{quotes,angles,arrows,decorations.markings,decorations.pathmorphing}
\usepackage{hyperref}
\hypersetup{colorlinks=true,breaklinks=true,citecolor=blue,linkcolor=[rgb]{0,0.5,0.9},urlcolor=blue}
\graphicspath{{figures/}}

\newcommand{\be}{\begin{equation}}
\newcommand{\ee}{\end{equation}}
\newcommand{\Be}{\begin{eqnarray}}
\newcommand{\Ee}{\end{eqnarray}}

\newcommand{\mincir}{\raise-3.truept\hbox{\rlap{\hbox{$\sim$}}\raise4.truept\hbox{$<$}\ }}
\newcommand{\magcir}{\raise-3.truept\hbox{\rlap{\hbox{$\sim$}}\raise4.truept\hbox{$>$}\ }}

\providecommand{\U}[1]{}
\newcommand{\ie}{\begin{equation}}
\newcommand{\fe}{\end{equation}}
\newcommand{\se}{\begin{eqnarray}}
\newcommand{\ff}{\end{eqnarray}}

\begin{document}
\emergencystretch=4em


\title{A general static and spherically symmetric black hole in traceless metric-affine bumblebee gravity}

\author{A. A. Ara\'{u}jo Filho}
\email{dilto@fisica.ufc.br}
\affiliation{Departamento de F\'isica, Universidade Federal da Para\'iba, Caixa Postal 5008, 58051--970, Jo\~ao Pessoa, Para\'iba, Brazil.}
\affiliation{Departamento de F\'isica, Universidade Federal de Campina Grande, Caixa Postal 10071, 58429--900 Campina Grande, Para\'iba, Brazil.}
\affiliation{Center for Theoretical Physics, Khazar University, 41 Mehseti Street, Baku, AZ-1096, Azerbaijan.}

\author{N. Heidari}
\email{heidari.n@gmail.com}
\affiliation{Center for Theoretical Physics, Khazar University, 41 Mehseti Street, Baku, AZ-1096, Azerbaijan.}
\affiliation{School of Physics, Damghan University, Damghan, 3671641167, Iran.}

\author{V. B. Bezerra}
\email{valdir@fisica.ufpb.br}
\affiliation{Departamento de F\'isica, Universidade Federal da Para\'iba, Caixa Postal 5008, 58051--970, Jo\~ao Pessoa, Para\'iba, Brazil.}

\author{Francisco S. N. Lobo} 
\email{fslobo@ciencias.ulisboa.pt}
\affiliation{Instituto de Astrof\'{i}sica e Ci\^{e}ncias do Espa\c{c}o, Faculdade de Ci\^{e}ncias da Universidade de Lisboa, Edifício C8, Campo Grande, P-1749-016 Lisbon, Portugal}
\affiliation{Departamento de F\'{i}sica, Faculdade de Ci\^{e}ncias da Universidade de Lisboa, Edif\'{i}cio C8, Campo Grande, P-1749-016 Lisbon, Portugal}

\date{\today}


\begin{abstract}
We construct and analyze a general exact static and spherically symmetric black hole solution in traceless \textit{metric--affine} bumblebee gravity by considering a vacuum expectation value with temporal and radial components. The temporal component remains constant and the auxiliary geometry is Ricci flat, while Lorentz symmetry breaking is transferred to the physical metric through the \textit{metric--affine} deformation. After introducing the areal radius and normalizing the Killing time at spatial infinity, the geometry is controlled by a single invariant deformation parameter. The resulting spacetime is asymptotically conical, possesses a regular Killing horizon, and retains the causal structure of the Schwarzschild black hole, with a curvature singularity located at the origin. Initially, we examine its thermodynamic properties, gravitational Doppler effect, and tidal forces. Lorentz symmetry breaking modifies the Hawking temperature and the magnitude of the tidal field, although the characteristic ratio between radial stretching and transverse compression is preserved. The thermodynamic--topology analysis reveals no critical point within the admissible black hole sector. Furthermore, massless scalar perturbations are also investigated through WKB and characteristic time domain solution, showing decaying profiles throughout the sampled parameter range and deformation--dependent quasinormal frequencies. Finally, light deflection and perihelion advance are used to establish stringent Solar System bounds on the invariant deformation, which, at leading order, depends on a combination of the temporal and norm sectors of the bumblebee field.
\end{abstract}

\maketitle

\tableofcontents


\section{Introduction }

Local Lorentz symmetry is built into the geometric formulation of general relativity and into the particle dynamics described by the Standard Model. Whether it remains exact in a more fundamental theory, however, is an empirical question. Several candidates for physics beyond these two frameworks permit tensor fields to acquire nonzero vacuum expectation values, leaving small Lorentz--violating coefficients in the low--energy description. The so-called Standard Model Extension organizes such effects within effective field theory and gives a natural language for confronting them with gravitational, astrophysical, and laboratory observations \cite{kostelecky1989spontaneous,colladay1997cpt,kostelecky2004gravity,Bailey:2006fd,kostelecky2011data}. Spontaneous breaking is especially suitable in gravity because the action may remain generally covariant even though its vacuum selects preferred spacetime directions, preserving Bianchi identities and other essential features.

The bumblebee model realizes this mechanism through a dynamical vector field $B_{\mu}$. A potential fixes the field at a nonvanishing vacuum configuration, while a nonminimal contraction between $B_{\mu}$ and the Ricci tensor transfers the symmetry breaking to the gravitational sector \cite{bluhm2005spontaneous,bluhm2008spontaneous,bluhm2008constraints,Bluhm:2019ato,Bluhm:2023kph,Ou:2026ohc}. Around the vacuum, the theory may contain Nambu--Goldstone excitations and a massive mode measuring departures from the minimum of the potential. Its constraint structure and propagating content depend on the kinetic terms and couplings adopted in the action \cite{Maluf:2013nva,Maluf:2014dpa}. The causal character and differential structure of the vacuum field are equally important. Timelike, spacelike, and lightlike expectation values need not generate equivalent geometries, while a closed configuration with $B_{\mu\nu}=0$ belongs to a different dynamical sector from a profile with nonzero field strength. Also, the extension with cloud strings \cite{Ahmed:2026fat,Ahmed:2026fat,Ahmed:2026zly,Belchior:2026mqk} and other scenarios \cite{Kanzi:2021cbg,Kanzi:2019gtu,Uniyal:2022xnq,Belchior:2026jrm,Belchior:2026poc} has been reported .

Black hole solutions have made these distinctions explicit. The early vacuum analysis of Bertolami and P\'aramos \cite{Bertolami:2005bh} was followed by the Schwarzschild-like geometry of Casana \textit{et al.} \cite{Casana:2017jkc}, obtained from a purely radial spacelike vacuum field (and its subsequent non--commutative version \cite{AraujoFilho:2025rvn}). Although its temporal potential coincides with the Schwarzschild one, the radial normalization changes the asymptotic spatial geometry and leaves observable corrections in light bending, time delay, and perihelion advance. The corresponding cosmological constant version \cite{Maluf:2020kgf} and topological defect \cite{Gullu:2020qzu} extensions subsequently enlarged this family of solutions. A later treatment of the general static ansatz separated the spherical vacuum sector into two families; one activates a temporal bumblebee profile and carries $B_{\mu\nu}\neq0$ \cite{Xu:2022frb}. Leaving the minimum of the potential enlarges the solution space still further, producing Schwarzschild--AdS and Reissner--Nordstr\"om limits together with repulsive domains, naked singularities, and steep profiles close to the source \cite{Bailey:2025qgm}. Purely timelike vacua admit nontrivial curved configurations only under a special norm condition \cite{Li:2025timelike}. Another family has recently yielded exact black holes that are asymptotically flat and support a temporal vector profile \cite{Yang:2026asymptotic}.

The static solution space has advanced further with vacuum fields carrying both temporal and radial components. Exact black holes with spacelike or lightlike expectation values were obtained in Ref.~\cite{Liu:2025oho}, and a complete classification of static spherical vacua with general causal character was presented in Ref.~\cite{Zhu:2025fiy}. The latter analysis also identified degenerate regions of parameter space and showed that apparently identical metrics may be accompanied by inequivalent bumblebee or matter configurations. Generalized curvature couplings produce additional black holes, wormholes, and naked singularity segments \cite{Zhu:2026vae}. Matter supported solutions now include charged and slowly rotating configurations, dyonic and Taub--NUT--like geometries, and black holes sourced by nonlinear electrodynamics \cite{Liu:2024charged,Li:2025dyonic,Li:2026tae}. The two component background relevant to the present work has already been examined through geodesics, shadows, lensing, quasinormal frequencies, time domain profiles, Solar-System tests, accretion observables, and quantum information \cite{AraujoFilho:2025zaj,Shi:2025hfe,AraujoFilho:2025nmc}.

These geometries have supported an extensive phenomenological scenario, ranging from weak and strong gravitational lensing to accretion, thermodynamics, wave propagation, and orbital dynamics \cite{Ovgun:2018ran,Li:2020wvn,Yang:2018zef,Gomes:2018oyd,Oliveira:2021abg}. Recent studies have moved beyond test field observables. Odd--parity perturbations of charged bumblebee black holes have supplied theoretical stability restrictions that may be stronger than shadow bounds in part of the parameter space \cite{Mai:2024dzy}, while the coupled metric--bumblebee perturbations of the Schwarzschild--like solution have been reduced to exact master equations in both parity sectors \cite{Liu:2026cxs}.  In nonminimally coupled bumblebee gravity, the horizon charge obtained from the covariant phase-space construction may contain contributions not captured by a direct area or Wald density identification \cite{An:2024twi,Liu:2025oho}. At the observational level, extreme mass ratio inspirals have recently been used to estimate the sensitivity of future space-based measurements to the bumblebee deformation \cite{Long:2026dcb}.

The scope of bumblebee gravity is not confined to black holes. Applications include anisotropic and accelerating cosmologies, relativistic stars, wormholes, black bounces, neutrino propagation, and early Universe scenarios \cite{Neves:2022qyb,Gonzalez-Espinoza:2025fmi,Neves:2024ggn,Magalhaes:2025nql,AraujoFilho:2024iox,Shi:2025plr,Shi:2025ywa}. Cosmological perturbations now provide stability and gravitational wave constraints on the allowed coupling space \cite{Lai:2025cosmo}, while primordial black hole production and neutron star tidal deformability have been considered in Refs.~\cite{Khodadi:2026nja,Banerjee:2026bav}.

A conceptually distinct realization arises when the metric and the affine connection are independent variables. In projectively invariant Ricci--based \textit{metric--affine} theories, the connection equation can be solved through an auxiliary metric, and the physical metric is recovered from an algebraic deformation map that carries the matter and Lorentz--
breaking dependence \cite{Afonso:2018mapping,BeltranJimenez:2019acz}. \textit{metric--affine} bumblebee gravity implements this structure in the gravitational sector of the Standard Model Extension \cite{Delhom:2021bumblebee}.

The \textit{metric--affine} theory has developed at both classical and quantum levels. Radiative corrections, the one-loop effective action, gauge field couplings, electron scattering and gravitational waves induced by nonmetricity have been studied in Refs.~\cite{Delhom:2021radiative,Delhom:2022oneLoop,Lehum:2024mab}. The first exact static black hole was derived in Ref.~\cite{Filho:2022yrk}, and its geodesic, optical, accretion, scattering, quasinormal, neutrino, and quantum emission properties were subsequently investigated \cite{Lambiase:2023zeo,AraujoFilho:2024ioxMA,Heidari:2024scattering,Gao:2024mab,Jha:2023mab,AraujoFilho:2025nonmetricity}. Exact axisymmetric solutions and their shadows extended the construction to rotation \cite{AraujoFilho:2024ykw,Nascimento:2026shadow}, while the propagation and generation of gravitational waves, as well as nonmetricity corrections to electron scattering, have recently been obtained directly from the underlying \textit{metric--affine} dynamics \cite{AraujoFilho:2026gwma,AraujoFilho:2026electron}.

The general two--component vacuum classification developed in the metric formulation has no corresponding exact construction in traceless \textit{metric--affine} bumblebee gravity. We address this problem by beginning with a general static and spherically symmetric configuration,
$b_{\mu}=\bigl(b_t(r),b_r(r),0,0\bigr)$, and isolating the exact stress--free closed--VEV branch for which the auxiliary geometry is Ricci flat. In this approach, $b_t'(r)=0$, while the radial component is fixed by the vacuum norm condition. Configurations with $b_t'(r)\neq0$ are not discarded: they possess a Maxwell--like bumblebee stress tensor and belong to a distinct sourced static sector.

After applying the inverse \textit{metric--affine} deformation, we express the physical spacetime in terms of its areal radius and normalize the timelike Killing vector at spatial infinity. These steps reduce the metric sector to a single invariant deformation, even though the original parametrization contains independent temporal and norm combinations of the bumblebee field. We then determine the curvature singularities, maximal causal extension, Killing horizon, and monopole--like asymptotics of the geometry, distinguishing invariant effects from auxiliary coordinate features. Its physical consequences are examined through the Hawking temperature and response functions, the gravitational Doppler effect, tidal forces, massless scalar perturbations, WKB quasinormal frequencies, and characteristic time-domain evolution. Finally, light deflection and perihelion advance are derived in the normalized areal variables and used to constrain the invariant deformation with Solar System measurements.

The paper is organized as follows. Section~\ref{general} presents the \textit{metric--affine} Standard-Model Extension and its field equations. Section~\ref{traceless} introduces the traceless bumblebee sector. In Sec.~\ref{Application}, we derive the closed VEV static solution and analyze its curvature, causal structure, horizon, asymptotic geometry, and invariant normalization. Sections~V--VII examine thermodynamics, frequency shifts, and tidal effects. Scalar perturbations and their quasinormal and time domain behavior are developed in Secs.~VIII and IX, respectively. Section~X establishes the Solar-System bounds. We summarize the results in Sec.~XI.


\section{Geometric structure of traceless \textit{metric--affine} bumblebee gravity  }
\label{general}

\subsection{\textit{metric--affine} Lorentz-violating sector }
\label{general_sector}

The gravitational sector of the minimal Standard--Model Extension can be formulated without identifying the affine connection with the Levi--Civita connection of the spacetime metric \cite{kostelecky2004gravity}. In this \textit{metric--affine} description, $g_{\mu\nu}$ and $\Gamma^{\lambda}{}_{\mu\nu}$ are independent variables, and the action may be written as
\begin{equation}
\begin{split}
\mathcal{S}_{\mathrm{SME}}={}& \frac{1}{2\kappa^{2}}\int\mathrm{d}^{4}x\sqrt{-g}\, \bigl[(1-u)R(\Gamma)+s^{\mu\nu}R_{\mu\nu}(\Gamma) +t^{\mu\nu\alpha\beta}R_{\mu\nu\alpha\beta}(\Gamma)\bigr] \\ &+\mathcal{S}_{\mathrm{mat}}(g_{\mu\nu},\psi) +\mathcal{S}_{\mathrm{coe}} (g_{\mu\nu},u,s^{\mu\nu},t^{\mu\nu\alpha\beta}),
\end{split}
\label{1}
\end{equation}
where $\kappa^{2}=8\pi G$ and $R(\Gamma)\equiv g^{\mu\nu}R_{\mu\nu}(\Gamma)$. The metric fixes the causal structure experienced by minimally coupled matter, whereas the independent connection determines parallel transport and the affine curvature. Throughout this section, ordinary matter is assumed to be described by $\mathcal{S}_{\mathrm{mat}}$, having no explicit dependence on $\Gamma^{\lambda}{}_{\mu\nu}$. Notice that spinorial matter, or any source coupled directly to torsion or nonmetricity, would modify the connection equation and must be considered separately.

The coefficient $s^{\mu\nu}$ is taken to be symmetric. It therefore selects only $R_{(\mu\nu)}(\Gamma)$, while $t^{\mu\nu\alpha\beta}$ has the algebraic symmetries associated with the Riemann tensor. We set the latter coefficient to zero. Apart from the $t$--puzzle encountered in the gravitational SME \cite{Bonder:2015maa}, a nontrivial $t$ sector prevents the connection equation from being encoded in the auxiliary metric employed below. The sector relevant to the present analysis is
\begin{equation}
\mathcal{S}= \frac{1}{2\kappa^{2}}\int\mathrm{d}^{4}x\sqrt{-g}\, \left[(1-u)R(\Gamma)+s^{\mu\nu}R_{\mu\nu}(\Gamma)\right] +\mathcal{S}_{\mathrm{mat}}+\mathcal{S}_{\mathrm{coe}}.
\label{S2}
\end{equation}
This theory lies within the projectively invariant Ricci--based \textit{metric--affine} class \cite{Afonso:2018mapping,BeltranJimenez:2019acz,BeltranJimenez:2020kjt}. A different realization of spontaneous Lorentz breaking in a \textit{metric--affine} construction is provided by \cite{Kostelecky:2009zr}; its nonlinear completion is not the one adopted here.

With the ordering of the connection indices used in this work, a projective transformation takes the form
\begin{equation}
\Gamma^{\lambda}{}_{\mu\nu} \longrightarrow \Gamma^{\lambda}{}_{\mu\nu}+\delta^{\lambda}_{\mu}A_{\nu},
\label{Proj}
\end{equation}
where $A_{\nu}$ is an arbitrary one-form. For the curvature convention adopted here,
\begin{equation}
R^{\lambda}{}_{\rho\mu\nu}(\Gamma) \longrightarrow R^{\lambda}{}_{\rho\mu\nu}(\Gamma) -2\delta^{\lambda}_{\rho}\partial_{[\mu}A_{\nu]}.
\label{projective_riemann}
\end{equation}
The corresponding change in $R_{\mu\nu}(\Gamma)$ is antisymmetric. Both $R(\Gamma)$ and $s^{\mu\nu}R_{\mu\nu}(\Gamma)$ are therefore invariant under Eq.~\eqref{Proj}. Projective symmetry also identifies the vectorial component of the connection that remains undetermined by the action, preventing it from being mistaken for an additional physical degree of freedom \cite{BeltranJimenez:2019acz,BeltranJimenez:2020kjt}.


\subsection{Connection equation and deformation map }
\label{connection_deformation}

We now derive the relation between the two geometries. Variation of Eq.~\eqref{S2} with respect to the independent connection gives
\begin{equation}
\nabla^{(\Gamma)}_{\alpha} \left(\sqrt{-h}\,h^{\mu\nu}\right) =\sqrt{-h}\left[ T^{\mu}{}_{\alpha\lambda}h^{\nu\lambda} +T^{\lambda}{}_{\lambda\alpha}h^{\mu\nu} -\frac{1}{3}T^{\lambda}{}_{\lambda\beta}h^{\nu\beta} \delta^{\mu}_{\alpha}\right],
\label{fieldequation}
\end{equation}
where $T^{\lambda}{}_{\mu\nu} \equiv2\Gamma^{\lambda}{}_{[\mu\nu]}$
is the torsion tensor. The auxiliary metric appearing in Eq.~\eqref{fieldequation} is introduced through
\begin{equation}
\sqrt{-h}\,h^{\mu\nu} =\sqrt{-g}\left[(1-u)g^{\mu\nu}+s^{\mu\nu}\right].
\label{hg}
\end{equation}
Notice that this definition is possible whenever the symmetric tensor density on the right--hand side is nondegenerate and possesses the appropriate Lorentzian signature. The torsion contained in Eq.~\eqref{fieldequation} belongs to the projective sector. Before a gauge is chosen, the solution reads
\begin{equation}
\Gamma^{\lambda}{}_{\mu\nu} =\left\{^{\,\lambda}_{\mu\nu}\right\}^{(h)} +\delta^{\lambda}_{\mu}A_{\nu},
\label{conn_general}
\end{equation}
where
\begin{equation}
\left\{^{\,\lambda}_{\mu\nu}\right\}^{(h)} =\frac{1}{2}h^{\lambda\rho} \left(\partial_{\mu}h_{\rho\nu} +\partial_{\nu}h_{\rho\mu} -\partial_{\rho}h_{\mu\nu}\right).
\label{christoffel_h}
\end{equation}
Choosing the projective gauge $A_{\nu}=0$, it removes the torsion trace and yields $\Gamma^{\lambda}{}_{\mu\nu} =\left\{^{\,\lambda}_{\mu\nu}\right\}^{(h)}$. The independent connection does not propagate an additional field on this part. It becomes the Levi--Civita connection of $h_{\mu\nu}$, although it is not, in general, compatible with the physical metric. The nonmetricity tensor
\begin{equation}
Q_{\alpha\mu\nu} \equiv\nabla^{(\Gamma)}_{\alpha}g_{\mu\nu}
\label{nonmetricity_definition}
\end{equation}
is therefore determined by the deformation relating $g_{\mu\nu}$ and $h_{\mu\nu}$ instead of by an independent propagating mode, which turns out to be simpler to deal with.

To obtain this deformation explicitly, define the mixed tensor
\begin{equation}
\Delta^{\mu}{}_{\nu} \equiv(1-u)\delta^{\mu}_{\nu}+s^{\mu}{}_{\nu}.
\label{DeltaDef}
\end{equation}
Eq. \eqref{hg} can then be written in matrix notation as
\begin{equation}
\sqrt{-h}\,\hat{h}^{-1} =\sqrt{-g}\,\hat{g}^{-1}\hat{\Delta}.
\label{matrix_hg}
\end{equation}
Taking the determinant of both sides in four dimensions gives
\begin{equation}
(\sqrt{-h})^{4}\det(\hat{h}^{-1}) =(\sqrt{-g})^{4}\det(\hat{g}^{-1})\det\hat{\Delta},
\label{det_step}
\end{equation}
so that we have
\begin{equation}
\sqrt{-h}=\sqrt{-g}\sqrt{\det\hat{\Delta}}.
\label{dethg}
\end{equation}
We select the branch that is continuously connected to $h_{\mu\nu}=g_{\mu\nu}$ when $u=s^{\mu\nu}=0$. Substitution of Eq.~\eqref{dethg} into Eq.~\eqref{matrix_hg} yields the contravariant relation
\begin{equation}
h^{\mu\nu} =\frac{1}{\sqrt{\det\hat{\Delta}}} g^{\mu\alpha}\Delta^{\nu}{}_{\alpha}.
\label{hcontra_general}
\end{equation}
Let $\hat{\Omega}\equiv\hat{\Delta}^{-1}$. Inverting Eq.~\eqref{hcontra_general} gives
\begin{equation}
h_{\mu\nu} =\sqrt{\det\hat{\Delta}}\, g_{\mu\alpha}\Omega^{\alpha}{}_{\nu}.
\label{hk_general}
\end{equation}
The condition that Eqs.~\eqref{hcontra_general} and \eqref{hk_general} define mutually inverse metrics is
\begin{equation}
\delta^{\mu}_{\nu} =\left[(1-u)\delta^{\mu}_{\alpha} +s^{\mu}{}_{\alpha}\right]\Omega^{\alpha}{}_{\nu},
\label{Omega_component}
\end{equation}
or, in an equivalent manner, we have
\begin{equation}
\hat{I}=(1-u)\hat{\Omega}+\hat{s}\,\hat{\Omega}.
\label{Io}
\end{equation}
For a generic symmetric coefficient, its determinant may be represented as
\begin{equation}
\det\hat{\Delta} =\det\left[(1-u)\hat{I}+\hat{s}\right] =\exp\left\{\operatorname{Tr} \ln\left[(1-u)\hat{I}+\hat{s}\right]\right\}.
\label{detformal}
\end{equation}
No universal simplification is available until the
eigenvalue structure of $s^{\mu}{}_{\nu}$ has been specified.

As an algebraic check, consider the rank one configuration $u=0$, and $s^{\mu}{}_{\nu}=\xi b^{\mu}b_{\nu}$. The matrix determinant lemma immediately gives
\begin{equation}
\det(\hat{I}+\xi\hat{b}\hat{b})=1+\xi b^{2}, \qquad b^{2}\equiv b^{\mu}b_{\mu},
\label{rankone_det}
\end{equation}
while the Sherman--Morrison identity gives
\begin{equation}
(\hat{I}+\xi\hat{b}\hat{b})^{-1} =\hat{I}-\frac{\xi}{1+\xi b^{2}}\hat{b}\hat{b}.
\label{rankone_inverse}
\end{equation}

The two metrics are then related by
\begin{align}
h^{\mu\nu} &=\frac{1}{\sqrt{1+\xi b^{2}}} \left(g^{\mu\nu}+\xi b^{\mu}b^{\nu}\right), \label{rankone_hcontra} \\ h_{\mu\nu} &=\sqrt{1+\xi b^{2}} \left(g_{\mu\nu}-\frac{\xi}{1+\xi b^{2}}b_{\mu}b_{\nu}\right).
\label{rankone_hcov}
\end{align}
This example is not yet the traceless bumblebee model. Its purpose is to show how the deformation becomes explicit once the coefficient has a finite--rank tensor structure.


\subsection{Metric equation for the general $u$--$s^{\mu\nu}$ sector }
\label{general_metric_equation}

We next vary Eq.~\eqref{S2} with respect to $g^{\mu\nu}$. At this stage, $u$ and the covariant representative $s_{\mu\nu}$ are regarded as independent coefficient fields; their own dynamical variations belong to $\mathcal{S}_{\mathrm{coe}}$. The metric equation becomes
\begin{equation}
\left(1-u\right)R_{(\mu\nu)}(\Gamma) -\frac{1}{2}g_{\mu\nu} \left[\left(1-u\right)R(\Gamma) +s^{\alpha\beta}R_{\alpha\beta}(\Gamma)\right] +2s^{\beta}{}_{(\mu}R_{\nu)\beta}(\Gamma) =\kappa^{2}T_{\mu\nu}.
\label{equationofmotion1}
\end{equation}
The complete stress--energy tensor is separated into
\begin{equation}
T_{\mu\nu}=T^{(\mathrm{mat})}_{\mu\nu} +T^{(\mathrm{coe})}_{\mu\nu},
\label{Tsplit}
\end{equation}
with
\begin{equation}
T^{(\mathrm{mat})}_{\mu\nu} =-\frac{2}{\sqrt{-g}} \frac{\delta\mathcal{S}_{\mathrm{mat}}}{\delta g^{\mu\nu}}.
\label{matt}
\end{equation}

Several relations required below follow directly from Eq.~\eqref{equationofmotion1}. Contracting it initially with $g^{\mu\nu}$, the terms proportional to $s^{\mu\nu}R_{\mu\nu}$ cancel, leaving
\begin{equation}
(1-u)R(\Gamma)=-\kappa^{2}T, \qquad T\equiv g^{\mu\nu}T_{\mu\nu}.
\label{RT1}
\end{equation}
A second contraction with $s^{\mu\nu}$ gives
\begin{equation}
\left(1-u-\frac{s}{2}\right) s^{\mu\nu}R_{\mu\nu}(\Gamma) +2s^{\beta}{}_{\mu}s^{\mu\nu}R_{\nu\beta}(\Gamma) =\kappa^{2}\left(T^{(s)}-\frac{sT}{2}\right),
\label{scontraction}
\end{equation}
where $s\equiv s^{\mu}{}_{\mu}$, and $T^{(s)}\equiv s^{\mu\nu}T_{\mu\nu}$. To display the next step without obscuring its origin, let us introduce
\begin{equation}
\mathcal{R}_{s} \equiv s^{\mu\nu}R_{\mu\nu}(\Gamma), \qquad \mathcal{R}_{s^{2}} \equiv s^{\beta}{}_{\alpha}s^{\alpha\lambda} R_{\lambda\beta}(\Gamma).
\label{Rs_definitions}
\end{equation}
Eq.~\eqref{scontraction} can then be solved for $\mathcal{R}_{s}$:
\begin{equation}
\mathcal{R}_{s} =\frac{2}{2-s-2u} \left[\kappa^{2}\left(T^{(s)}-\frac{sT}{2}\right) -2\mathcal{R}_{s^{2}}\right],
\label{Rs_solution}
\end{equation}
provided $2-s-2u\neq0$. Substituting Eqs.~\eqref{RT1} and \eqref{Rs_solution} back into Eq.~\eqref{equationofmotion1} yields
\begin{equation}
\begin{split}
&(1-u)R_{(\mu\nu)}(\Gamma) +2s^{\alpha}{}_{(\mu}R_{\nu)\alpha}(\Gamma) +\frac{2}{2-s-2u}g_{\mu\nu} s^{\beta}{}_{\alpha}s^{\alpha\lambda}R_{\lambda\beta}(\Gamma) \\ &\hspace{0.8cm} =\kappa^{2}\left[ T_{\mu\nu}-\frac{1}{2}g_{\mu\nu}T +\frac{g_{\mu\nu}}{2-s-2u} \left(T^{(s)}-\frac{sT}{2}\right)\right].
\end{split}
\label{re}
\end{equation}
Every curvature tensor on the left-hand side of Eq.~\eqref{re} is constructed from the connection; in this manner, $R_{\mu\nu}(\Gamma)=R_{\mu\nu}(h)$. Once $\hat{\Omega}$ is known, Eq.~\eqref{re} becomes a closed equation for $h_{\mu\nu}$ coupled algebraically to $u$, $s^{\mu\nu}$, and the sources. The independent equations obtained by varying the coefficient sector must still be imposed; the metric and connection equations alone do not determine an admissible Lorentz--breaking profile.


\subsection{Einstein-frame representation and weak-breaking limit }
\label{einstein_general}

The definition \eqref{hg} implies the exact density identity
\begin{equation}
\sqrt{-g}\left[(1-u)g^{\mu\nu}+s^{\mu\nu}\right] R_{\mu\nu}(\Gamma) =\sqrt{-h}\,h^{\mu\nu}R_{\mu\nu}(h).
\label{density_identity_general}
\end{equation}
After the connection has been eliminated, the theory can therefore be written as
\begin{equation}
\widetilde{\mathcal{S}} =\frac{1}{2\kappa^{2}}\int\mathrm{d}^{4}x\sqrt{-h}\,R(h) +\widetilde{\mathcal{S}}_{\mathrm{mat}} (h_{\mu\nu},u,s_{\mu\nu},\psi) +\widetilde{\mathcal{S}}_{\mathrm{coe}} (h_{\mu\nu},u,s_{\mu\nu}).
\label{ck}
\end{equation}
The Einstein--Hilbert form of the gravitational term does not eliminate the Lorentz--violating coefficients. Their effect is transferred to the transformed matter and coefficient sectors through the nonlinear metric map \eqref{hcontra_general}--\eqref{hk_general}. The auxiliary metric governs the affine curvature, while minimally coupled matter continues to identify $g_{\mu\nu}$ as the physical metric.

The perturbative relation with the metric SME follows by expanding
\begin{equation}
\hat{\Delta}=\hat{I}+\hat{\varepsilon}, \qquad \hat{\varepsilon}\equiv\hat{s}-u\hat{I}.
\label{epsilon_def}
\end{equation}
Since $\operatorname{Tr}\hat{\varepsilon} = s-4u$, we obtain
\begin{align}
\det\hat{\Delta} &=1+s-4u+\mathcal{O}(u^{2},us,s^{2}), \label{detweak} \\ \sqrt{\det\hat{\Delta}} &=1+\frac{s}{2}-2u+\mathcal{O}(u^{2},us,s^{2}),
\label{sqrtdetweak} \\
\hat{\Delta}^{-1} &=\hat{I}+u\hat{I}-\hat{s} +\mathcal{O}(u^{2},us,s^{2}).
\label{inverseweak}
\end{align}
Eqs.~\eqref{hcontra_general} and \eqref{hk_general} become
\begin{align}
h^{\mu\nu} &=\left(1+u-\frac{s}{2}\right)g^{\mu\nu}+s^{\mu\nu} +\mathcal{O}(u^{2},us,s^{2}), \label{weak_hcontra} \\ h_{\mu\nu} &=\left(1-u+\frac{s}{2}\right)g_{\mu\nu}-s_{\mu\nu} +\mathcal{O}(u^{2},us,s^{2}).
\label{weak_hcov}
\end{align}
At the same order, replacing $\Gamma$ by the Levi--Civita connection of $h_{\mu\nu}$ and discarding the resulting boundary term gives
\begin{equation}
\mathcal{S}_{\mathrm{grav}} =\frac{1}{2\kappa^{2}}\int\mathrm{d}^{4}x\sqrt{-g}\, \left[(1-u)R(g)+s^{\mu\nu}R_{\mu\nu}(g)\right] +\mathcal{O}(u^{2},us,s^{2}).
\label{metricSMEweak}
\end{equation}
The metric and \textit{metric--affine} formulations therefore agree in the gravitational sector at linear order in the coefficients. Their inequivalence remains in the exact deformation and in the nonlinear couplings generated beyond this order.


\section{The traceless \textit{metric--affine} bumblebee sector }
\label{traceless}

\subsection{Dynamical realization of the Lorentz-violating coefficients }
\label{bumblebee_realization}

We now specialize the \textit{metric--affine} construction developed in the preceding section to a vector realization of spontaneous Lorentz symmetry breaking. The coefficient $s^{\mu\nu}$ is generated by the traceless combination
\begin{equation}
s^{\mu\nu} =\xi\left(B^{\mu}B^{\nu}-\frac{1}{4}B^{2}g^{\mu\nu}\right), \qquad B^{2}\equiv g^{\mu\nu}B_{\mu}B_{\nu},
\label{tracel}
\end{equation}
which satisfies $s^{\mu}{}_{\mu}=0$. The coupling $\xi$ controls the strength of the interaction between the bumblebee direction and the symmetric affine Ricci tensor. Unlike the rank one model considered in \cite{Delhom:2021bumblebee,Filho:2022yrk}, the trace of $B^{\mu}B^{\nu}$ has been removed explicitly from Eq.~\eqref{tracel}.

The complete action is
\begin{equation}
\begin{split}
\mathcal{S}_{B}={}& \int\mathrm{d}^{4}x\sqrt{-g}\left\{ \frac{1}{2\kappa^{2}} \left[R(\Gamma) +\xi\left(B^{\mu}B^{\nu}-\frac{1}{4}B^{2}g^{\mu\nu}\right) R_{\mu\nu}(\Gamma)\right] -\frac{1}{4}B_{\mu\nu}B^{\mu\nu} -V(\mathcal{X})\right\} \\ &+\int\mathrm{d}^{4}x\sqrt{-g}\, \mathcal{L}_{\mathrm{mat}}(g_{\mu\nu},\psi),
\end{split}
\label{bumb}
\end{equation}
where
\begin{equation}
B_{\mu\nu}\equiv(\mathrm{d}B)_{\mu\nu} =2\partial_{[\mu}B_{\nu]}, \qquad \mathcal{X}\equiv B^{\mu}B_{\mu}\mp b^{2}.
\label{bumblebee_definitions}
\end{equation}
The use of the exterior derivative in Eq.~\eqref{bumblebee_definitions} ensures that the kinetic term carries no direct dependence on the independent connection. The potential possesses a minimum at $\mathcal{X}=0$, where the vector develops the vacuum value $\langle B_{\mu}\rangle=b_{\mu}$, and  $g^{\mu\nu}b_{\mu}b_{\nu}=\pm b^{2}$. The preferred direction is therefore selected dynamically. General covariance is preserved by the action even though the vacuum is not locally Lorentz invariant.

For comparison with the gravitational sector of the SME
\cite{kostelecky2004gravity}, Eq.~\eqref{bumb} may be written as
\begin{equation}
\begin{split}
\mathcal{S}_{B}={}& \int\mathrm{d}^{4}x\sqrt{-g}\left\{ \frac{1}{2\kappa^{2}} \left[(1-u)R(\Gamma)+s^{\mu\nu}R_{\mu\nu}(\Gamma)\right] -\frac{1}{4}B_{\mu\nu}B^{\mu\nu}-V(\mathcal{X})\right\} \\ &+\int\mathrm{d}^{4}x\sqrt{-g}\, \mathcal{L}_{\mathrm{mat}}(g_{\mu\nu},\psi).
\end{split}
\label{bumblebee_SME_form}
\end{equation}
There are two equivalent ways of distributing the trace between the SME coefficients. The explicitly traceless parametrization is
\begin{equation}
u=0, \qquad s^{\mu\nu}=\xi\left(B^{\mu}B^{\nu} -\frac{1}{4}B^{2}g^{\mu\nu}\right),
\label{traceless_identification}
\end{equation}
whereas the same curvature interaction is obtained from
\begin{equation}
u=\frac{\xi B^{2}}{4}, \qquad s^{\mu\nu}=\xi B^{\mu}B^{\nu}.
\label{rankone_identification}
\end{equation}
The second representation is particularly convenient for calculating the deformation matrix. It does not introduce another coupling: the scalar part removed in Eq.~\eqref{traceless_identification} has only been reassigned to $u$.


\subsection{Exact deformation of the auxiliary metric }
\label{exact_deformation}

Applying the general construction to Eq.~\eqref{rankone_identification}, the inverse deformation matrix becomes
\begin{equation}
\hat{\Omega}^{-1} =\left(1-\frac{\xi B^{2}}{4}\right)\hat{I} +\xi\hat{B}\hat{B}.
\label{Omega_inverse_bumblebee}
\end{equation}
To keep the algebra transparent, let us define
\begin{equation}
X\equiv\xi B^{2}, \qquad \mathcal{A}\equiv1-\frac{X}{4}, \qquad \mathcal{C}\equiv1+\frac{3X}{4}, \qquad \xi'\equiv\frac{\xi}{\mathcal{A}}.
\label{deformation_parameters}
\end{equation}
Eq.~\eqref{Omega_inverse_bumblebee} can then be factorized as
\begin{equation}
\hat{\Omega}^{-1} =\mathcal{A}\left(\hat{I}+\xi'\hat{B}\hat{B}\right),
\label{Omega_factorized}
\end{equation}
so that
\begin{equation}
\det\hat{\Omega}^{-1} =\mathcal{A}^{4} \det\left(\hat{I}+\xi'\hat{B}\hat{B}\right).
\label{ww}
\end{equation}

The remaining determinant can be evaluated either by the matrix determinant lemma or through the trace expansion used in the general formalism. To display the rank one structure explicitly, we consider
\begin{equation}
\det\left(\hat{I}+\xi'\hat{B}\hat{B}\right) =\exp\left\{\operatorname{Tr} \ln\left(\hat{I}+\xi'\hat{B}\hat{B}\right)\right\}.
\label{wi}
\end{equation}
The powers of $\hat{B}\hat{B}$ obey
\begin{align}
\operatorname{Tr}(\xi'\hat{B}\hat{B}) &=\xi'B^{2}, \\ \operatorname{Tr}\left[(\xi'\hat{B}\hat{B})^{2}\right] &=(\xi'B^{2})^{2}, \\ \operatorname{Tr}\left[(\xi'\hat{B}\hat{B})^{n}\right] &=(\xi'B^{2})^{n}.
\label{we}
\end{align}
In other words, we have
\begin{equation}
\begin{split}
\operatorname{Tr} \ln\left(\hat{I}+\xi'\hat{B}\hat{B}\right) &=\sum_{n=1}^{\infty} \frac{(-1)^{n+1}}{n}(\xi'B^{2})^{n} =\ln(1+\xi'B^{2}),
\end{split}
\label{trace_log_sum}
\end{equation}
and the determinant assumes the closed form
\begin{equation}
\det\hat{\Omega}^{-1} =\mathcal{A}^{4}(1+\xi'B^{2}) =\mathcal{A}^{3}\mathcal{C} =\left(1-\frac{\xi B^{2}}{4}\right)^{3} \left(1+\frac{3\xi B^{2}}{4}\right).
\label{Omega_determinant}
\end{equation}

The contravariant auxiliary metric now follows immediately from the general map:
\begin{equation}
h^{\mu\nu} =\frac{1}{\sqrt{\mathcal{A}\mathcal{C}}} \left(g^{\mu\nu} +\frac{\xi}{\mathcal{A}}B^{\mu}B^{\nu}\right).
\label{metric1}
\end{equation}
For completeness, the inverse deformation can also be derived without invoking a matrix identity. Since Eq.~\eqref{Omega_inverse_bumblebee} contains only the identity and $\hat{B}\hat{B}$, take
\begin{equation}
\hat{\Omega}=A\hat{I}+C\hat{B}\hat{B}.
\label{Omega_ansatz}
\end{equation}
The condition $\hat{\Omega}^{-1}\hat{\Omega}=\hat{I}$ gives two independent algebraic equations,
$\mathcal{A}A=1$, and   $\mathcal{A}C+\xi A+\xi B^{2}C=0$. Their solution is
$A= 1/\mathcal{A}$, and $C=-\xi/\mathcal{A}\mathcal{C}$,
and therefore
\begin{equation}
\hat{\Omega} =\frac{1}{\mathcal{A}}\hat{I} -\frac{\xi}{\mathcal{A}\mathcal{C}} \hat{B}\hat{B}.
\label{Omega_explicit}
\end{equation}
The covariant auxiliary metric reads
\begin{equation}
h_{\mu\nu} =\sqrt{\mathcal{A}\mathcal{C}}\,g_{\mu\nu} -\xi\sqrt{\frac{\mathcal{A}}{\mathcal{C}}} B_{\mu}B_{\nu},
\label{metric2}
\end{equation}
while the inverse relation is
\begin{equation}
g_{\mu\nu} =\frac{1}{\sqrt{\mathcal{A}\mathcal{C}}}\,h_{\mu\nu} +\frac{\xi}{\mathcal{C}}B_{\mu}B_{\nu}.
\label{physical_metric_reconstruction}
\end{equation}
Eqs.~\eqref{metric1} and \eqref{metric2} exhibit the disformal character of the metric map. The conformal factors depend only on the bumblebee norm, whereas the tensor $B_{\mu}B_{\nu}$ retains the orientation selected by the Lorentz--violating vacuum.

Notice that the map is nondegenerate only if
$\mathcal{A}\neq 0$, and $ \mathcal{C}\neq 0$. The sufficient requirements $\mathcal{A}>0$ and $\mathcal{C}>0$ preserve the Lorentzian signature on the branch connected continuously to general relativity, restricting the invariant combination $X$ to
$-\frac{4}{3}<X<4$. A null vector provides a useful limiting case. Although $B^{2}=0$ makes $\det\hat{\Omega}^{-1}=1$, the disformal term remains in Eqs.~\eqref{metric1} and \eqref{metric2}. The null norm sector is therefore not identical to the Lorentz--symmetric limit unless $\xi B_{\mu}B_{\nu}$ also vanishes.


\subsection{Metric equation and its independent projections }
\label{projected_metric_equations}

Because the coefficient in Eq.~\eqref{tracel} depends explicitly on the physical metric, the variation must include the metric dependence carried by $B^{\mu}$ and $B^{2}$. A direct variation of Eq.~\eqref{bumb} gives
\begin{equation}
\begin{split}
&\left(1-\frac{\xi B^{2}}{4}\right)R_{(\mu\nu)}(\Gamma) -\frac{1}{2}g_{\mu\nu}R(\Gamma) +2\xi B^{\alpha}B_{(\mu}R_{\nu)\alpha}(\Gamma) -\frac{\xi}{4}B_{\mu}B_{\nu}R(\Gamma) \\ &\hspace{1.0cm} -\frac{\xi}{2}g_{\mu\nu}B^{\alpha}B^{\beta} R_{\alpha\beta}(\Gamma) +\frac{\xi}{8}B^{2}g_{\mu\nu}R(\Gamma) =\kappa^{2}T_{\mu\nu}.
\end{split}
\label{kjh}
\end{equation}
The total stress--energy tensor is
\begin{equation}
T_{\mu\nu} = T^{(\mathrm{mat})}_{\mu\nu}+T^{(B)}_{\mu\nu},
\label{stress_split}
\end{equation}
where $T^{(\mathrm{mat})}_{\mu\nu}$ is defined as in Eq.~\eqref{matt} and
\begin{equation}
T^{(B)}_{\mu\nu} =B_{\mu\alpha}B_{\nu}{}^{\alpha} -\frac{1}{4}g_{\mu\nu}B_{\alpha\beta}B^{\alpha\beta} -Vg_{\mu\nu}+2V'B_{\mu}B_{\nu},
\label{bumblebee_stress}
\end{equation}
with $V'\equiv\mathrm{d}V/\mathrm{d}\mathcal{X}$.

The independent curvature projections can be isolated without choosing a spacetime symmetry. Introduce
\begin{equation}
\mathcal{V}_{\nu} \equiv B^{\mu}R_{\mu\nu}(\Gamma), \qquad \mathcal{Q} \equiv B^{\mu}B^{\nu}R_{\mu\nu}(\Gamma), \qquad \mathcal{T}_{B} \equiv B^{\mu}B^{\nu}T_{\mu\nu}.
\label{projection_definitions}
\end{equation}
The trace of Eq.~\eqref{kjh} is independent of $\xi$ and gives
\begin{equation}
R(\Gamma)=-\kappa^{2}T, \qquad T\equiv g^{\mu\nu}T_{\mu\nu}.
\label{RT}
\end{equation}
Contracting Eq.~\eqref{kjh} twice with $B^{\mu}$ produces
\begin{equation}
\left(1+\frac{5\xi B^{2}}{4}\right)\mathcal{Q} -\frac{B^{2}}{8}\left(4+\xi B^{2}\right)R(\Gamma) =\kappa^{2}\mathcal{T}_{B}.
\label{double_projection_intermediate}
\end{equation}
Using Eq.~\eqref{RT}, we get
\begin{equation}
B^{\mu}B^{\nu}R_{\mu\nu}(\Gamma) =\frac{4\kappa^{2}}{4+5\xi B^{2}} \left[ B^{\mu}B^{\nu}T_{\mu\nu} -\frac{B^{2}T}{8}\left(4+\xi B^{2}\right) \right].
\label{b2}
\end{equation}
The single projection of the metric equation may be arranged as
\begin{equation}
\left(1+\frac{3\xi B^{2}}{4}\right)\mathcal{V}_{\nu} +\frac{\xi}{2}B_{\nu}\mathcal{Q} -\left(\frac{1}{2}+\frac{\xi B^{2}}{8}\right) B_{\nu}R(\Gamma) =\kappa^{2}T_{\mu\nu}B^{\mu}.
\label{single_projection_intermediate}
\end{equation}
Substitution of Eqs.~\eqref{RT} and \eqref{b2} yields
\begin{equation}
\begin{split}
B^{\mu}R_{\mu\nu}(\Gamma) =\frac{4\kappa^{2}}{4+3\xi B^{2}} \Bigg\{&T_{\mu\nu}B^{\mu}-\frac{1}{2}B_{\nu}T \\ &-\frac{2\xi B_{\nu}}{4+5\xi B^{2}} \left[ B^{\alpha}B^{\beta}T_{\alpha\beta} -\frac{B^{2}T}{4} \left(1-\frac{3\xi B^{2}}{4}\right) \right]\Bigg\}.
\end{split}
\label{b1}
\end{equation}
For later use, define the source vector appearing as
\begin{equation}
\begin{split}
\mathcal{F}_{\nu}\equiv{}& T_{\mu\nu}B^{\mu}-\frac{1}{2}B_{\nu}T -\frac{2\xi B_{\nu}}{4+5\xi B^{2}} \left[ B^{\alpha}B^{\beta}T_{\alpha\beta} -\frac{B^{2}T}{4} \left(1-\frac{3\xi B^{2}}{4}\right) \right].
\end{split}
\label{F_source}
\end{equation}
Eqs.~\eqref{b1} and \eqref{F_source} then imply
\begin{equation}
\mathcal{V}_{\nu} =\frac{4\kappa^{2}}{4+3\xi B^{2}}\mathcal{F}_{\nu}.
\label{V_in_terms_F}
\end{equation}

Substituting the three projections back into Eq.~\eqref{kjh}, and recalling that $R_{\mu\nu}(\Gamma)=R_{\mu\nu}(h)$, gives the Einstein-like equation
\begin{equation}
\begin{split}
R_{\mu\nu}(h) =\kappa_{\mathrm{eff}}^{2}\Bigg\{& T_{\mu\nu}-\frac{1}{2}g_{\mu\nu}T +\frac{2\xi g_{\mu\nu}}{4+5\xi B^{2}} \left[ B^{\alpha}B^{\beta}T_{\alpha\beta} -\frac{B^{2}T}{16}\left(4-3\xi B^{2}\right) \right] \\ &-\frac{\xi}{4}B_{\mu}B_{\nu}T -\frac{8\xi}{4+3\xi B^{2}} B_{(\mu}\mathcal{F}_{\nu)} \Bigg\},
\end{split}
\label{RR}
\end{equation}
where
$\kappa_{\mathrm{eff}}^{2} \equiv\frac{\kappa^{2}}{1-\frac{\xi B^{2}}{4}}$. Eq.~\eqref{RR} is written entirely in terms of the physical metric, $B_{\mu}$, and the matter variables on its right-hand side. Using Eqs.~\eqref{metric1} and \eqref{metric2}, all occurrences of $g_{\mu\nu}$ can instead be expressed through $h_{\mu\nu}$ and the bumblebee field, yielding a closed Einstein frame equation for the auxiliary geometry.

The divisions used in Eqs.~\eqref{b1}, \eqref{b2}, and \eqref{RR} require
$4+3\xi B^{2}\neq 0$, $4+5\xi B^{2}\neq0$, and $4-\xi B^{2} \neq 0$. The first and third conditions coincide with the invertibility of the metric map. The value $4+5\xi B^{2}=0$ defines a distinct algebraic branch: the double projection \eqref{double_projection_intermediate} remains meaningful there, but it cannot be solved for $\mathcal{Q}$ by division. Any solution lying on this branch must be studied directly from the unsolved field equations.


\subsection{Bumblebee dynamics and the effective current }
\label{bumblebee_dynamics}

Variation of Eq.~\eqref{bumb} with respect to $B_{\mu}$ gives
\begin{equation}
\nabla^{(g)}_{\mu}B^{\mu\alpha} =-\frac{\xi}{\kappa^{2}} g^{\nu\alpha}B^{\mu}R_{\mu\nu}(\Gamma) +\frac{\xi}{4\kappa^{2}}B^{\alpha}R(\Gamma) +2V'B^{\alpha},
\label{bumb2}
\end{equation}
where $\nabla^{(g)}$ is the Levi--Civita derivative of the physical metric. Inserting Eqs.~\eqref{RT} and \eqref{b1} into Eq.~\eqref{bumb2} produces a Proca--like equation,
\begin{equation}
\nabla^{(g)}_{\mu}B^{\mu\alpha} =\mathcal{M}^{\alpha}{}_{\nu}B^{\nu},
\label{proca}
\end{equation}
with the effective mass squared tensor
\begin{equation}
\begin{split}
\mathcal{M}^{\alpha}{}_{\nu} =\Bigg\{&2V' +\frac{\xi T(4-3\xi B^{2})} {4(4+3\xi B^{2})} \\ &+\frac{8\xi^{2}} {(4+3\xi B^{2})(4+5\xi B^{2})} \left[ B^{\mu}B^{\lambda}T_{\mu\lambda} -\frac{B^{2}T}{4} \left(1-\frac{3\xi B^{2}}{4}\right) \right]\Bigg\}\delta^{\alpha}_{\nu} \\ &-\frac{4\xi}{4+3\xi B^{2}}T^{\alpha}{}_{\nu}.
\end{split}
\label{Mat}
\end{equation}
Naturally, as it is straightforward to conclude, the tensor $\mathcal{M}^{\alpha}{}_{\nu}$ is not a constant Proca mass. It is a local quantity determined by the potential, the bumblebee norm, and the matter distribution. The last term makes the response sensitive to the principal directions of $T^{\alpha}{}_{\nu}$, while the terms proportional to $T$ and $B^{\mu}B^{\nu}T_{\mu\nu}$ distinguish the trace and preferred--vector projections of the source.

This structure permits a mechanism analogous to spontaneous vectorization near sufficiently compact matter configurations \cite{Ramazanoglu:2017xbl,Ramazanoglu:2019jrr}. In a local frame where $\mathcal{M}^{\alpha}{}_{\nu}$ is diagonalizable with a real spectrum, a negative eigenvalue in a physical vector sector produces a tachyonic--like mode. A negative determinant can indicate an odd number of negative eigenvalues under these assumptions, but it should not by itself be regarded as a complete stability criterion. A definitive analysis also requires the kinetic operator, the background profile, the boundary conditions and so forth.

The antisymmetry of $B^{\mu\alpha}$ implies the differential identity
\begin{equation}
\nabla^{(g)}_{\alpha}\nabla^{(g)}_{\mu}B^{\mu\alpha}=0.
\label{antisymmetric_identity}
\end{equation}
Taking the divergence of Eq.~\eqref{proca} therefore gives
\begin{equation}
\nabla^{(g)}_{\mu}J^{\mu}=0,
\label{fg}
\end{equation}
where the effective current is
\begin{equation}
J^{\mu}\equiv\mathcal{M}^{\mu}{}_{\nu}B^{\nu}.
\label{current_definition}
\end{equation}
Eq.~\eqref{fg} is an integrability condition inherited from the bumblebee equation. It is particularly useful once spacetime symmetries are imposed. For example, if a static and spherically symmetric configuration carries only a radial current, then
\begin{equation}
\nabla^{(g)}_{\mu}J^{\mu}=0 \quad\Longrightarrow\quad \partial_{r}\left(\sqrt{-g}\,J^{r}\right)=0, \qquad \sqrt{-g}\,J^{r}=\mathcal{Q}_{B},
\label{radial_current}
\end{equation}
where $\mathcal{Q}_{B}$ is constant. Regularity at a center or at a regular horizon frequently selects the zero--current $\mathcal{Q}_{B}=0$, though this conclusion must be checked for the geometry under consideration.

At the minimum of a smooth potential, with $V=V'=0$, $B_{\mu\nu}=0$, and no ordinary matter, we have $T_{\mu\nu}=0$, $\mathcal{M}^{\alpha}{}_{\nu}=0$, and $J^{\mu}=0$. The projected equations then reduce to
\begin{equation}
R(\Gamma)=0, \qquad B^{\mu}R_{\mu\nu}(\Gamma)=0, \qquad B^{\mu}B^{\nu}R_{\mu\nu}(\Gamma)=0,
\label{vacuum_projections}
\end{equation}
and Eq.~\eqref{kjh} gives $R_{\mu\nu}(h)=0$. This final relation determines the most efficient route to exact solutions: we solve the Ricci flat problem for the auxiliary metric and subsequently reconstructs the physical spacetime through Eq.~\eqref{physical_metric_reconstruction}. If the potential is implemented with a Lagrange multiplier, $V'$ need not vanish on the norm constraint, and the vacuum reduction must be reconsidered with its associated effective source.


\section{General static and spherically symmetric solution }
\label{Application}

\subsection{The new solution }

This section is devoted to obtaining a static and spherically symmetric solution for the metric-affine traceless bumblebee model discussed above. We restrict our attention to vacuum configurations, so that $T_{\mu\nu}^{(\mathrm{mat})}=0$. In addition, we assume that the bumblebee field is frozen at its vacuum expectation value, namely, $\langle B_\mu\rangle=b_\mu$, which implies $V=0$ and $V'=0$.

As in the previous discussion, it is more convenient to work in the Einstein frame, since Eq.~(\ref{RR}) is the dynamical equation for the auxiliary metric $h_{\mu\nu}$. For static and spherically symmetric geometries, we take
\begin{equation}
\mathrm{d}s^2_{(h)} = -e^{2\sigma(r)}\mathrm{d}t^2 + e^{-2\rho(r)}\mathrm{d}r^2 + r^2\left(\mathrm{d}\theta^2+\sin^2\theta\,\mathrm{d}\phi^2\right),
\end{equation}
where $\sigma(r)$ and $\rho(r)$ are the metric functions.

Instead of assuming from the outset a purely radial vacuum expectation value, let us begin with a general static and spherically symmetric configuration compatible with the spacetime symmetries,
\begin{equation}
b_\mu=\bigl[b_t(r),b_r(r),0,0\bigr].
\label{bumbp_general}
\end{equation}
For this ansatz, the field strength associated with $b_\mu$ is
\begin{equation}
b_{\mu\nu}=(\mathrm{d}b)_{\mu\nu}, \qquad b_{tr}=-b_t'(r),
\end{equation}
so that the only potentially nonvanishing component is governed by the radial derivative of the temporal piece. Therefore, whenever $b_t'(r)\neq 0$, the Maxwell--like contribution of the bumblebee sector does not vanish. In particular, the vacuum stress--energy tensor becomes
\begin{equation}
T_{\mu\nu} = b_{\mu\alpha}b_{\nu}{}^{\alpha} -\frac{1}{4}g_{\mu\nu}b_{\alpha\beta}b^{\alpha\beta},
\end{equation}
which is traceless, $T=0$, but nonzero if $b_t'(r)\neq 0$. In this case, the Einstein--frame field equations no longer reduce to the vacuum condition $R_{\mu\nu}(h)=0$. In order to remain within the vacuum, which leads to a Schwarzschild solution for $h_{\mu\nu}$, we must impose
\begin{equation}
b_t'(r)=0, \qquad \Longrightarrow\qquad b_t(r)=b_0=\mathrm{constant}.
\end{equation}
This restriction should be interpreted precisely. It is required in order to remain on the closed VEV, stress--free case for which $b_{\mu\nu}=0$ and the auxiliary geometry is Ricci flat. It does not constitute a no--go theorem for static solutions with $b_t'(r)\neq0$; those configurations source the auxiliary geometry through the Maxwell--like bumblebee stress tensor and require a separate analysis. In this manner, the admissible general case is
\begin{equation}
b_\mu=\bigl[b_0,b_r(r),0,0\bigr].
\label{bumbp_branch}
\end{equation}

For this case, $b_{\mu\nu}=0$, since the temporal component is constant and the radial one depends only on $r$. Thereby, the bumblebee stress--energy tensor vanishes identically, and so does the conserved current:
\begin{equation}
T_{\mu\nu}=0, \qquad J_\mu=0.
\end{equation}
In this way, the vacuum field equations in the Einstein frame reduce to $R_{\mu\nu}(h)=0$. The corresponding solution is the Schwarzschild line element,
\begin{equation}
\mathrm{d}s^2_{(h)} = -f(r)\,\mathrm{d}t^2 + \frac{\mathrm{d}r^2}{f(r)} + r^2\left(\mathrm{d}\theta^2+\sin^2\theta\,\mathrm{d}\phi^2\right),
\label{hm_general}
\end{equation}
where $f(r)=1-\frac{2M}{r}$. At this stage, it is convenient to introduce the norm in the Einstein frame,
\begin{equation}
\tilde b^2=h^{\mu\nu}b_\mu b_\nu = -\frac{b_0^2}{f(r)}+f(r)\,b_r^2,
\end{equation}
which must remain constant throughout the spacetime. This condition yields
\begin{equation}
b_r(r) = \frac{\sqrt{b_0^2+\tilde b^2 f(r)}}{f(r)}.
\end{equation}
The vacuum expectation value can be written as
\begin{equation}
b_\mu = \left[ b_0,\, \frac{\sqrt{b_0^2+\tilde b^2 f(r)}}{f(r)}, \,0,\,0 \right].
\label{vev_general}
\end{equation}

As before, $b^2$ and $\tilde b^2$ are algebraically related by
\begin{equation}
\tilde b^2 = b^2 \frac{\left(1+\frac{3\xi b^2}{4}\right)^{1/2}} {\left(1-\frac{\xi b^2}{4}\right)^{3/2}}.
\end{equation}
It is then convenient to define
\begin{equation}
X=\xi b^2, \qquad Y=\xi b_0^2, \qquad \Delta(X)=\sqrt{\left(1+\frac{3X}{4}\right)\left(1-\frac{X}{4}\right)}, \qquad \Gamma(X)=\sqrt{\frac{1+\frac{3X}{4}}{\left(1-\frac{X}{4}\right)^3}},
\end{equation}
so that
\begin{equation}
\tilde b^2=\frac{X}{\xi}\,\Gamma(X).
\end{equation}

The crucial point is to invert Eq.~(\ref{metric2}) before substituting the ansatz for $b_\mu$. Using Eq.~(\ref{metric2}), we obtain
\begin{equation}
g_{\mu\nu} = \frac{1}{\Delta(X)}\,h_{\mu\nu} + \frac{\xi}{1+\frac{3X}{4}}\,b_\mu b_\nu.
\label{ginv_from_h}
\end{equation}
Substituting Eqs.~(\ref{hm_general}) and (\ref{vev_general}) into Eq.~(\ref{ginv_from_h}), the physical metric $g_{\mu\nu}$ takes the form
\begin{equation}
\mathrm{d}s^2_{(g)} = -A(r)\,\mathrm{d}t^2 + 2\,C(r)\,\mathrm{d}t\,\mathrm{d}r + B(r)\,\mathrm{d}r^2 + \frac{r^2}{\Delta(X)}\left(\mathrm{d}\theta^2+\sin^2\theta\,\mathrm{d}\phi^2\right),
\label{metric_nondiag}
\end{equation}
where
\begin{equation}
A(r) = \frac{f(r)}{\Delta(X)} - \frac{Y}{1+\frac{3X}{4}},
\label{Afunc_general}
\end{equation}
\begin{equation}
C(r) = \frac{\xi b_0}{1+\frac{3X}{4}}\, \frac{\sqrt{b_0^2+\tilde b^2 f(r)}}{f(r)},
\label{Cfunc_general}
\end{equation}
and
\begin{equation}
B(r) = \frac{\Gamma(X)}{f(r)} + \frac{Y}{\left(1+\frac{3X}{4}\right)f(r)^2}.
\label{Bfunc_general}
\end{equation}

The dependence on the temporal component of the vacuum expectation value is entirely encoded in the new constant $Y=\xi b_0^2$. Moreover, the sign of $b_0$ affects only the sign of $C(r)$ and can be absorbed by reversing the time orientation. Therefore, after diagonalization, only $Y$ remains relevant. To remove the mixed term, we introduce the new time coordinate
\begin{equation}
\mathrm{d}T = \mathrm{d}t-\frac{C(r)}{A(r)}\,\mathrm{d}r.
\label{time_redef}
\end{equation}
In terms of $T$, the metric assumes the diagonal form
\begin{equation}
\mathrm{d}s^2_{(g)} = -A(r)\,\mathrm{d}T^2 + \frac{\mathrm{d}r^2}{\left(1-\frac{X}{4}\right)^2A(r)} + \frac{r^2}{\Delta(X)}\left(\mathrm{d}\theta^2+\sin^2\theta\,\mathrm{d}\phi^2\right).
\label{metric_general}
\end{equation}
The radial coefficient above follows from the exact identity
\begin{equation}
B(r)+\frac{C(r)^2}{A(r)} = \frac{1}{\left(1-\frac{X}{4}\right)^2A(r)}.
\label{compact_B}
\end{equation}
It is also convenient to rewrite $A(r)$ as
\begin{equation}
A(r) = \frac{1}{\Delta(X)} \left[ f(r) - Y\sqrt{\frac{1-\frac{X}{4}}{1+\frac{3X}{4}}} \right].
\label{Afunc_compact}
\end{equation}

In other words, the generic vacuum configuration of the bumblebee field still gives rise to a Schwarzschild solution in the Einstein frame, whereas in the \textit{metric--affine} frame the geometry is described by the diagonal line element in Eq.~(\ref{metric_general}). The purely radial part (of the bumblebee field, i.e., only with $b_{\mu}= (0,b_{r}(r),0,0)$) is immediately recovered for $Y=0$, in which case $C(r)=0$ and the previous solution is obtained.

Accordingly, the general vacuum is characterized by the three constants $(M,X,Y)$. In addition, the reality of the solution requires $1-\frac{X}{4}>0$, and $1+\frac{3X}{4}>0$, that is, $-\frac{4}{3}<X<4$. Moreover, under the choice $\xi>0$, the radial component $b_r(r)$ remains real provided that $Y+X\Gamma(X)f(r)\ge 0$ in the region under consideration.

In addition, Ref.~\cite{Filho:2022yrk} derived a bumblebee black hole in the \textit{metric--affine} approach by considering the simplest vacuum expectation value of the bumblebee field, namely, $b_{\mu}=(0,b_{r}(r),0,0)$. In the present work, we extend this construction by adopting the more general configuration $b_{\mu}=(b_{t}(r),b_{r}(r),0,0)$ (motivated by a recent result reported in the literature \cite{Zhu:2025fiy}). As discussed previously, the temporal component must be constant in order to ensure current conservation and the validity of the vacuum conditions.


\subsection{Kretschmann invariant and spacetime singularities }

A curvature analysis is essential because the horizon discussion below assumes that the zero of $A(r)$ is a coordinate singularity instead of a curvature singularity. The most transparent expressions follow after the physical normalizations introduced explicitly below. Let $R=r/\sqrt{\Delta}$ be the areal radius, let $\mathcal M=M/[\sqrt{\Delta}\,\mathcal N]$, and define
\begin{equation}
D(X,Y)=\frac{1+3X/4}{(1-X/4)\mathcal N(X,Y)},\qquad \mathcal N(X,Y)=1-\chi(X,Y).
\label{Dpredef}
\end{equation}
With the asymptotically unit-normalized Killing time the metric takes the form displayed later in Eq.~\eqref{normalizedmetric}. Its Levi--Civita curvature invariants are
\begin{align}
R[g]&=\frac{2(D-1)}{DR^2},\label{RicciScalarPhys}\\ R_{\mu\nu}R^{\mu\nu}&=\frac{2(D-1)^2}{D^2R^4},\\ \mathcal K_g\equiv R_{\mu\nu\rho\sigma}R^{\mu\nu\rho\sigma} &=\frac{48\mathcal M^2}{D^2R^6} +\frac{16(D-1)\mathcal M}{D^2R^5} +\frac{4(D-1)^2}{D^2R^4}.
\label{curvatureinv}
\end{align}
These invariants are finite at the Killing horizon $R=2\mathcal M$ and diverge only at $R=0$. The center is therefore a genuine curvature singularity, whereas the horizon is regular. The invariants decay at large radius even though the spatial geometry retains a finite conical defect for $D\neq1$. In a \textit{metric--affine} theory this physical metric curvature should also be distinguished from curvature built from the independent connection, which on the present case is the Levi--Civita connection of the auxiliary Schwarzschild metric.


\subsection{Penrose diagram }

The causal structure of the spacetime can be obtained from the radial sector of Eq.~(\ref{metric_general}). Since the metric function has a single positive root at $r=r_h$, the hypersurface $r=r_h$ defines a simple Killing horizon. In the black hole sector considered here, no second positive root appears; therefore, the geometry does not contain an inner Cauchy horizon.

The radial null curves are determined by
\begin{equation}
0 = -A(r)\,\mathrm{d}T^2 + \frac{\mathrm{d}r^2} {\left(1-\frac{X}{4}\right)^2 A(r)} ,
\end{equation}
which gives
\begin{equation}
\frac{\mathrm{d}T}{\mathrm{d}r} = \pm \frac{1} {\left(1-\frac{X}{4}\right) A(r)}.
\end{equation}
In this manner, the tortoise coordinate is introduced through
\begin{equation}
\mathrm{d}r_\ast = \frac{\mathrm{d}r} {\left(1-\frac{X}{4}\right)A(r)}.
\end{equation}
Using the explicit form of $A(r)$, we obtain
\begin{equation}
r_\ast = \frac{\Delta(X)} {\left(1-\frac{X}{4}\right)\left[1-\chi(X,Y)\right]} \left[ r + r_h \ln\left| \frac{r}{r_h}-1 \right| \right] .
\label{tortoise_bumblebee}
\end{equation}
As usual, the logarithmic term sends $r_\ast\to -\infty$ at the horizon, showing that $r=r_h$ is only a coordinate singularity in the diagonal chart.

The corresponding null coordinates are
\begin{equation}
u=T-r_\ast, \qquad v=T+r_\ast.
\end{equation}
In terms of these coordinates, the two dimensional radial sector becomes
\begin{equation}
\mathrm{d}s^2_{(2)} = -A(r)\,\mathrm{d}u\,\mathrm{d}v .
\end{equation}

The surface gravity associated with the Killing horizon is
\begin{equation}
\kappa = \frac{1}{2} \left(1-\frac{X}{4}\right)A'(r_h) = \frac{ \left(1-\frac{X}{4}\right) \left[1-\chi(X,Y)\right] } {2\Delta(X)r_h} = \frac{ \left(1-\frac{X}{4}\right) \left[1-\chi(X,Y)\right]^2 } {4M\Delta(X)} .
\label{kappa_bumblebee}
\end{equation}
We may define Kruskal--like coordinates by
\begin{equation}
U=-e^{-\kappa u}, \qquad  V=e^{\kappa v}.
\end{equation}
They satisfy
\begin{equation}
UV = -\left( \frac{r}{r_h}-1 \right) e^{r/r_h} = \left( 1-\frac{r}{r_h} \right) e^{r/r_h}.
\label{UV_relation_bumblebee}
\end{equation}
Therefore, the horizon is located at $U=0$ or $V=0$, while the curvature singularity at $r=0$ is mapped into $UV=1$. Since this boundary is spacelike, the singularity has the same causal character as in the Schwarzschild diagram.

The compactified null coordinates may be chosen as
\begin{equation}
U=\tan\tilde{U}, \qquad V=\tan\tilde{V}, \qquad -\frac{\pi}{2}<\tilde{U}<\frac{\pi}{2}, \qquad -\frac{\pi}{2}<\tilde{V}<\frac{\pi}{2}.
\end{equation}
Accordingly, the singularity condition $UV=1$ becomes $\tan\tilde{U}\tan\tilde{V}=1$, or, equivalently, $\tilde{U}+\tilde{V} = \pm\frac{\pi}{2}$. In this case, the singularities are represented by horizontal spacelike boundaries in the Penrose diagram.

The diagram is divided into four regions. Region $\mathrm{I}$ is the right asymptotic exterior, where $r>r_h$ and static observers can remain at fixed radial coordinate. Region $\mathrm{III}$ is the second asymptotic exterior, obtained after the maximal analytic extension. Region $\mathrm{II}$ corresponds to the black-hole interior, where $0<r<r_h$ and every future-directed causal trajectory reaches the spacelike singularity at $r=0$. Finally, region $\mathrm{IV}$ is the white-hole interior, representing the time-reversed sector: causal curves emerge from the past spacelike singularity and may enter either exterior region. In addition, the remaining labels identify the conformal boundaries of the extended spacetime. The symbols $\mathscr{I}^{+}_{\mathrm{R}}$ and $\mathscr{I}^{-}_{\mathrm{R}}$ denote, respectively, future and past null infinity in the right exterior region, while $\mathscr{I}^{+}_{\mathrm{L}}$ and $\mathscr{I}^{-}_{\mathrm{L}}$ play the same role in the left exterior region. Thus, outgoing null rays in the exterior regions end at $\mathscr{I}^{+}$, whereas incoming null rays originate from $\mathscr{I}^{-}$. The points $i^{+}_{\mathrm{R}}$ and $i^{+}_{\mathrm{L}}$ represent future timelike infinity, reached by future-directed timelike observers that remain in the corresponding exterior region. Similarly, $i^{-}_{\mathrm{R}}$ and $i^{-}_{\mathrm{L}}$ denote past timelike infinity. The labels $i^{0}_{\mathrm{R}}$ and $i^{0}_{\mathrm{L}}$ indicate spatial infinity in each asymptotic exterior. The surfaces marked by $r=r_h$ are the future and past event horizons, separating the exterior regions from the black-hole and white-hole interiors. Finally, the zigzag spacelike boundaries labeled $r=0$ represent the curvature singularities.

Therefore, the conformal diagram is Schwarzschild--like. The Lorentz--violating parameters modify the position of the horizon and the normalization of the Kruskal coordinates, but they do not change the global causal structure as long as the solution remains in the one-horizon black hole sector.

\begin{figure}[t]
\centering
\includegraphics[width=.66\linewidth]{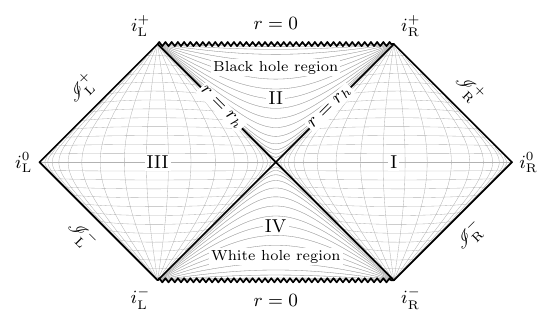}
\caption{Penrose diagram of the maximally extended bumblebee black hole in the black hole sector $a>0$, $c>0$, and $\mathcal{N}(X,Y)>0$. The Lorentz--violating parameters modify the horizon radius, $r_h=2M/\mathcal{N}$, and the Kruskal normalization, while preserving the Schwarzschild--type causal structure.}
\label{fig:penrose_bumblebee}
\end{figure}

One important fact is worthy to be commented: although the angular sector carries the constant factor $\Delta^{-1}$, this contribution changes the area of the two--spheres but not the radial null cones. In other words, the conformal diagram is controlled by the $T$--$r$ sector and remains Schwarzschild--like whenever $\mathcal{N}>0$.


\subsection{The event horizon analysis }

Let us now discuss the event horizon of the generalized solution. Since the metric in Eq.~(\ref{metric_general}) is written in diagonal form, we have
\begin{equation}
g^{rr} = \left(1-\frac{X}{4}\right)^2A(r).
\end{equation}
Thereby, the event horizon is determined by the condition $A(r_h)=0$, which yields
\begin{equation}
f(r_h) = Y\sqrt{\frac{1-\frac{X}{4}}{1+\frac{3X}{4}}}.
\end{equation}
Using $f(r)=1-\frac{2M}{r}$, we get
\begin{equation}
r_h = \frac{2M}{ 1- Y\sqrt{\frac{1-\frac{X}{4}}{1+\frac{3X}{4}}} }.
\label{rh_correct}
\end{equation}

This expression shows that the horizon of the physical metric is shifted with respect to the Schwarzschild radius. In the particular case $Y=0$, we obtain $r_h=2M$, which corresponds to the purely radial part discussed previously. On the other hand, for $Y\neq 0$, the horizon is no longer located at $r=2M$. For the nonnegative $Y$ sector emphasized in the numerical examples, $0\leq f(r)<1$ in the exterior region and the existence of a black hole horizon requires that
\begin{equation}
0\leq Y\sqrt{\frac{1-\frac{X}{4}}{1+\frac{3X}{4}}} <1.
\label{bh_condition_correct}
\end{equation}

\begin{figure}
    \centering
    \includegraphics[scale=0.55]{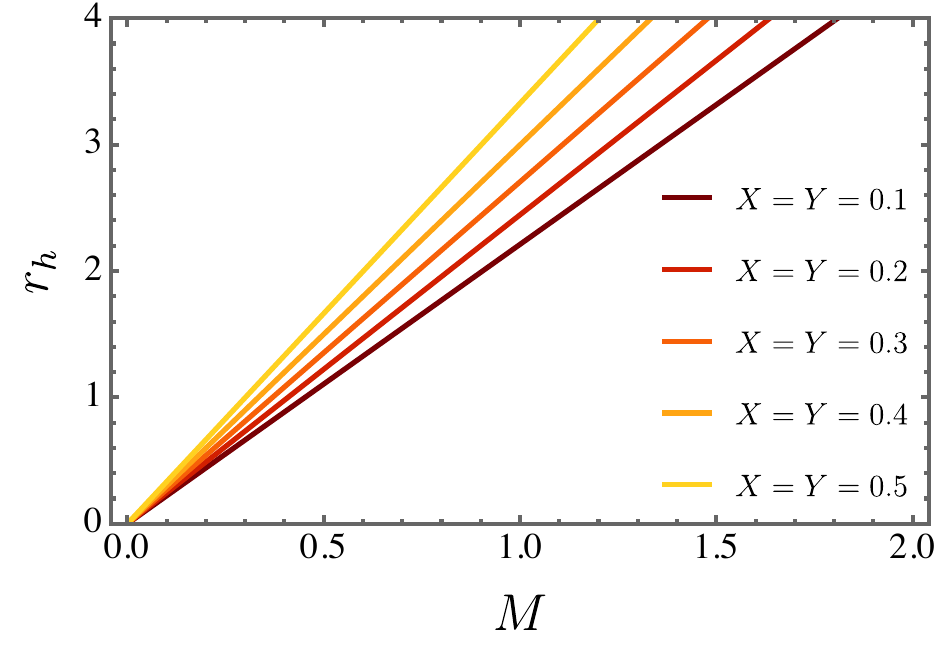}
    \caption{Event horizon radius $r_{h}$ as a function of the mass parameter $M$ for different equal values of the deformation parameters, $X=Y$. }
    \label{eventhorizonrh}
\end{figure}

In Fig.~\ref{eventhorizonrh}, we have the behavior of the event horizon radius $r_{h}$ versus the mass parameter $M$ for different choices of $X=Y$. Note that the curves indicate that the event horizon increases linearly with $M$, whereas higher values of $X=Y$ correspond to larger horizon radii. In addition, a central aspect associated with the event horizon concerns the thermodynamic properties of the system. It is also worth emphasizing that, to the best of our knowledge, the bumblebee black hole solutions available in the literature do not modify the position of the event horizon \cite{Casana:2017jkc,Filho:2022yrk,Zhu:2025fiy}. By contrast, in the present case, the event horizon is indeed shifted, as it is shown in Eq.~\eqref{rh_correct}.

Eq.~\eqref{rh_correct} gives the exact relation, in the original radial coordinate, between the physical horizon position $r_h$ and the auxiliary Schwarzschild integration constant $M$. This coordinate dependence does not represent an invariant departure from the Schwarzschild horizon relation. Indeed, after introducing the areal radius and normalizing the Killing time as described in Sec.~\ref{subsec:physical_normalization}, the same horizon satisfies $R_h=2\mathcal M$. The figure therefore illustrates the horizon behavior within the original parametrization.


\subsection{Monopole--like structure of the Lorentz--violating solution }

It is worth emphasizing that the diagonal solution obtained in Eq.~\eqref{metric_general} has a close structural resemblance to the Schwarzschild black hole endowed with a global monopole charge. In the latter case, the monopole contribution enters the Schwarzschild potential as a constant subtraction, so that the metric function takes the form
\begin{equation}
F_{\rm GM}(r)=1-8\pi\eta^2-\frac{2M}{r},
\end{equation}
where $\eta$ denotes the global monopole charge \cite{Dadhich:1997mh}. In our case, the same type of radial dependence appears in the physical metric through
\begin{equation}
A(r) = \frac{1}{\Delta(X)} \left[ 1-\chi(X,Y)-\frac{2M}{r} \right], \qquad \chi(X,Y) \equiv Y \sqrt{ \frac{1-\frac{X}{4}}      {1+\frac{3X}{4}}} .
\end{equation}
At the level of the lapse function, the Lorentz--violating sector produces an effective monopole--like contribution according to the formal correspondence
\begin{equation}
8\pi\eta^2 \quad \longleftrightarrow \quad \chi(X,Y).
\end{equation}

This analogy explains why the horizon radius obtained from Eq.~\eqref{rh_correct} has the same scaling behavior as the Schwarzschild--global--monopole solution, namely
\begin{equation}
r_h = \frac{2M}{1-\chi(X,Y)} .
\end{equation}
In this way, the constant contribution generated by the temporal component of the bumblebee vacuum expectation value shifts the horizon in the same way that the global monopole charge shifts the Schwarzschild horizon. This is precisely the feature that distinguishes the present configuration from the purely radial bumblebee scenario \cite{Filho:2022yrk}, for which the horizon remains located at the Schwarzschild value.

The resemblance also extends to the semiclassical sector. For the Schwarzschild black hole with global monopole charge, the Hawking temperature is reduced by a factor $(1-8\pi\eta^2)^2$ with respect to the Schwarzschild result \cite{Dadhich:1997mh}. In the present geometry, the temperature can be written as
\begin{equation}
T_H = \sqrt{ \frac{1-\frac{X}{4}}      {1+\frac{3X}{4}} } \, \frac{\left[1-\chi(X,Y)\right]^2}{8\pi M}.
\end{equation}
Notice that the factor $\left[1-\chi(X,Y)\right]^2$ reproduces the same monopole--like scaling, while the additional prefactor comes from the \textit{metric--affine} deformation of the radial sector. In the limit in which this extra radial normalization is removed, the thermal behavior reduces directly to the global monopole pattern under the identification $8\pi\eta^2\leftrightarrow \chi(X,Y)$.

Despite this formal correspondence, the physical origin of the two geometries is different. In the Schwarzschild--global--monopole case, the constant term is associated with the stress--energy distribution of a topological defect and with the loss of asymptotic flatness. Here, on the other hand, no topological charge is introduced. Instead, the constant shift arises from the \textit{metric--affine} relation between the Einstein--frame and physical frame metrics after the bumblebee field acquires a vacuum expectation value with a nonvanishing temporal component. In other words, the present solution may be interpreted as a Lorentz--violating analogue of the Schwarzschild black hole with global monopole charge, with $\chi(X,Y)$ playing the role of an effective monopole--like parameter at the geometrical level.


The monopole analogy admits an invariant formulation. In the normalized variables of Eq.~\eqref{normalizedmetric}, the proper radial distance is asymptotically $\rho\simeq\sqrt D\,R$, and it reads
\begin{equation}
\frac{A(\rho)}{4\pi\rho^2}\longrightarrow\frac{1}{D}.
\end{equation}
Thereby, the case where $D>1$ corresponds to a solid angle deficit and $0<D<1$ to a surplus. This asymptotic conical factor is the invariant content of the
global-monopole-like resemblance \cite{Barriola:1989hx,Ono:2018ybw}. Again, as argued before, the analogy remains purely geometrical: no topological monopole charge is present in the bumblebee construction.


\subsection{Comparison with the purely radial bumblebee configuration: mass and radial normalizations }

Before proceeding, it is useful to clarify a technical point concerning the relation between the present solution and the earlier \textit{metric--affine} bumblebee black hole reported in Ref.~\cite{Filho:2022yrk}. In that case, the vacuum was obtained in the Einstein frame from the Schwarzschild line element
\begin{equation}
\mathrm{d}s^2_{(h)} = -f(r)\mathrm{d}t^2 + \frac{\mathrm{d}r^2}{f(r)} + r^2 \mathrm{d}\Omega^2, \qquad f(r)=1-\frac{2M}{r}.
\label{old_auxiliary_schwarzschild}
\end{equation}
The constant $M$ appearing in Eq.~\eqref{old_auxiliary_schwarzschild} is the integration constant associated with the auxiliary geometry $h_{\mu\nu}$. Since the physical metric $g_{\mu\nu}$ is obtained only after applying the deformation map, this parameter should not be identified, without further qualifications, with the mass parameter appearing after a radial normalization of the physical line element.

Indeed, by taking the purely radial limit of the present solution, namely $Y=0$, the physical metric obtained from the inverse deformation map reads
\begin{equation}
\mathrm{d}s^2_{(g)}\big|_{Y=0} = -\frac{f(r)}{\Delta(X)}\,\mathrm{d}T^2 + \frac{\Gamma(X)}{f(r)}\,\mathrm{d}r^2 + \frac{r^2}{\Delta(X)}\,\mathrm{d}\Omega^2 .
\label{corrected_radial_branch}
\end{equation}
In this manner, the angular sector is not written in the areal--radius form $r^2\mathrm{d}\Omega^2$. If we want to cast Eq.~\eqref{corrected_radial_branch} into a form closer to the one used in Ref.~\cite{Filho:2022yrk}, the radial coordinate must be rescaled according to
\begin{equation}
\bar r=\frac{r}{\sqrt{\Delta(X)}} , \qquad \bar M=\frac{M}{\sqrt{\Delta(X)}} .
\label{radial_mass_redefinitions_old_branch}
\end{equation}
With these redefinitions, we are able to write
\begin{equation}
f(r) = 1-\frac{2M}{r} = 1-\frac{2\bar M}{\bar r}, \qquad \mathrm{d}r^2=\Delta(X)\,\mathrm{d}\bar r^2,
\end{equation}
so that Eq.~\eqref{corrected_radial_branch} becomes
\begin{equation}
\mathrm{d}s^2_{(g)}\big|_{Y=0} = -\frac{1}{\Delta(X)} \left(1-\frac{2\bar M}{\bar r}\right)\mathrm{d}T^2 + \frac{c}{a} \frac{\mathrm{d}\bar r^2} {1-\frac{2\bar M}{\bar r}} + \bar r^2\mathrm{d}\Omega^2 ,
\label{corrected_eq55_like_form}
\end{equation}
where $a = 1-\frac{X}{4}$ and $c = 1+\frac{3X}{4}$, so that $\Delta(X)=\sqrt{ac}$. This expression shows explicitly that the mass parameter in the Schwarzschild factor of the areal--radius form is $\bar M$, not the original $M$ appearing in the auxiliary metric \eqref{old_auxiliary_schwarzschild}. Notice that if the same symbol $M$ is used before and after the transformation, a relabeling has implicitly been performed, as it was done in Ref.~\cite{Filho:2022yrk}.

Eq.~\eqref{corrected_eq55_like_form} also shows that, even after the radial and mass redefinitions, the radial component does not coincide with the expression displayed in Ref.~\cite{Filho:2022yrk}. In the corrected form, the constant prefactor of $g_{\bar r\bar r}$ is $c/a$, whereas the form written directly in terms of the original radial coordinate contains instead $\Gamma(X)$. The difference comes from the transformation of the radial differential, $\mathrm{d}r^2=\Delta(X)\mathrm{d}\bar r^2$, and therefore it cannot be ignored when comparing the two parametrizations.

Analogously, the same point also applies to the full solution obtained in the present work. Starting from the diagonal form
\begin{equation}
\mathrm{d}s^2_{(g)} = -A(r)\mathrm{d}T^2 + \frac{\mathrm{d}r^2}{a^2A(r)} + \frac{r^2}{\Delta(X)}\mathrm{d}\Omega^2,
\end{equation}
and using the same areal--radius redefinition $\bar r=r/\sqrt{\Delta(X)}$, together with
\begin{equation}
\bar M=\frac{M}{\sqrt{\Delta(X)}}, \qquad \mathcal{M} = \frac{\bar M}{\mathcal{N}(X,Y)} = \frac{M}{\sqrt{\Delta(X)}\,\mathcal{N}(X,Y)},
\end{equation}
we have
\begin{equation}
\mathrm{d}s^2_{(g)} = -\frac{\mathcal{N}(X,Y)}{\Delta(X)} \left( 1-\frac{2\mathcal{M}}{\bar r} \right)\mathrm{d}T^2 + \frac{c}{a\,\mathcal{N}(X,Y)} \frac{\mathrm{d}\bar r^2} { 1-\frac{2\mathcal{M}}{\bar r} } + \bar r^2\mathrm{d}\Omega^2.
\label{general_areal_radius_form}
\end{equation}
Accordingly, the horizon position in the areal radial coordinate is
\begin{equation}
\bar r_h=2\mathcal{M} = \frac{2M}{\sqrt{\Delta(X)}\,\mathcal{N}(X,Y)}.
\end{equation}
Again, this makes clear that the mass parameter inherited from the Einstein--frame Schwarzschild solution and the mass parameter appearing after the areal--radius normalization are related, but they are not the same quantity. This distinction is important when comparing the present geometry with the earlier \textit{metric--affine} solution, as well as when using quantities such as horizon radius, shadow radius, or weak--field observables.


\subsection{Physical areal radius, normalized Killing time, and invariant deformation }
\label{subsec:physical_normalization}

The preceding comparison shows that the auxiliary Schwarzschild integration constant and the mass coefficient of an areal--radius representation must not be conflated. This point can be taken one step further and used to expose the invariant metric content of the solution. Let us write
\begin{equation}
a_X=1-\frac{X}{4},\qquad c_X=1+\frac{3X}{4},\qquad \mathcal N(X,Y)=1-\chi(X,Y),
\end{equation}
with $\chi$ defined above, and identify the areal radius with
\begin{equation}
R\equiv\bar r=\frac{r}{\sqrt{\Delta(X)}},\qquad \mathcal M=\frac{M}{\sqrt{\Delta(X)}\,\mathcal N(X,Y)}.
\label{physicalvars}
\end{equation}
The asymptotic norm of the original Killing vector is not unity. We introduce
\begin{equation}
t_\infty=\sqrt{\frac{\mathcal N}{\Delta}}\,T.
\label{tnorm}
\end{equation}
The physical line element then reduces exactly to
\begin{equation}
\mathrm{d}s^2=-F(R)\mathrm{d}t_\infty^2+D\frac{\mathrm{d}R^2}{F(R)}+R^2\mathrm{d}\Omega^2, \qquad F(R)=1-\frac{2\mathcal M}{R},
\label{normalizedmetric}
\end{equation}
where
\begin{equation}
D(X,Y)=\frac{c_X}{a_X\mathcal N(X,Y)} =\frac{1+3X/4}{(1-X/4)\left[1-Y\sqrt{(1-X/4)/(1+3X/4)}\right]}.
\label{Ddef}
\end{equation}
At fixed normalized $1/R$ mass coefficient $\mathcal M$, every observable constructed solely from the physical metric depends on the Lorentz--violating sector through the single combination $D$. The underlying bumblebee configuration still contains the separate information encoded in $X$ and $Y$, so direct matter, connection or bumblebee observables may break this metric degeneracy.

Furthermore, Eq.~\eqref{normalizedmetric} also makes the asymptotics precise:
\begin{equation}
\mathrm{d}s^2\sim-\mathrm{d}t_\infty^2+D\,\mathrm{d}R^2+R^2\mathrm{d}\Omega^2.
\end{equation}
For $D\neq1$ the spacetime is asymptotically conical instead of asymptotically Minkowskian in the usual ADM sense. With proper radial distance $\rho\simeq\sqrt D\,R$, a large sphere has area $4\pi\rho^2/D$. Accordingly, $\mathcal M$ is used below as the normalized lapse mass coefficient; a global conserved mass would require the corresponding asymptotically conical prescription \cite{Nucamendi:1996,Barriola:1989hx}. Finally, the apparently shifted coordinate horizon becomes the invariant areal--radius relation $R_h=2\mathcal M$. The temporal VEV therefore changes the map between auxiliary and physical parameters, while the normalized lapse zero retains the Schwarzschild form.


\section{Thermodynamics }

In this section, we address the thermodynamic properties of the black hole solution. Since the bumblebee parameters modify the location of the event horizon, they also affect the corresponding thermodynamic quantities. In particular, the Hawking temperature, the geometrical horizon--area feature, and its associated thermal response allow us to examine how the Lorentz--violating background changes the stationary thermal behavior of the geometry. The status of the entropy itself requires additional care in the present \textit{metric--affine} nonminimally coupled theory, as discussed below. In this manner, the thermodynamic analysis follows naturally from the horizon discussion and makes it possible to relate the deformation of the solution to its semiclassical physical consequences \cite{Bekenstein:1973ur,Hawking:1975vcx,Wald:1999vt}.


The thermodynamic discussion is most transparent (and evident naturally) in the normalized variables introduced in Sec.~\ref{subsec:physical_normalization}. In this representation, the horizon satisfies $R_h=2\mathcal M$, whereas the Lorentz--violating sector enters the physical metric through the invariant deformation $D(X,Y)$. This distinction is essential because the original coordinate $r$, the Killing coordinate $T$, and the auxiliary integration constant $M$ are not, respectively, the areal radius, the unit normalized time, and the normalized lapse mass coefficient.


\subsection{Hawking temperature }

\begin{figure}
    \centering
    \includegraphics[scale=0.55]{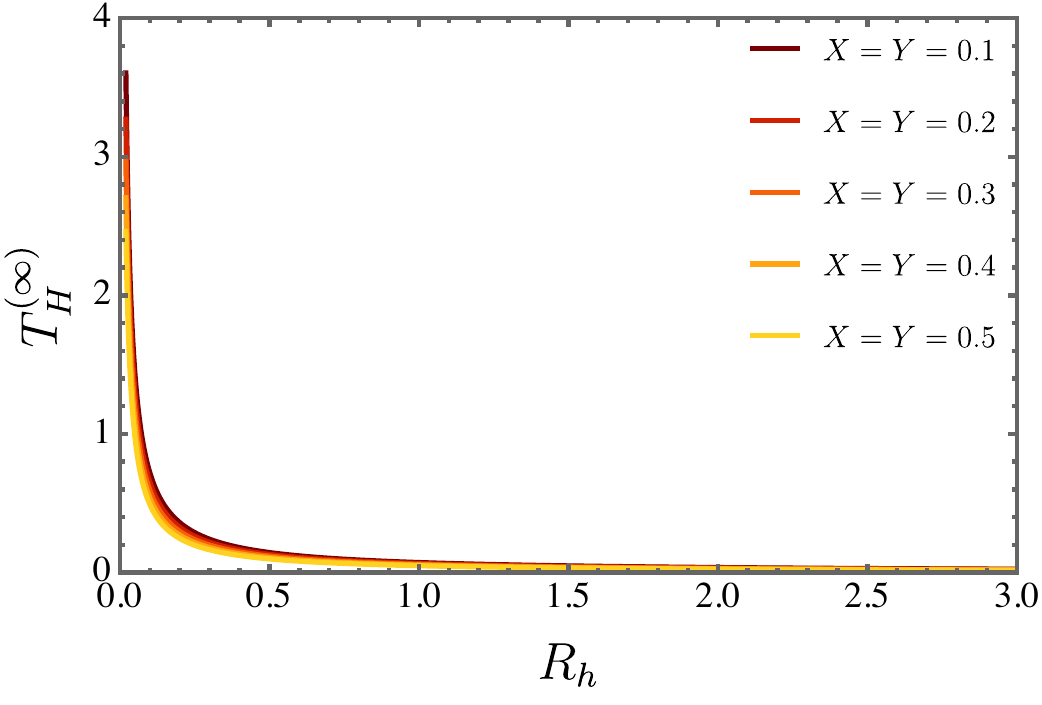}
    \caption{Unit normalized Hawking temperature $T_H^{(\infty)}$ as a function of the areal horizon radius $R_h$ for $X=Y=0.1,0.2,0.3,0.4,0.5$.}
    \label{hakingtevent}
\end{figure}

For the unit normalized line element in Eq.~\eqref{normalizedmetric}, the surface gravity associated with the Killing vector $\partial_{t_\infty}$ is
\begin{equation}
\kappa_\infty=\left.\frac{1}{2}\frac{\partial_R F(R)}{\sqrt{F(R)D/F(R)}}\right|_{R=R_h}=\frac{F'(R_h)}{2\sqrt{D}}=\frac{1}{2R_h\sqrt{D}}=\frac{1}{4\mathcal M\sqrt{D}}.
\label{previsttt}
\end{equation}
The corresponding Hawking temperature is therefore
\begin{equation}
T_H^{(\infty)}=\frac{\kappa_\infty}{2\pi}=\frac{1}{4\pi R_h\sqrt{D}}=\frac{1}{8\pi\mathcal M\sqrt{D}}.
\label{THphys}
\end{equation}
This result may be checked directly in the original parametrization. Writing $a_X=1-X/4$, $c_X=1+3X/4$, $\Delta(X)=\sqrt{a_Xc_X}$, and $\mathcal N(X,Y)=1-Y\sqrt{a_X/c_X}$, the coordinate horizon is $r_h=2M/\mathcal N$. Since $t_\infty=\sqrt{\mathcal N/\Delta}\,T$, the temperature associated with the unnormalized Killing coordinate $T$ is
\begin{equation}
T_H^{(T)}=\sqrt{\frac{\mathcal N}{\Delta}}\,T_H^{(\infty)}=\frac{\mathcal N}{4\pi r_h}\sqrt{\frac{a_X}{c_X}}=\frac{\mathcal N^2}{8\pi M}\sqrt{\frac{a_X}{c_X}}.
\label{HawTem}
\end{equation}
Eq.~\eqref{HawTem} is algebraically equivalent to the expressions obtained directly from the metric written in the $(T,r)$ coordinates, but Eq.~\eqref{THphys} is the quantity measured with respect to the time (normalized) at infinity. In the weak deformation regime, $D=1+X+Y+\mathcal O(X^2,XY,Y^2)$, and Eq.~\eqref{THphys} becomes
\begin{equation}
T_H^{(\infty)}=\frac{1}{4\pi R_h}\left[1-\frac{X+Y}{2}+\mathcal O(X^2,XY,Y^2)\right].
\label{THweak}
\end{equation}
For the illustrative choice $X=Y>0$, the function $D(X,Y)$ increases throughout the interval considered in Fig.~\ref{hakingtevent}; consequently, the temperature decreases as the Lorentz--violating parameters increase at fixed $R_h$. It also decreases monotonically as $R_h$ grows and approaches zero only in the infinite radius limit. Within the admissible sector $a_X>0$, $c_X>0$, and $\mathcal N>0$, one has $D>0$ and $F'(R_h)=1/R_h\neq0$, so the horizon is nondegenerate and its temperature remains positive for every finite $R_h$. The stationary family therefore contains no finite mass extremal configuration and does not indicate a zero temperature remnant. This conclusion concerns the fixed background geometry alone; the late stage of evaporation would require backreaction and the possible evolution of the bumblebee vacuum.

\subsection{Horizon area and area-law feature }

The angular sector of the physical metric determines the horizon area without reference to the radial normalization. Using $R=r/\sqrt{\Delta(X)}$ and $R_h=2\mathcal M$, we obtain
\begin{equation}
A_H=4\pi R_h^2=\frac{4\pi r_h^2}{\Delta(X)}=16\pi\mathcal M^2.
\label{AH}
\end{equation}
In units $G=1$, the corresponding area law quantity is
\begin{equation}
S_A\equiv\frac{A_H}{4}=\pi R_h^2=\frac{\pi r_h^2}{\Delta(X)}=4\pi\mathcal M^2.
\label{SA}
\end{equation}
The factor $\Delta^{-1}$ is required only when the area is written in terms of the original coordinate $r$. At fixed physical mass coefficient $\mathcal M$, Eq.~\eqref{SA} contains no explicit dependence on $D$, $X$, or $Y$; the parameter dependence obtained by plotting $S_A$ against the auxiliary constant $M$ originates entirely from the map $\mathcal M=M/[\sqrt{\Delta}\,\mathcal N]$ and does not represent an invariant correction to the horizon area. Moreover, the present action is first order and contains a nonminimal bumblebee--Ricci coupling. In addition, it it worthy to be mentionned that the complete stationary entropy must therefore be derived from the Noether charge of the full \textit{metric--affine} theory \cite{Wald:1993nt,Iyer:1994ys,Wald:1999vt,Vollick:2007wn}. We use $S_A$ only as a geometric area law feature until that derivation is carried out.


\subsection{Area-law thermal response }

The same qualification applies to the thermal response. A canonical heat capacity requires the conserved energy and the Noether charge entropy to be defined within the same variational prescription. A geometric response may nevertheless be constructed from Eqs.~\eqref{THphys} and \eqref{SA}. Varying $\mathcal M$ while keeping $D$, or equivalently $X$ and $Y$, fixed gives
\begin{equation}
C_A\equiv T_H^{(\infty)}\frac{\mathrm dS_A}{\mathrm dT_H^{(\infty)}}=T_H^{(\infty)}\frac{\mathrm dS_A/\mathrm d\mathcal M}{\mathrm dT_H^{(\infty)}/\mathrm d\mathcal M}=-8\pi\mathcal M^2=-2\pi R_h^2=-\frac{2\pi r_h^2}{\Delta(X)}=-2S_A.
\label{CA}
\end{equation}
The negative sign reproduces the Schwarzschild--like thermal response: a reduction of the mass scale increases the temperature. At fixed $D$, the surrogate first law $\mathrm dE_A=T_H^{(\infty)}\mathrm dS_A$ gives $E_A=\mathcal M/\sqrt D$, up to an additive constant, and Eq.~\eqref{CA} may equivalently be written as $C_A=\mathrm dE_A/\mathrm dT_H^{(\infty)}$. This relation does not replace the conserved charge of the complete theory, but it shows that the area law signature is internally consistent. As in the entropy analysis, plotting $C_A$ against $\mathcal M$ produces a single curve, whereas the separation obtained in terms of the auxiliary parameter $M$ reflects only the parameter transformation. The response remains negative throughout the admissible domain and exhibits neither a divergence nor a change of sign, so no second order transition or locally stable interval arises within this geometric approximation.


%
%

%


\subsection{Thermodynamic topology }

Thermodynamic topology provides a geometric classification of critical phenomena through the zeros of a vector field defined on an auxiliary parameter space \cite{wei2022black,wu2025novel,yerra2022topology,wu2023topological1,zhang2023bulk,sadeghi2024thermodynamic,Wu:2023sue,Wu:2023fcw,Zhu:2024zcl,fang2023revisiting,Afshar:2024bgi}. Two related constructions should be distinguished. The temperature based prescription assigns a winding number to each thermodynamic critical point, with $w=-1$ conventionally associated with a conventional critical point and $w=+1$ with a novel one, whereas the off--shell free energy prescription treats equilibrium black hole configurations as topological defects and relates their local winding numbers to thermal stability. Since the present analysis is constructed directly from the Hawking temperature, we employ the first prescription, based on Duan's $\phi$--mapping topological current theory \cite{Duan:1984,wei2022black}. Applications of this method to different gravitational systems have shown that the resulting topological charge supplies information that is not contained in the location of the critical point alone \cite{gashti2025thermodynamic,zafar2026thermodynamic,rathi2025topology,huang2025interaction,Gashti2025}.

We treat $X$ and $Y$ as fixed parameters of the theory and use the normalized Hawking temperature obtained in Eq.~\eqref{THphys}. To keep the auxiliary parameter space dimensionless, let
\begin{equation}
x\equiv\frac{R_h}{R_0}>0,
\label{dimensionlessRh}
\end{equation}
where $R_0$ is an arbitrary reference scale. The thermodynamic potential is then defined by
\begin{equation}
\Phi(x,\theta)=\frac{T_H^{(\infty)}}{\sin\theta}=\frac{\csc\theta}{4\pi R_0 x\sqrt{D}},
\label{thermotoppotential}
\end{equation}
with $0<\theta<\pi$. The associated vector field in the $(x,\theta)$ plane is
\begin{equation}
\phi^a=\left(\phi^x,\phi^\theta\right)=\left(\partial_x\Phi,\partial_\theta\Phi\right),
\label{thermovector}
\end{equation}
whose components are
\begin{align}
\phi^x&=-\frac{\csc\theta}{4\pi R_0x^2\sqrt D},
\label{phi1}\\
\phi^\theta&=-\frac{\cot\theta\,\csc\theta}{4\pi R_0x\sqrt D}.
\label{phi2}
\end{align}
The normalized field $n^a=\phi^a/\|\phi\|$, with $\|\phi\|=\sqrt{(\phi^x)^2+(\phi^\theta)^2}$, therefore takes the explicit form
\begin{equation}
n^a=-\frac{\left(1,x\cot\theta\right)}{\sqrt{1+x^2\cot^2\theta}}.
\label{normalizedthermovector}
\end{equation}
Eq.~\eqref{normalizedthermovector} shows that the common factor $(4\pi R_0\sqrt D)^{-1}$ cancels identically. The direction of the vector field is independent of $X$ and $Y$ throughout the admissible sector $D>0$.

A thermodynamic critical point would require the simultaneous conditions $\phi^x=0$ and $\phi^\theta=0$. However, Eq.~\eqref{phi1} is strictly negative for every finite $x>0$ and $0<\theta<\pi$. Although $\phi^\theta$ vanishes at $\theta=\pi/2$, the radial component remains nonzero there, and the vector field possesses no finite isolated zero. Both components approach zero only at the asymptotic boundary $x\rightarrow\infty$, which does not define an isolated critical point in the thermodynamic parameter space. The absence of finite zeros follows directly from the monotonic dependence $T_H^{(\infty)}\propto R_h^{-1}$ and cannot be altered by the positive multiplicative factor $D^{-1/2}$.

For a closed contour $C$ contained in the regular domain, the winding number is
\begin{equation}
w(C)=\frac{1}{2\pi}\oint_C\epsilon_{ab}n^a\,\mathrm dn^b.
\label{windingnumber}
\end{equation}
Since the normalized field is smooth and nonvanishing throughout this domain, every such contour can be continuously contracted without crossing a zero, and Eq.~\eqref{windingnumber} gives $w(C)=0$. The total topological charge is therefore zero, and neither a conventional nor a novel thermodynamic critical point occurs within the fixed--$(X,Y)$ family. This conclusion does not depend on the illustrative values assigned to the Lorentz--violating parameters. If $X$ or $Y$ were promoted to thermodynamic variables, their conjugate quantities and the corresponding enlarged thermodynamic potential would have to be established before repeating the topological classification.

\begin{figure}
    \centering
    \includegraphics[scale=0.5]{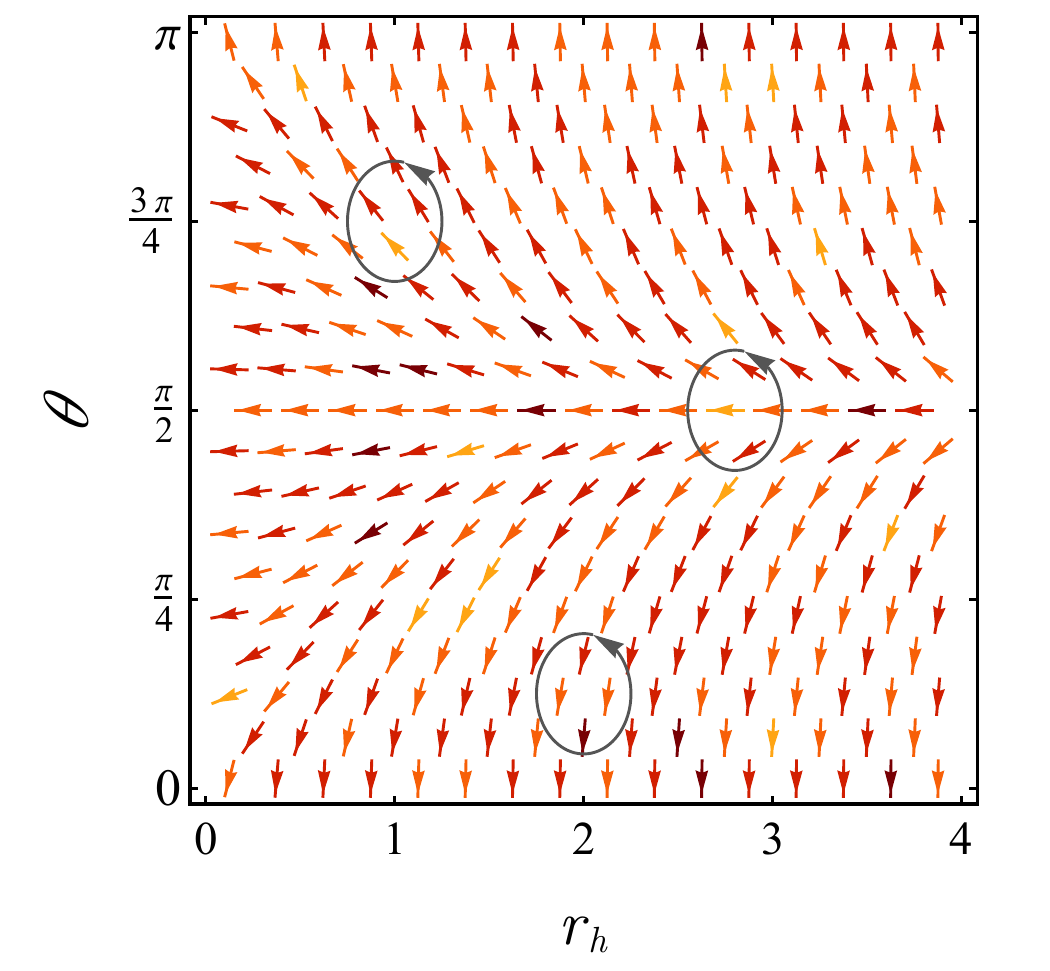}
    \caption{Normalized thermodynamic vector field in the $(x,\theta)$ plane. The same vector portrait applies to every admissible pair $(X,Y)$ because the factor $D^{-1/2}$ cancels from $n^a$. The displayed closed contours enclose no zero and have vanishing winding number.}
    \label{fig:Topology}
\end{figure}


\section{Gravitational Doppler effect}
\label{sec:grav_doppler}

The radial motion and the signal exchange are most clearly described in the normalized variables of Eq.~\eqref{normalizedmetric}, since $t_\infty$ is the Killing time measured at infinity and $R$ is the areal radius. This formulation also separates locally measured quantities from effects produced by the radial normalization \cite{augousti2018speed,radosz2019inside,cordeiro2025free}. For a neutral massive particle following a radial timelike geodesic, the normalization condition is
\begin{equation}
-F(R)\dot t_\infty^{\,2}+\frac{D}{F(R)}\dot R^{\,2}=-1,
\label{norm_radial_doppler}
\end{equation}
where the dot denotes differentiation with respect to the particle's proper time. The conserved energy per unit rest mass associated with $\partial_{t_\infty}$ is
\begin{equation}
\mathcal E=F(R)\dot t_\infty,
\label{energy_doppler}
\end{equation}
and Eq.~\eqref{norm_radial_doppler} becomes
\begin{equation}
D\dot R^{,2}+F(R)=\mathcal E^2.
\label{radial_energy_doppler}
\end{equation}
For an inward trajectory, we have
\begin{equation}
\dot R=-\frac{1}{\sqrt D}\sqrt{\mathcal E^2-F(R)}.
\label{rdot_doppler}
\end{equation}
If the particle is released from rest at the areal radius $R=B>R_h$, its energy is fixed by
\begin{equation}
\mathcal E^2=F(B)=1-\frac{2\mathcal M}{B}.
\label{energy_b_doppler}
\end{equation}
In particular, release from the asymptotic region gives $\mathcal E=1$. The parameter dependence previously obtained for the asymptotic energy in the original $(T,r)$ coordinates was therefore a consequence of the nonunit norm of $\partial_T$ and does not survive the normalization of the Killing time.

The local radial speed measured by a static observer at $R$ is the ratio between the proper radial displacement and the proper time of the static frame,
\begin{equation}
v^2=\frac{D}{F(R)^2}\left(\frac{\mathrm dR}{\mathrm dt_\infty}\right)^2=1-\frac{F(R)}{\mathcal E^2}.
\label{velocity_general_doppler}
\end{equation}
For release from rest at $B$, this gives
\begin{equation}
v^2(R;B)=\frac{2\mathcal M\left(R^{-1}-B^{-1}\right)}{1-2\mathcal M/B}=\frac{2\mathcal M(B-R)}{R(B-2\mathcal M)}.
\label{velocity_b_doppler}
\end{equation}
The particle starts with $v(B;B)=0$, while the family of static observers approaching the horizon measures $v(R;B)\rightarrow1$ as $R\rightarrow R_h^{+}$. No timelike static observer exists on the horizon itself, so this statement must be understood as an exterior limit. For release from the asymptotic region, Eq.~\eqref{velocity_b_doppler} reduces to
\begin{equation}
v_\infty^2(R)=\frac{2\mathcal M}{R}=\frac{R_h}{R}.
\label{velocity_infty_doppler}
\end{equation}
At fixed $(\mathcal M,R,B)$, the local velocity is therefore identical to the Schwarzschild expression and contains no explicit dependence on $D$. The deformation remains present in the relation between the local velocity and the normalized coordinate velocity,
\begin{equation}
\frac{\mathrm dR}{\mathrm dt_\infty}=-\frac{F(R)}{\sqrt D}\,v(R;B),
\label{coordinate_velocity_general_doppler}
\end{equation}
which vanishes as $R\rightarrow R_h^{+}$. It also modifies the elapsed proper time of the fall and the Killing time interval of a radial photon propagating between two fixed areal radii:
\begin{equation}
\Delta\tau(B\rightarrow R)=\sqrt D\int_R^B\frac{\mathrm dR'}{\sqrt{\mathcal E^2-F(R')}},\qquad \Delta t_\infty^{(\gamma)}(R\rightarrow B)=\sqrt D\int_R^B\frac{\mathrm dR'}{F(R')}.
\label{radial_intervals_doppler}
\end{equation}
The second integral applies to an outgoing radial signal and diverges when its emission point approaches the horizon. These relations show that $D$ changes the radial duration and propagation time even though it cancels from the instantaneous speed measured by a static observer.

We now consider electromagnetic signals exchanged between the freely falling particle and the static observer that remains at the release radius $B$. For a radial null ray with wave vector $k^\mu$, conservation of the photon Killing frequency $\omega_\infty=-k_{t_\infty}$ gives
\begin{equation}
k^{t_\infty}=\frac{\omega_\infty}{F(R)},\qquad k^R=\pm\frac{\omega_\infty}{\sqrt D},
\label{wave_vector_doppler}
\end{equation}
where the upper and lower signs correspond to outgoing and ingoing propagation, respectively. The relevant four velocities are
\begin{equation}
u_{\rm ff}^{\mu}=\left(\frac{\mathcal E}{F(R)},-\frac{\sqrt{\mathcal E^2-F(R)}}{\sqrt D},0,0\right),\qquad u_B^{\mu}=\left(\frac{1}{\sqrt{F(B)}},0,0,0\right).
\label{observers_doppler}
\end{equation}
Using $\omega=-k_\mu u^\mu$ and $\mathcal E=\sqrt{F(B)}$, the frequency of an outgoing signal emitted by the freely falling observer at $R$ and received by the static observer at $B$ obeys
\begin{equation}
\frac{\omega_B^{(r)}}{\omega_{\rm ff}^{(s)}}=1-v(R;B)=1-\sqrt{\frac{2\mathcal M\left(R^{-1}-B^{-1}\right)}{1-2\mathcal M/B}}.
\label{freq_out_general_doppler}
\end{equation}
For an ingoing signal emitted by the static observer and received by the freely falling particle, the corresponding ratio is
\begin{equation}
\frac{\omega_{\rm ff}^{(r)}}{\omega_B^{(s)}}=\frac{1}{1+v(R;B)}=\left[1+\sqrt{\frac{2\mathcal M\left(R^{-1}-B^{-1}\right)}{1-2\mathcal M/B}}\right]^{-1}.
\label{freq_in_general_doppler}
\end{equation}
The two ratios equal unity at the release point and approach
\begin{equation}
\frac{\omega_B^{(r)}}{\omega_{\rm ff}^{(s)}}\longrightarrow0,\qquad \frac{\omega_{\rm ff}^{(r)}}{\omega_B^{(s)}}\longrightarrow\frac{1}{2},\qquad R\rightarrow R_h^{+}.
\label{freq_horizon_limits_doppler}
\end{equation}
The first limit represents the infinite redshift of an outgoing signal emitted arbitrarily close to the horizon and received at $B$ after an arbitrarily large Killing time interval. The finite ingoing limit results from the compensation between the gravitational blueshift acquired by the inward photon and the kinematic redshift measured by an observer falling in the same direction. Notice that the asymmetry of the two processes is therefore physical feature, although neither limiting ratio depends on the Lorentz--violating deformation.

For release from the asymptotic region, the frequency ratios take the simpler form
\begin{equation}
\frac{\omega_\infty^{(r)}}{\omega_{\rm ff}^{(s)}}=1-\sqrt{\frac{2\mathcal M}{R}},\qquad \frac{\omega_{\rm ff}^{(r)}}{\omega_\infty^{(s)}}=\left(1+\sqrt{\frac{2\mathcal M}{R}}\right)^{-1}.
\label{freq_infty_doppler}
\end{equation}
The cancellation of $D$ from Eqs.~\eqref{velocity_b_doppler}, \eqref{freq_out_general_doppler}, and \eqref{freq_in_general_doppler} occurs because the same constant radial normalization enters both the timelike and null radial equations. In other words, the parameter dependent curves obtained by holding the auxiliary quantities $(M,r,b)$ fixed compare different physical masses and areal radii. They should not be interpreted as Lorentz--violating corrections to the local Doppler ratios. A dependence on $D$ reappears when the comparison is made at fixed proper radial distances, since
\begin{equation}
\mathcal L(R_1,R_2)=\sqrt D\int_{R_1}^{R_2}\frac{\mathrm dR}{\sqrt{F(R)}},
\label{proper_distance_doppler}
\end{equation}
and it is also present in the fall and signal propagation intervals in Eq.~\eqref{radial_intervals_doppler}. The invariant effect of the Lorentz--violating sector is therefore a radial stretching or compression of the geometry itself.

\begin{figure}[t]
    \centering
    \includegraphics[width=.48\linewidth]{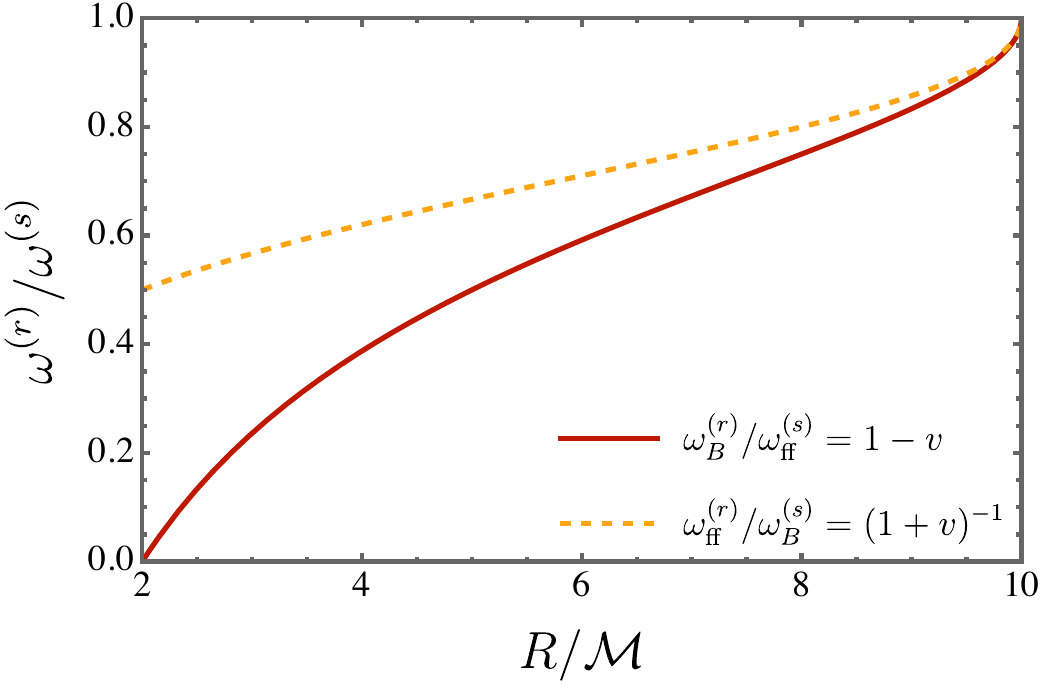}
    \includegraphics[width=.48\linewidth]{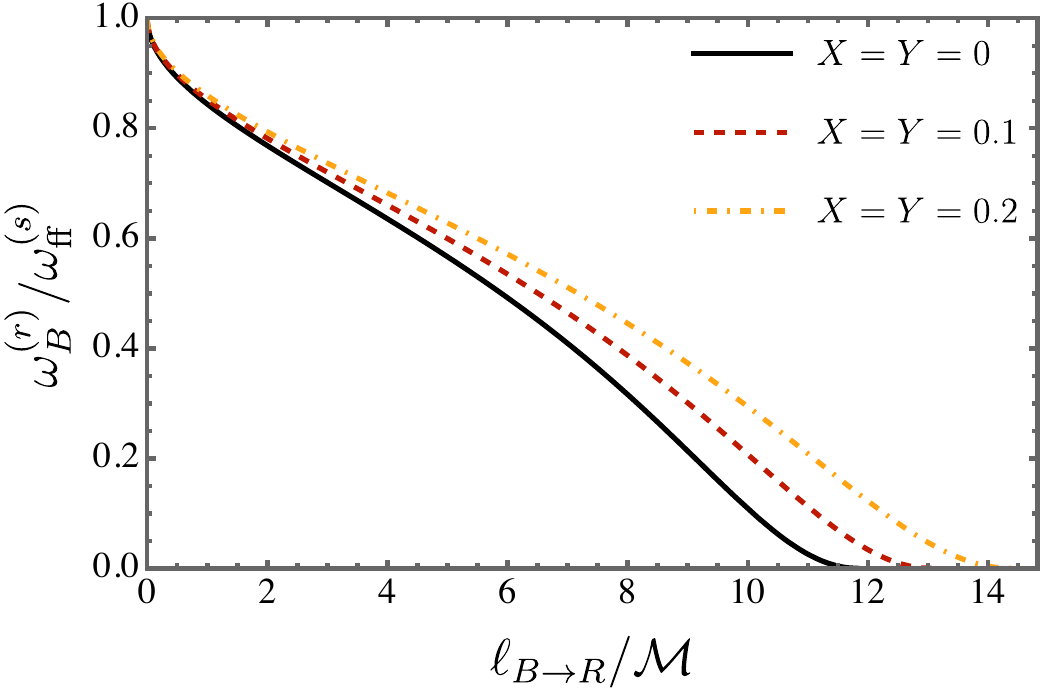}
    \includegraphics[width=.48\linewidth]{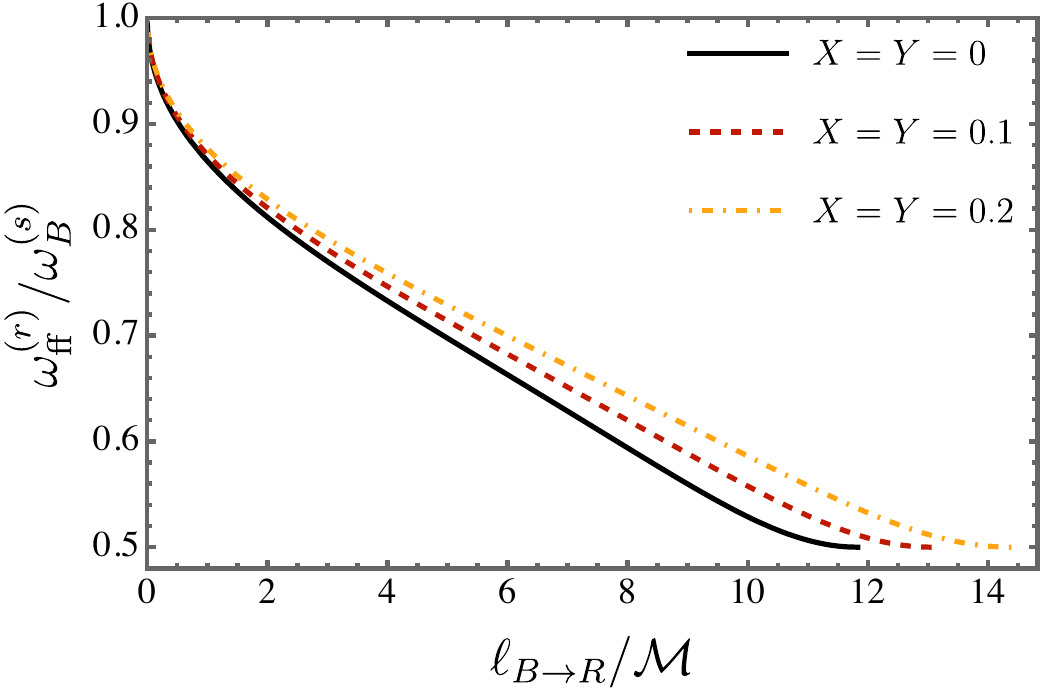}
    \caption{Gravitational Doppler ratios for a particle released from rest at $B=10\mathcal M$. The left panel shows the outgoing ratio $\omega_B^{(r)}/\omega_{\rm ff}^{(s)}=1-v$ and the ingoing ratio $\omega_{\rm ff}^{(r)}/\omega_B^{(s)}=(1+v)^{-1}$ as functions of $R/\mathcal M$. At fixed physical parameters, both quantities are independent of $X$ and $Y$, approaching $0$ and $1/2$ at the horizon, respectively. The right and bottom panels display the outgoing and ingoing ratios as functions of the proper radial distance $\ell_{B\rightarrow R}/\mathcal M$ for representative values of $X=Y$. The separation of these curves originates exclusively from the factor $\sqrt{\mathcal D(X,Y)}$.}
    \label{fig:doppler_ratios}
\end{figure}

In Fig.~\ref{fig:doppler_ratios}, we see that both frequency ratios decrease monotonically as the freely falling particle moves from $B=10\mathcal M$ toward the horizon. The outgoing ratio vanishes in the limit $R\rightarrow2\mathcal M$, whereas the ingoing ratio approaches $1/2$. Their independence from $X$ and $Y$ at fixed $R/\mathcal M$, displayed in the left panel, confirms that the Lorentz--violating deformation does not alter the local Doppler relations. A different behavior emerges when the same quantities are expressed in terms of the proper radial distance. For the representative positive values of $X=Y$ considered in the right and bottom panels, $\mathcal D>1$ stretches the radial geometry, and a given frequency ratio is attained after a larger proper radial interval than in the Schwarzschild case.

\section{Tidal forces }
\label{sec:tidal_forces}

The tidal field measured by a freely falling observer is determined by the components of the Riemann tensor projected onto a parallelly transported orthonormal frame. We initially derive these components in the original parametrization of Eq.~\eqref{metric_general}, which allows a direct comparison with Figs.~\ref{fig:tidalradial} and \ref{fig:tidalang}, and then express the result in the normalized areal variables, where its invariant dependence on the Lorentz--violating sector becomes evident.


\subsection{Tidal tensor in the freely falling frame }
\label{subsec:tidal_general}

Let us consider a neutral test body following a radial timelike geodesic. The relative acceleration between two neighboring worldlines is governed by
\begin{equation}
\frac{D^{2}\eta^{\mu}}{D\tau^{2}}+R^{\mu}{}_{\nu\alpha\beta}u^{\nu}\eta^{\alpha}u^{\beta}=0,
\label{geodesic_deviation_tidal}
\end{equation}
where $u^{\mu}$ is the four velocity and $\eta^{\mu}$ is the deviation vector. For a spatial deviation vector in the comoving orthonormal frame, $\eta^{\hat{0}}=0$, Eq.~\eqref{geodesic_deviation_tidal} becomes
\begin{equation}
\frac{D^{2}\eta^{\hat{a}}}{D\tau^{2}}=-R_{\hat{0}\hat{a}\hat{0}\hat{b}}\eta^{\hat{b}},\qquad \hat{a},\hat{b}=\hat{1},\hat{2},\hat{3}.
\label{geodesic_deviation_orthonormal}
\end{equation}

For radial motion, $\dot{\theta}=\dot{\phi}=0$, the Killing symmetry generated by $\partial_{T}$ gives the conserved energy per unit mass
\begin{equation}
E=A(r)\dot{T}.
\label{energy_tidal}
\end{equation}
The normalization \(g_{\mu\nu}u^{\mu}u^{\nu}=-1\) then yields
\begin{equation}
\frac{\dot{r}^{2}}{\left(1-\frac{X}{4}\right)^{2}}+A(r)=E^{2},
\label{radial_equation_tidal}
\end{equation}
and the inward solution is
\begin{equation}
\dot{r}=-\left(1-\frac{X}{4}\right)\sqrt{E^{2}-A(r)}.
\label{rdot_tidal}
\end{equation}
A freely falling orthonormal tetrad adapted to this trajectory is
\begin{equation}
\begin{aligned}
e^{\mu}_{\hat{0}}&=\left(\frac{E}{A(r)},-\left(1-\frac{X}{4}\right)\sqrt{E^{2}-A(r)},0,0\right),\\
e^{\mu}_{\hat{1}}&=\left(-\frac{\sqrt{E^{2}-A(r)}}{A(r)},\left(1-\frac{X}{4}\right)E,0,0\right),\\
e^{\mu}_{\hat{2}}&=\left(0,0,\frac{\sqrt{\Delta(X)}}{r},0\right),\qquad
e^{\mu}_{\hat{3}}=\left(0,0,0,\frac{\sqrt{\Delta(X)}}{r\sin\theta}\right).
\end{aligned}
\label{tetrad_tidal}
\end{equation}
Besides satisfying
\begin{equation}
g_{\mu\nu}e^{\mu}_{\hat{a}}e^{\nu}_{\hat{b}}=\eta_{\hat{a}\hat{b}},\qquad \eta_{\hat{a}\hat{b}}=\mathrm{diag}(-1,1,1,1),\qquad e^{\mu}_{\hat{0}}=u^{\mu},
\label{tetrad_orthonormal_tidal}
\end{equation}
this tetrad is parallelly transported along the geodesic,
\begin{equation}
\frac{D e^{\mu}_{\hat{a}}}{D\tau}=0.
\label{tetrad_parallel_tidal}
\end{equation}
In this manner, its components contain no inertial contributions to the relative acceleration.

Projection of the Riemann tensor onto Eq.~\eqref{tetrad_tidal} gives
\begin{equation}
R_{\hat{0}\hat{1}\hat{0}\hat{1}}=\frac{1}{2}\left(1-\frac{X}{4}\right)^{2}A''(r),
\label{riemann_radial_tidal}
\end{equation}
and
\begin{equation}
R_{\hat{0}\hat{i}\hat{0}\hat{i}}=\frac{1}{2r}\left(1-\frac{X}{4}\right)^{2}A'(r),\qquad
R_{\hat{1}\hat{i}\hat{1}\hat{i}}=-\frac{1}{2r}\left(1-\frac{X}{4}\right)^{2}A'(r),\qquad \hat{i}=\hat{2},\hat{3},
\label{riemann_angular_tidal}
\end{equation}
with no summation over $\hat{i}$. The remaining independent component is
\begin{equation}
R_{\hat{2}\hat{3}\hat{2}\hat{3}}=\frac{\Delta(X)-\left(1-\frac{X}{4}\right)^{2}A(r)}{r^{2}}.
\label{riemann_pure_angular_tidal}
\end{equation}
Although the tetrad depends on $E$, the projected tidal components do not. This cancellation follows from $R_{\hat{0}\hat{i}\hat{0}\hat{i}}=-R_{\hat{1}\hat{i}\hat{1}\hat{i}}$, which makes the transverse components invariant under the radial Lorentz boost relating different freely falling frames.


\subsection{Radial tidal stretching }
\label{subsec:radial_tidal}

For the metric function of the present solution,
\begin{equation}
A'(r)=\frac{2M}{\Delta(X)r^{2}},\qquad A''(r)=-\frac{4M}{\Delta(X)r^{3}}.
\label{A_derivatives_tidal}
\end{equation}
The radial component of Eq.~\eqref{geodesic_deviation_orthonormal} therefore reads
\begin{equation}
\frac{D^{2}\eta^{\hat{1}}}{D\tau^{2}}=\lambda_{r}(r)\eta^{\hat{1}},\qquad
\lambda_{r}(r)=\frac{2M}{r^{3}}\frac{\left(1-\frac{X}{4}\right)^{2}}{\Delta(X)}.
\label{radial_tidal_final}
\end{equation}
In the admissible black hole sector with $M>0$, $\lambda_{r}>0$ at every finite radius. Neighboring freely falling worldlines are consequently separated along the radial direction. This eigenvalue has no zero away from the excluded limiting values of the parameters: it vanishes only for $M=0$ or asymptotically as $r\rightarrow\infty$, while it diverges as $r^{-3}$ when the curvature singularity is approached.

Using $r_{h}=2M/[1-\chi(X,Y)]$, the radial eigenvalue at the horizon is
\begin{equation}
\lambda_{r}(r_{h})=\frac{\left(1-\frac{X}{4}\right)^{2}}{4\Delta(X)M^{2}}\left[1-\chi(X,Y)\right]^{3}.
\label{radial_tidal_horizon}
\end{equation}
At fixed values of the original variables $(M,r)$, the radial profile depends explicitly only on $X$, because the temporal contribution to $A(r)$ is constant and disappears upon differentiation. The dependence on $Y$ in Eq.~\eqref{radial_tidal_horizon} originates from evaluating the eigenvalue at the parameter dependent coordinate position $r_{h}$. Since the shift of $r_{h}$ is not itself invariant, the physical content of this dependence is more accurately described in the normalized variables below.

\begin{figure}[t]
\centering
\includegraphics[width=.52\linewidth]{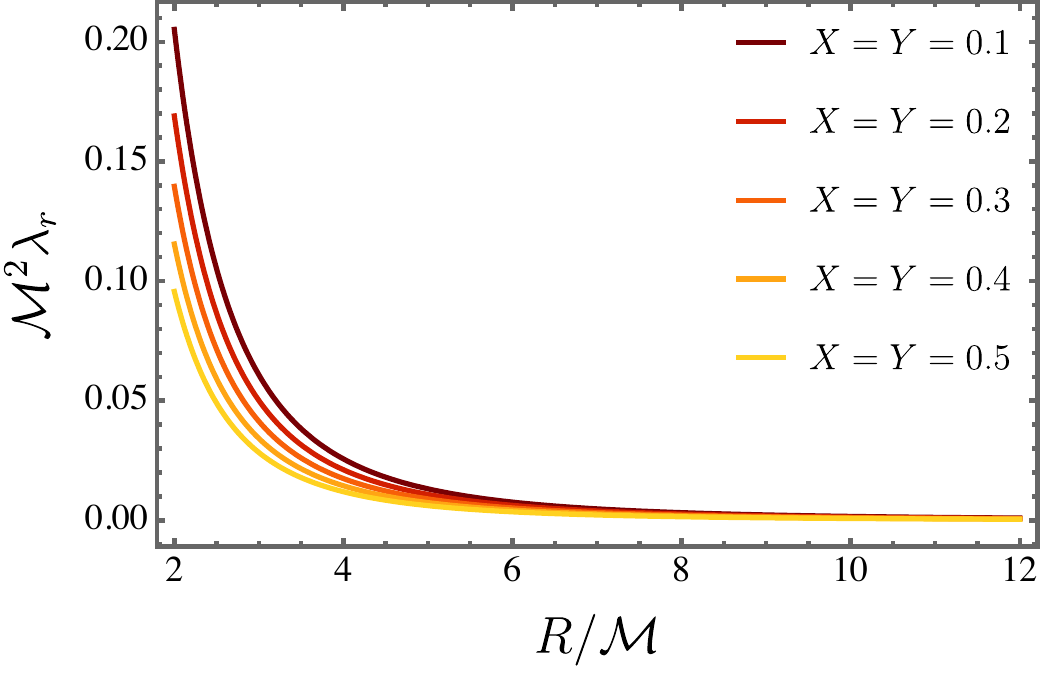}
\caption{Dimensionless radial tidal eigenvalue $\mathcal M^{2}\lambda_{r}$ as a function of $R/\mathcal M$ for $\mathcal M=1$ and representative values of $X=Y$. All curves begin at the invariant event horizon $R_{h}/\mathcal M=2$.}
\label{fig:tidalradial}
\end{figure}


\subsection{Transverse tidal compression }
\label{subsec:angular_tidal}

The two degenerate transverse components follow from Eq.~\eqref{riemann_angular_tidal},
\begin{equation}
\frac{D^{2}\eta^{\hat{i}}}{D\tau^{2}}=\lambda_{\perp}(r)\eta^{\hat{i}},\qquad
\lambda_{\perp}(r)=-\frac{M}{r^{3}}\frac{\left(1-\frac{X}{4}\right)^{2}}{\Delta(X)},\qquad \hat{i}=\hat{2},\hat{3}.
\label{angular_tidal_final}
\end{equation}
For $M>0$, $\lambda_{\perp}<0$, and both angular directions are compressed throughout the radial infall. No finite radius separates compressive and stretching regimes. The transverse eigenvalue approaches zero from below at spatial infinity and diverges negatively as $r^{-3}$ near the singularity. Its horizon value is
\begin{equation}
\lambda_{\perp}(r_{h})=-\frac{\left(1-\frac{X}{4}\right)^{2}}{8\Delta(X)M^{2}}\left[1-\chi(X,Y)\right]^{3}.
\label{angular_tidal_horizon}
\end{equation}

\begin{figure}[t]
\centering
\includegraphics[width=.52\linewidth]{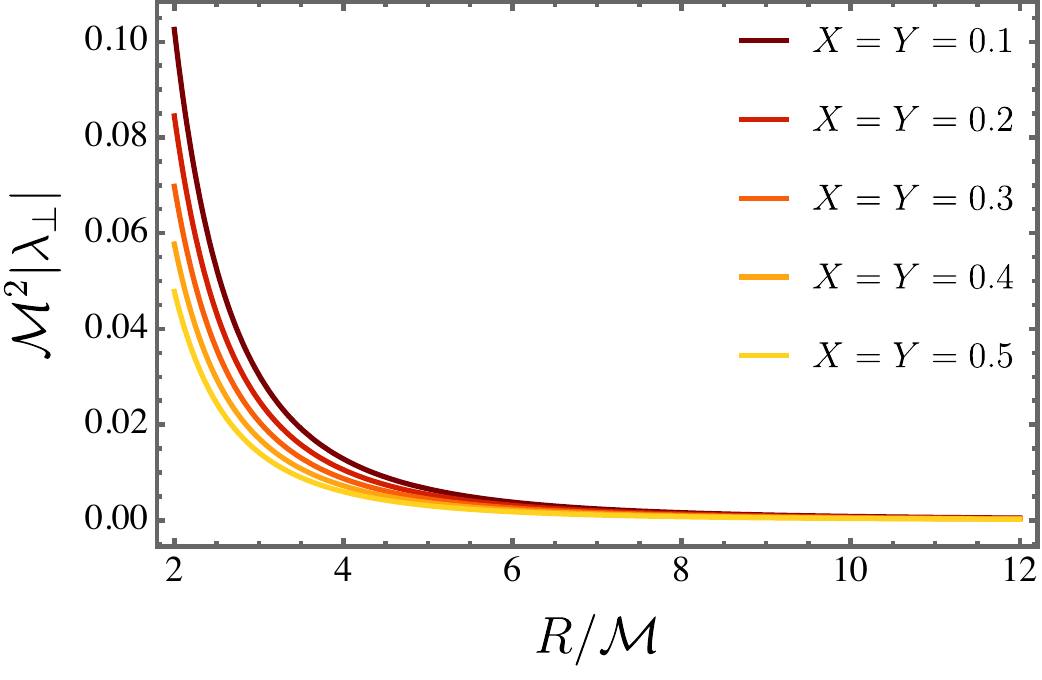}
\caption{Dimensionless magnitude \(\mathcal{M}^{2}|\lambda_{\perp}|\) of the transverse tidal compression eigenvalue for the same configurations considered in Fig.~\ref{fig:tidalradial}.}
\label{fig:tidalang}
\end{figure}

The radial and transverse eigenvalues satisfy
\begin{equation}
\lambda_{r}=-2\lambda_{\perp},\qquad \lambda_{r}+2\lambda_{\perp}=0.
\label{tidal_eigenvalue_relation}
\end{equation}
The first identity establishes the Schwarzschild stretching compression ratio without introducing the arbitrary relative magnitudes of the deviation vector components. The second shows that the electric tidal tensor measured along a radial geodesic is trace free. This property does not imply that the physical spacetime is Ricci flat for a nontrivial deformation, because the conical contribution also appears in the purely spatial angular curvature.


\subsection{Invariant form in normalized variables }
\label{subsec:tidal_normalized}

Passing to the areal radius and the unit normalized Killing time of Eq.~\eqref{normalizedmetric}, the original tidal coefficients obey
\begin{equation}
\frac{2M}{r^{3}}\frac{\left(1-\frac{X}{4}\right)^{2}}{\Delta(X)}
=\frac{2\mathcal{M}}{D(X,Y)R^{3}}.
\label{tidal_parameter_identity}
\end{equation}
The geodesic deviation equations assume the form
\begin{equation}
\frac{D^{2}\eta^{\hat{1}}}{D\tau^{2}}=\frac{2\mathcal{M}}{D(X,Y)R^{3}}\eta^{\hat{1}},\qquad
\frac{D^{2}\eta^{\hat{i}}}{D\tau^{2}}=-\frac{\mathcal{M}}{D(X,Y)R^{3}}\eta^{\hat{i}},\qquad \hat{i}=\hat{2},\hat{3}.
\label{tidal_normalized_summary}
\end{equation}
At the invariant areal horizon $R_{h}=2\mathcal{M}$, these eigenvalues reduce to
\begin{equation}
\lambda_{r}(R_{h})=\frac{1}{4D(X,Y)\mathcal{M}^{2}},\qquad
\lambda_{\perp}(R_{h})=-\frac{1}{8D(X,Y)\mathcal{M}^{2}}.
\label{tidal_normalized_horizon}
\end{equation}
Eqs.~\eqref{tidal_normalized_summary} and \eqref{tidal_normalized_horizon} show that, at fixed physical $(\mathcal{M},R)$, the metric tidal field depends on the bumblebee configuration only through the invariant deformation $D(X,Y)$. Relative to Schwarzschild, $D>1$ weakens both radial stretching and transverse compression, whereas $0<D<1$ increases them, without changing their signs or their relative factor of two.

In Figs.~\ref{fig:tidalradial} and \ref{fig:tidalang}, we see that the radial stretching eigenvalue and the magnitude of the transverse compression decrease as $R^{-3}$ from their maximum values at the invariant horizon $R/\mathcal M=2$ and vanish asymptotically. For the representative positive values of $X=Y$, increasing $\mathcal D$ suppresses both tidal eigenvalues at every fixed $R/\mathcal M$. Nevertheless, $\lambda_{r}$ remains positive and $\lambda_{\perp}$ remains negative throughout the exterior region, while the relation $\lambda_{r}=2|\lambda_{\perp}|$ is preserved. Lorentz symmetry breaking therefore changes the strength of the tidal field without modifying its stretching and compression pattern or its radial decay.

The conical curvature is retained by the purely angular component,
\begin{equation}
R_{\hat{2}\hat{3}\hat{2}\hat{3}} = \frac{D(X,Y)-F(R)}{D(X,Y)R^{2}}
=\frac{D(X,Y)-1}{D(X,Y)R^{2}}+\frac{2\mathcal{M}}{D(X,Y)R^{3}}.
\label{riemann_angular_normalized}
\end{equation}
The term proportional to $R^{-2}$ characterizes the asymptotically conical geometry but does not enter the radial tidal eigenvalues. Notice that, in the Schwarzschild limit, $D\rightarrow1$, Eqs.~\eqref{tidal_normalized_summary} and \eqref{riemann_angular_normalized} recover the standard radial stretching, transverse compression, and angular curvature of the Schwarzschild spacetime.


\section{Quasinormal modes }

\subsection{The effective potential }

To derive the radial equation without introducing notation that overlaps with the metric functions defined in the previous sections, we consider the generic static and spherically symmetric line element
\begin{equation}
\mathrm{d}s^{2}=-p(r)\,\mathrm{d}t^{2}+q(r)\,\mathrm{d}r^{2}+s^{2}(r)\left(\mathrm{d}\theta^{2}+\sin^{2}\theta\,\mathrm{d}\phi^{2}\right),
\label{general_sss_metric}
\end{equation}
where $p(r)>0$, $q(r)>0$, and $s(r)>0$ in the exterior region. The metric determinant and its associated volume element are
\begin{equation}
g=-p(r)q(r)s^{4}(r)\sin^{2}\theta,
\qquad
\sqrt{-g}=s^{2}(r)\sqrt{p(r)q(r)}\sin\theta.
\end{equation}
A minimally coupled massless scalar field obeys the Klein--Gordon equation
\begin{equation}
\frac{1}{\sqrt{-g}}\partial_{\mu}\left(\sqrt{-g}\,g^{\mu\nu}\partial_{\nu}\Phi\right)=0.
\label{KG_massless}
\end{equation}
For the geometry in Eq.~\eqref{general_sss_metric}, this equation becomes
\begin{equation}
-\frac{1}{p(r)}\frac{\partial^{2}\Phi}{\partial t^{2}}+\frac{1}{s^{2}(r)\sqrt{p(r)q(r)}}\frac{\partial}{\partial r}\left[s^{2}(r)\sqrt{\frac{p(r)}{q(r)}}\frac{\partial\Phi}{\partial r}\right]+\frac{1}{s^{2}(r)}\Delta_{\mathbb{S}^{2}}\Phi=0,
\label{KG_expanded}
\end{equation}
where $\Delta_{\mathbb{S}^{2}}$ denotes the Laplacian on the unit two--sphere. We separate the field according to
\begin{equation}
\Phi(t,r,\theta,\phi)=e^{-i\omega t}Y_{\ell m}(\theta,\phi)\frac{\psi(r)}{s(r)},
\label{scalar_ansatz}
\end{equation}
with the spherical harmonics satisfying
\begin{equation}
\Delta_{\mathbb{S}^{2}}Y_{\ell m}(\theta,\phi)=-\ell(\ell+1)Y_{\ell m}(\theta,\phi).
\end{equation}
Substituting Eq.~\eqref{scalar_ansatz} into Eq.~\eqref{KG_expanded}, we obtain
\begin{equation}
\frac{1}{s^{2}(r)\sqrt{p(r)q(r)}}\frac{\mathrm{d}}{\mathrm{d}r}\left[s^{2}(r)\sqrt{\frac{p(r)}{q(r)}}\frac{\mathrm{d}}{\mathrm{d}r}\left(\frac{\psi(r)}{s(r)}\right)\right]+\left[\frac{\omega^{2}}{p(r)}-\frac{\ell(\ell+1)}{s^{2}(r)}\right]\frac{\psi(r)}{s(r)}=0.
\label{radial_before_tortoise}
\end{equation}
The tortoise coordinate is introduced through
\begin{equation}
\frac{\mathrm{d}r_{*}}{\mathrm{d}r}=\sqrt{\frac{q(r)}{p(r)}}.
\label{tortoise_general}
\end{equation}
The normalization adopted in Eq.~\eqref{scalar_ansatz} removes the first derivative of the radial wave function and casts Eq.~\eqref{radial_before_tortoise} into the Schr\"odinger--like form
\begin{equation}
\frac{\mathrm{d}^{2}\psi}{\mathrm{d}r_{*}^{2}}+\left[\omega^{2}-V_{\mathrm{eff}}(r)\right]\psi=0,
\label{schrodinger_like}
\end{equation}
where
\begin{equation}
V_{\mathrm{eff}}(r)=p(r)\frac{\ell(\ell+1)}{s^{2}(r)}+\frac{1}{s(r)}\frac{\mathrm{d}^{2}s(r)}{\mathrm{d}r_{*}^{2}}.
\label{Veff_compact}
\end{equation}
Writing the derivatives entirely in terms of $r$, the effective potential assumes the equivalent form
\begin{equation}
V_{\mathrm{eff}}(r)=p(r)\frac{\ell(\ell+1)}{s^{2}(r)}+\frac{p(r)}{q(r)}\frac{s''(r)}{s(r)}+\frac{s'(r)}{2s(r)}\left(\frac{p(r)}{q(r)}\right)'.
\label{Veff_expanded}
\end{equation}

The physical interpretation is most transparent in the normalized variables introduced in Eq.~\eqref{normalizedmetric}. In this representation, the metric functions entering Eq.~\eqref{general_sss_metric} are
\begin{equation}
p(R)=F(R),
\qquad
q(R)=\frac{D}{F(R)},
\qquad
s(R)=R,
\qquad
F(R)=1-\frac{2\mathcal M}{R}.
\end{equation}
Equations~\eqref{tortoise_general} and \eqref{Veff_compact} then reduce to
\begin{equation}
\frac{\mathrm{d}r_{*}}{\mathrm{d}R}=\frac{\sqrt{D}}{F(R)},
\qquad
r_{*}=\sqrt{D}\left[R+2\mathcal M\ln\left|\frac{R}{2\mathcal M}-1\right|\right],
\label{tortoise_normalized}
\end{equation}
where an irrelevant additive constant has been omitted. The normalized effective potential is
\begin{equation}
V_{\mathrm{eff}}(R)=F(R)\left[\frac{\ell(\ell+1)}{R^{2}}+\frac{2\mathcal M}{D R^{3}}\right].
\label{Veff_normalized}
\end{equation}
At fixed $\mathcal M$, the Lorentz--violating dependence of the metric perturbation problem is entirely encoded in $D$. Although the centrifugal term does not contain $D$ explicitly, the deformation changes the radial propagation through the stretching $r_{*}\propto\sqrt{D}$. The potential is nonnegative throughout the exterior region and satisfies
\begin{equation}
\lim_{r_{*}\rightarrow-\infty}V_{\mathrm{eff}}=0,
\qquad
\lim_{r_{*}\rightarrow+\infty}V_{\mathrm{eff}}=0.
\label{Veff_asymptotics}
\end{equation}
It forms a single barrier without an exterior negative well or a secondary trapping region. In this manner, the effective geometry does not exhibit the potential structure required for scalar echoes or a test field bound state instability. The quasinormal boundary conditions are
\begin{equation}
\psi\sim
\begin{cases}
e^{-i\omega r_{*}}, & r_{*}\rightarrow-\infty,\\[1mm]
e^{+i\omega r_{*}}, & r_{*}\rightarrow+\infty,
\end{cases}
\label{qnm_boundary_conditions}
\end{equation}
corresponding to a purely ingoing wave at the event horizon and a purely outgoing wave at spatial infinity.

For comparison with the numerical results obtained in the original parametrization, Eq.~\eqref{Veff_expanded} gives
\begin{equation}
V_{\mathrm{eff}}^{(T)}(r)=\left(\mathcal N-\frac{2M}{r}\right)\left[\frac{\ell(\ell+1)}{r^{2}}+\frac{2Ma_{X}}{c_{X}r^{3}}\right],
\label{Veff_original}
\end{equation}
where $a_{X}$, $c_{X}$, $\mathcal N$, and $\Delta$ were defined previously. Eq.~\eqref{Veff_original} reproduces the scalar potential of the purely radial \textit{metric--affine} bumblebee solution when $Y\rightarrow0$ \cite{Filho:2023etf}, while the Schwarzschild result is recovered for $X,Y\rightarrow0$. The potentials and tortoise coordinates in the two parametrizations are related by
\begin{equation}
V_{\mathrm{eff}}(R)=\frac{\Delta}{\mathcal N}V_{\mathrm{eff}}^{(T)}(r),
\qquad r_{*}=\sqrt{\frac{\mathcal N}{\Delta}}\,r_{*}^{(T)}.
\label{potential_normalization}
\end{equation}
A mode written as $e^{-i\omega_{T}T}$ therefore has the unit normalized frequency
\begin{equation}
\omega_{\infty}=\sqrt{\frac{\Delta}{\mathcal N}}\,\omega_{T}, \qquad \mathcal M\omega_{\infty}=\frac{M\omega_{T}}{\mathcal N^{3/2}}.
\label{freqnorm}
\end{equation}

\begin{figure}[t]
\centering
\begin{minipage}{0.48\linewidth}
\centering
\includegraphics[width=\linewidth]{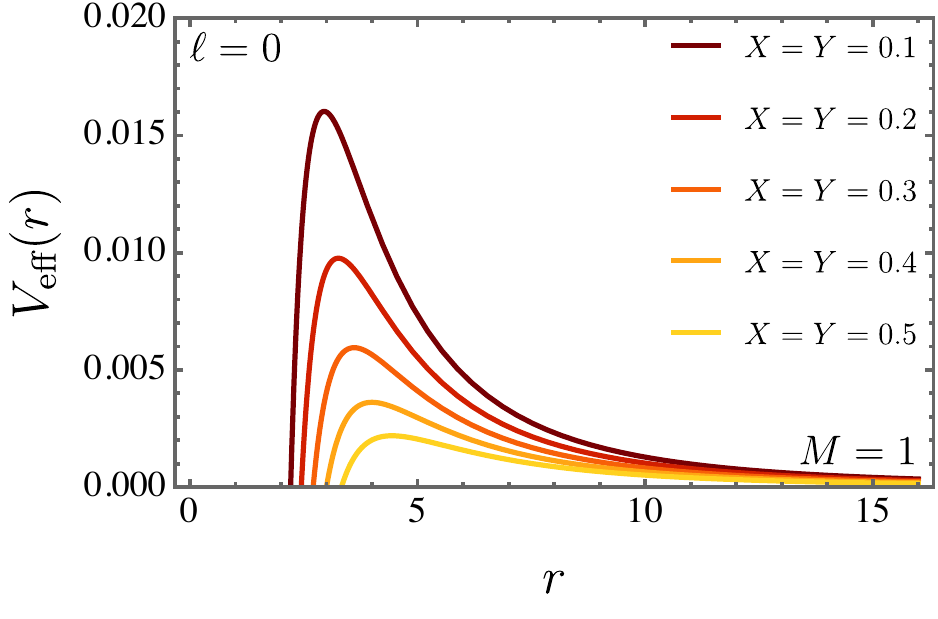}\\[-0.5ex]
{\small (a) $\ell=0$}
\end{minipage}
\hfill
\begin{minipage}{0.48\linewidth}
\centering
\includegraphics[width=\linewidth]{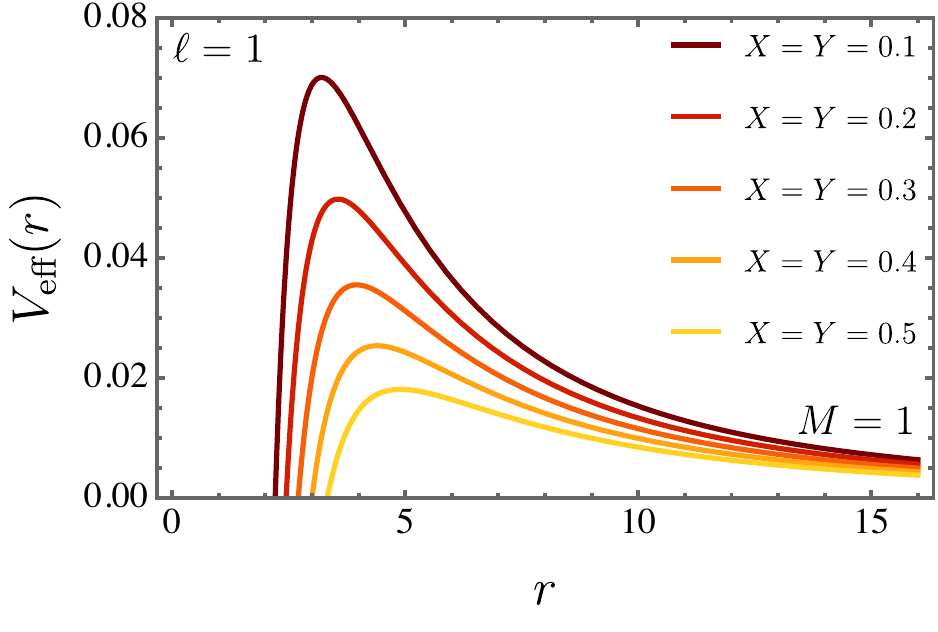}\\[-0.5ex]
{\small (b) $\ell=1$}
\end{minipage}

\vspace{0.8em}

\begin{minipage}{0.48\linewidth}
\centering
\includegraphics[width=\linewidth]{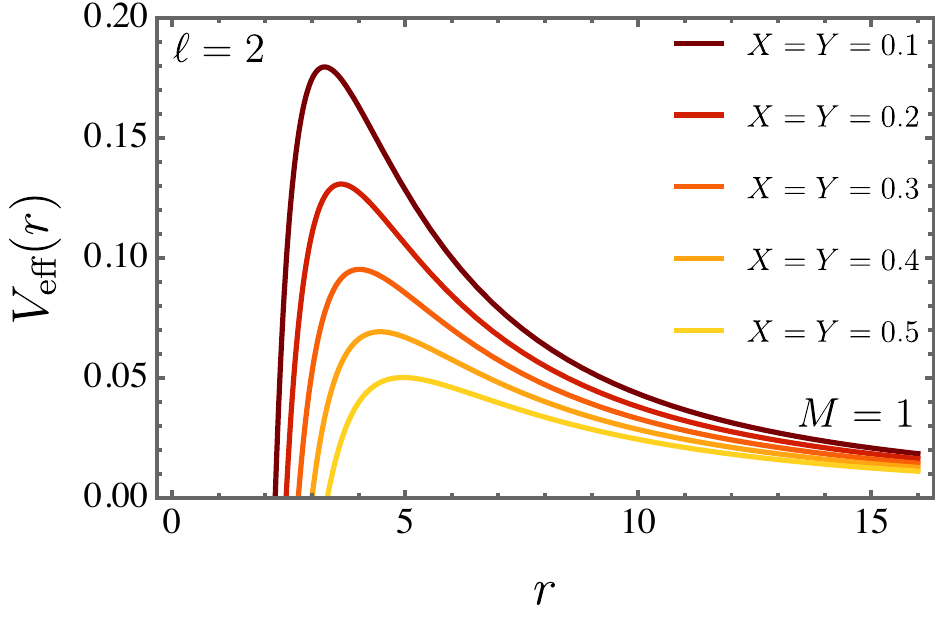}\\[-0.5ex]
{\small (c) $\ell=2$}
\end{minipage}
\caption{Effective scalar potential $V_{\mathrm{eff}}^{(T)}(r)$ in the original parametrization for $M=1$ and the displayed values of $X=Y$. Panels (a), (b), and (c) correspond to $\ell=0$, $\ell=1$, and $\ell=2$, respectively.}
\label{effecpotl0}
\end{figure}

Fig.~\ref{effecpotl0} displays $V_{\mathrm{eff}}^{(T)}(r)$ for $M=1$ along the illustrative choice $X=Y$. The barrier becomes lower and its onset moves toward larger values of $r$ as $X=Y$ increases. This behavior reflects both the displacement of the coordinate horizon and the normalization associated with the original variables. In the normalized representation, the relevant change is governed by $D$, which reduces the curvature contribution in Eq.~\eqref{Veff_normalized} and stretches the tortoise coordinate.

\begin{figure}[t]
\centering
\includegraphics[width=.58\linewidth]{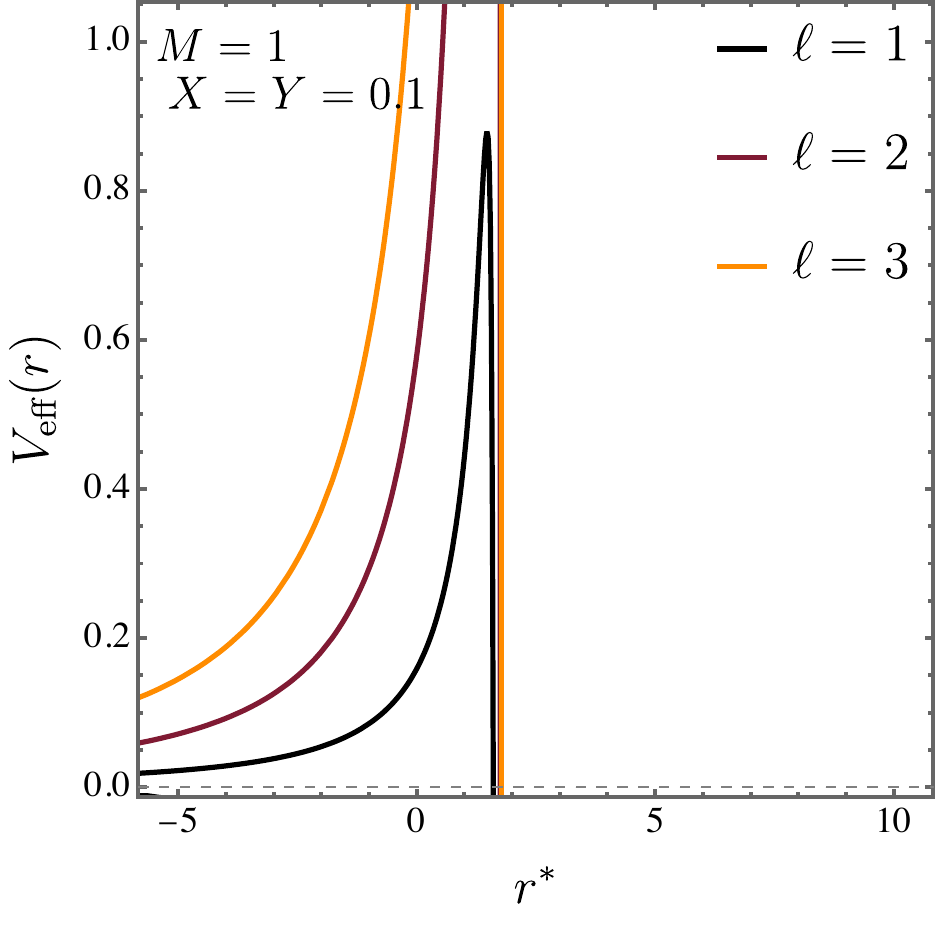}
\caption{Effective scalar potential $V_{\mathrm{eff}}(R(r_{*}))$ as a function of the normalized tortoise coordinate for $\mathcal M=1$, $X=Y=0.1$, and the displayed angular multipoles.}
\label{vefftortoise}
\end{figure}

The effective potential expressed in terms of the tortoise coordinate is displayed in Fig.~\ref{vefftortoise}. Each multipole produces a single smooth maximum, with the potential vanishing at both asymptotic ends. The absence of an additional barrier or potential well confirms that no echo producing cavity arises for the minimally coupled scalar field.

\subsection{Damped frequencies }

For a smooth effective potential with a single maximum, the quasinormal frequencies can be approximated by connecting the WKB solutions across the corresponding turning points. We employ the sixth--order prescription \cite{SchutzWill:1985,IyerWill:1987,Konoplya:2011qq}, for which
\begin{equation}
\frac{i\left(\omega_{n}^{2}-V_{0}\right)}{\sqrt{-2V_{0}^{(2)}}}-\sum_{j=2}^{6}\Lambda_{j}=n+\frac{1}{2},
\qquad
n=0,1,2,\ldots.
\label{WKB_condition}
\end{equation}
Here, $V_{0}=V_{\mathrm{eff}}(r_{0})$, where $r_{0}$ denotes the position of the potential maximum, while
\begin{equation}
V_{0}^{(k)}=\left.\frac{\mathrm{d}^{k}V_{\mathrm{eff}}}{\mathrm{d}r_{*}^{k}}\right|_{r=r_{0}}.
\end{equation}
The quantities $\Lambda_{j}$ contain the higher-order WKB corrections constructed from the derivatives of the effective potential evaluated at $r_{0}$. We adopt the convention $\operatorname{Im}\omega_{n}<0$, so that the corresponding perturbation decays in time.

Tables~\ref{tabqmnsl0}, \ref{tabqmnsl1}, and \ref{tabqmnsl2} present the frequencies calculated in the original coordinates for $\ell=0,1,2$, respectively, with $M=1$ and $X=Y$. The entries therefore correspond to $\omega_{T}$, and Eq.~\eqref{freqnorm} must be applied before comparing different geometries at fixed physical mass.

\begin{table*}[t]
\caption{Sixth--order WKB estimates of the scalar quasinormal frequencies $\omega_{T,n}$ for $\ell=0$, $M=1$, and $X=Y$ in the original parametrization.}
\label{tabqmnsl0}
\begin{ruledtabular}
\begin{tabular}{cccc}
$X=Y$ & $\omega_{T,0}$ & $\omega_{T,1}$ & $\omega_{T,2}$ \\
0.010 & $0.107183-0.098580\,\mathrm{i}$ & $0.086123-0.337562\,\mathrm{i}$ & $0.189396-0.463926\,\mathrm{i}$ \\
0.025 & $0.102434-0.095306\,\mathrm{i}$ & $0.081940-0.327322\,\mathrm{i}$ & $0.186043-0.445394\,\mathrm{i}$ \\
0.050 & $0.094913-0.090112\,\mathrm{i}$ & $0.075292-0.311274\,\mathrm{i}$ & $0.180170-0.417379\,\mathrm{i}$ \\
0.075 & $0.087995-0.085089\,\mathrm{i}$ & $0.069279-0.295358\,\mathrm{i}$ & $0.176048-0.387928\,\mathrm{i}$ \\
0.100 & $0.081496-0.080368\,\mathrm{i}$ & $0.063615-0.280606\,\mathrm{i}$ & $0.171618-0.361696\,\mathrm{i}$ \\
\end{tabular}
\end{ruledtabular}
\end{table*}

\begin{table*}[t]
\caption{Sixth--order WKB estimates of the scalar quasinormal frequencies $\omega_{T,n}$ for $\ell=1$, $M=1$, and $X=Y$ in the original parametrization.}
\label{tabqmnsl1}
\begin{ruledtabular}
\begin{tabular}{cccc}
$X=Y$ & $\omega_{T,0}$ & $\omega_{T,1}$ & $\omega_{T,2}$ \\
0.010 & $0.287942-0.095773\,\mathrm{i}$ & $0.260087-0.300222\,\mathrm{i}$ & $0.227128-0.531001\,\mathrm{i}$ \\
0.025 & $0.280672-0.092867\,\mathrm{i}$ & $0.253672-0.291023\,\mathrm{i}$ & $0.221454-0.514691\,\mathrm{i}$ \\
0.050 & $0.269020-0.088221\,\mathrm{i}$ & $0.243398-0.276320\,\mathrm{i}$ & $0.212399-0.488601\,\mathrm{i}$ \\
0.075 & $0.257918-0.083811\,\mathrm{i}$ & $0.233614-0.262364\,\mathrm{i}$ & $0.203812-0.463818\,\mathrm{i}$ \\
0.100 & $0.247329-0.079623\,\mathrm{i}$ & $0.224286-0.249113\,\mathrm{i}$ & $0.195661-0.440273\,\mathrm{i}$ \\
\end{tabular}
\end{ruledtabular}
\end{table*}

\begin{table*}[t]
\caption{Sixth--order WKB estimates of the scalar quasinormal frequencies $\omega_{T,n}$ for $\ell=2$, $M=1$, and $X=Y$ in the original parametrization.}
\label{tabqmnsl2}
\begin{ruledtabular}
\begin{tabular}{cccc}
$X=Y$ & $\omega_{T,0}$ & $\omega_{T,1}$ & $\omega_{T,2}$ \\
0.010 & $0.476075-0.094829\,\mathrm{i}$ & $0.456739-0.289666\,\mathrm{i}$ & $0.423996-0.498336\,\mathrm{i}$ \\
0.025 & $0.464962-0.091997\,\mathrm{i}$ & $0.446295-0.280952\,\mathrm{i}$ & $0.414604-0.483189\,\mathrm{i}$ \\
0.050 & $0.447051-0.087464\,\mathrm{i}$ & $0.429451-0.267012\,\mathrm{i}$ & $0.399445-0.458968\,\mathrm{i}$ \\
0.075 & $0.429867-0.083155\,\mathrm{i}$ & $0.413276-0.253765\,\mathrm{i}$ & $0.384875-0.435964\,\mathrm{i}$ \\
0.100 & $0.413372-0.079057\,\mathrm{i}$ & $0.397736-0.241173\,\mathrm{i}$ & $0.370863-0.414108\,\mathrm{i}$ \\
\end{tabular}
\end{ruledtabular}
\end{table*}

In the original coordinates, both $\operatorname{Re}\omega_{T}$ and $\lvert\operatorname{Im}\omega_{T}\rvert$ decrease as $X=Y$ increases. For the fundamental $\ell=1$ mode, changing $X=Y$ from $0.01$ to $0.10$ reduces the coordinate frequency from $0.287942-0.095773\,\mathrm{i}$ to $0.247329-0.079623\,\mathrm{i}$. A substantial part of this variation originates from the nonunit normalization of $T$. After applying Eq.~\eqref{freqnorm}, the corresponding dimensionless frequencies are
\begin{equation}
\left.\mathcal M\omega_{\infty}\right|_{X=Y=0.01}\simeq0.292294-0.097220\,\mathrm{i},
\qquad
\left.\mathcal M\omega_{\infty}\right|_{X=Y=0.10}\simeq0.287390-0.092519\,\mathrm{i}.
\end{equation}
The physically normalized oscillation frequency therefore changes more mildly than the unnormalized table entries suggest. The reduction of the damping magnitude remains present and is consistent with the radial stretching produced by the increase of $D$.

The accuracy of the WKB expansion depends strongly on the relation between the angular multipole and the overtone number. Its most reliable regime is $\ell>n$. All $\ell=0$ entries should therefore be regarded as qualitative estimates, while the fundamental $\ell=1$ mode and the $n=0,1$ modes with $\ell=2$ provide the most reliable values in the present tables. The $\ell=1$ modes with $n\geq1$ and the $\ell=2$, $n=2$ result lie outside the optimal WKB regime and should not be assigned the precision suggested by the displayed number of digits. Their reliability must be assessed through the convergence across successive WKB orders or by comparison with the time domain profiles.

All tabulated fundamental modes have negative imaginary parts, in agreement with the positive single barrier potential derived above and with the decaying scalar profiles obtained in the time domain. This establishes stability only within the minimally coupled test scalar sector and does not determine the behavior of the coupled perturbations of the metric, the independent connection, and the bumblebee field. Exact master equations for the coupled metric--bumblebee perturbations have recently been obtained in both parity sectors of the metric theory \cite{Liu:2026cxs}. Extending such a construction to the present \textit{metric--affine} geometry is required before any conclusion concerning the stability of the complete gravitational system can be drawn.


\section{Time-domain evolution }
\label{sec:time_domain}

The frequency domain analysis determines the complex resonances of the scalar field, while a direct evolution reveals how these modes are excited by a localized perturbation and how the signal passes from the prompt response to the quasinormal ringing and late time regimes. We thereby solve the characteristic initial value problem associated with Eq.~\eqref{schrodinger_like} by employing the double null integration method introduced in Ref.~\cite{Gundlach:1993tp} and subsequently applied to several black hole perturbation problems \cite{Skvortsova:2024wly,AraujoFilho:2024xhm,Bolokhov:2024ixe,Guo:2023nkd,Baruah:2023rhd,Shao:2023qlt,Lutfuoglu:2025kqp,AraujoFilho:2024lsi,Yang:2024rms}.

The numerical profiles retained in this section were obtained in the original parametrization. Denoting the corresponding time coordinate and tortoise coordinate by $T$ and $r_{*}^{(T)}$, respectively, the scalar master equation is
\begin{equation}
\left[\frac{\partial^{2}}{\partial T^{2}}-\frac{\partial^{2}}{\partial \left(r_{*}^{(T)}\right)^{2}}+V_{\mathrm{eff}}^{(T)}(r)\right]\widetilde{\psi}\left(T,r_{*}^{(T)}\right)=0.
\label{time_master_equation}
\end{equation}
We introduce the null coordinates
\begin{equation}
u=T-r_{*}^{(T)}, \qquad v=T+r_{*}^{(T)},
\label{null_coordinates_time}
\end{equation}
for which
\begin{equation}
T=\frac{u+v}{2}, \qquad r_{*}^{(T)}=\frac{v-u}{2}.
\end{equation}
Eq.~\eqref{time_master_equation} then assumes the characteristic form
\begin{equation}
\left[4\frac{\partial^{2}}{\partial u\,\partial v}+V(u,v)\right]\widetilde{\psi}(u,v)=0,
\label{double_null_equation}
\end{equation}
where
\begin{equation}
V(u,v)=V_{\mathrm{eff}}^{(T)}\left[r\left(\frac{v-u}{2}\right)\right].
\end{equation}
The dependence of the potential on both null coordinates is therefore inherited entirely from the relation between $r$ and $r_{*}^{(T)}$.

To integrate Eq.~\eqref{double_null_equation}, the $(u,v)$ plane is divided into elementary cells with equal spacing $h$ along both directions. The four vertices of each cell are defined by
\begin{equation}
S=(u,v),
\qquad
E=(u,v+h),
\qquad
W=(u+h,v),
\qquad
N=(u+h,v+h).
\label{null_grid_vertices}
\end{equation}
Integrating the master equation across one cell and expanding the field and the potential in powers of $h$ gives the update rule
\begin{equation}
\widetilde{\psi}(N)=\widetilde{\psi}(W)+\widetilde{\psi}(E)-\widetilde{\psi}(S)-\frac{h^{2}}{8}V(S)\left[\widetilde{\psi}(W)+\widetilde{\psi}(E)\right]+\mathcal{O}(h^{4}).
\label{characteristic_update}
\end{equation}
Once the values at $S$, $E$, and $W$ are known, Eq.~\eqref{characteristic_update} determines the field at $N$ and propagates the solution throughout the numerical domain. The term $\mathcal{O}(h^{4})$ denotes the local truncation error of the cell update, while the accumulated evolution is expected to exhibit second order convergence for sufficiently smooth data and potential.

Characteristic data are prescribed on the intersecting null segments $u=u_{0}$ and $v=v_{0}$. We introduce a localized Gaussian pulse along $u=u_{0}$ and impose vanishing data on the second segment,
\begin{equation}
\widetilde{\psi}(u_{0},v)=A_{0}\exp\left[-\frac{(v-v_{c})^{2}}{2\sigma^{2}}\right],
\qquad
\widetilde{\psi}(u,v_{0})=0.
\label{characteristic_initial_data}
\end{equation}
The center $v_{c}$ is placed inside the numerical domain and sufficiently far from the intersection $(u_{0},v_{0})$ that $\widetilde{\psi}(u_{0},v_{0})$ is negligible at the numerical precision of the calculation. This separation is required for the two prescriptions in Eq.~\eqref{characteristic_initial_data} to be mutually compatible. The amplitude $A_{0}$ affects only the overall normalization of the waveform because the perturbation equation is linear, whereas $\sigma$ controls the frequency content of the initial pulse without altering the quasinormal spectrum.

The evolutions were performed for massless scalar perturbations with $M=1$, using $A_{0}=\sigma=1$ and the same pulse center and extraction radius for every value of $X=Y$. The numerical domain was taken as $0\leq u,v\leq1000$ with uniform spacing $h=0.1$. The waveform measured by a static observer is obtained along the line
\begin{equation}
r_{*}^{(T)}=r_{*,\mathrm{obs}}^{(T)},
\qquad
v-u=2r_{*,\mathrm{obs}}^{(T)},
\label{extraction_line}
\end{equation}
and is displayed as a function of $T=(u+v)/2$. The value of $r_{*,\mathrm{obs}}^{(T)}$ must be kept fixed throughout the comparison. The adopted resolution is sufficient for the qualitative profiles shown below, although a numerical convergence order cannot be inferred from a single grid spacing. Such a determination requires at least two additional evolutions, for example with $h/2$ and $h/4$, followed by a Richardson comparison of the resulting waveforms.

\begin{figure}[t]
\centering
\begin{minipage}{0.48\linewidth}
\centering
\includegraphics[width=\linewidth]{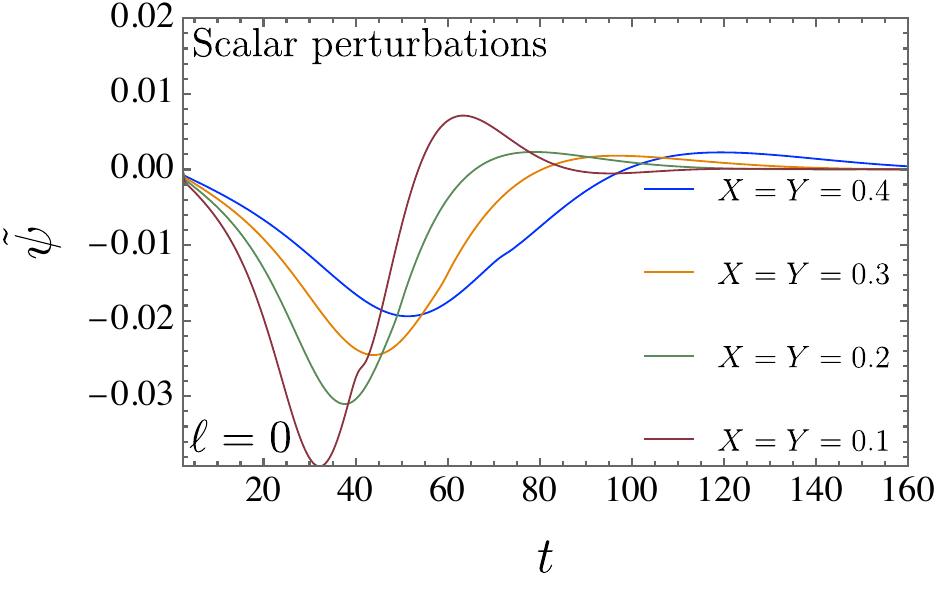}\\[-0.5ex]
{\small (a) $\ell=0$}
\end{minipage}
\hfill
\begin{minipage}{0.48\linewidth}
\centering
\includegraphics[width=\linewidth]{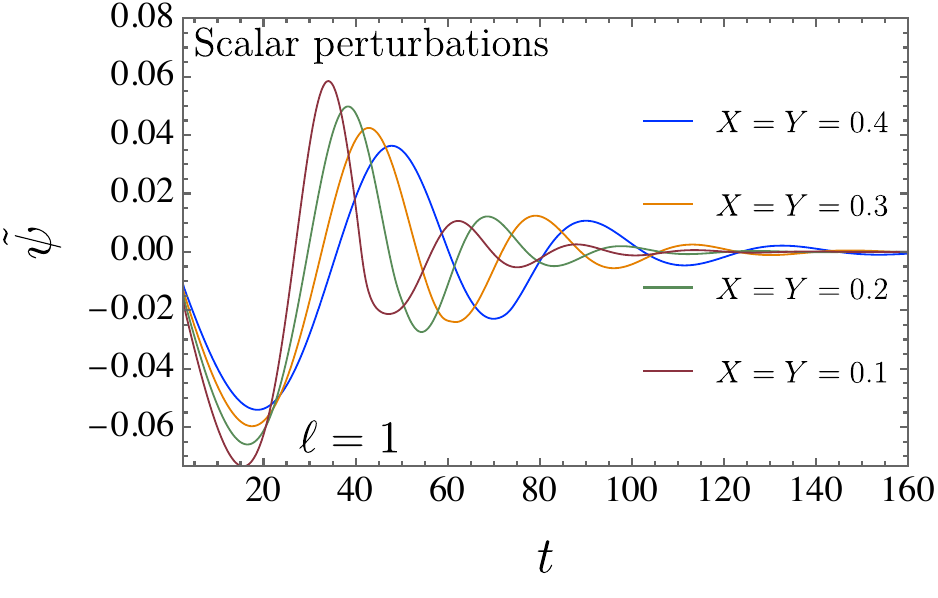}\\[-0.5ex]
{\small (b) $\ell=1$}
\end{minipage}

\vspace{0.8em}

\begin{minipage}{0.48\linewidth}
\centering
\includegraphics[width=\linewidth]{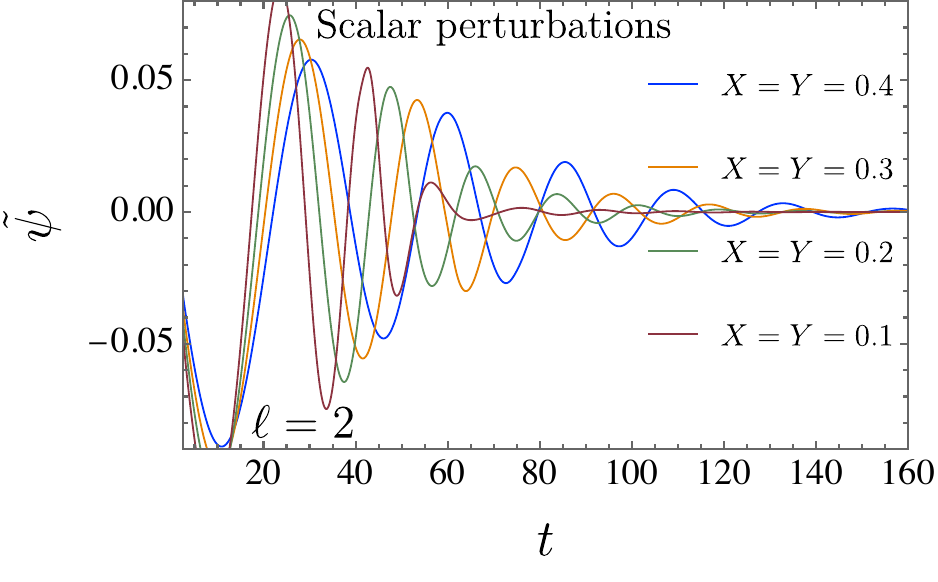}\\[-0.5ex]
{\small (c) $\ell=2$}
\end{minipage}
\caption{Time-domain scalar master field $\widetilde{\psi}(T,r_{*,\mathrm{obs}}^{(T)})$ in the original parametrization for $M=1$ and the displayed values of $X=Y$. Panels (a), (b), and (c) correspond to $\ell=0$, $\ell=1$, and $\ell=2$, respectively.}
\label{psitimedomain_l0}
\end{figure}

Fig.~\ref{psitimedomain_l0} displays the scalar master field in the original time coordinate. The $\ell=1$ and $\ell=2$ sectors show the expected sequence of prompt response and exponentially damped oscillations. As $X=Y$ increases, the oscillation period becomes longer and the decay becomes slower, consistently with the reduction of $\operatorname{Re}\omega_{T}$ and $\lvert\operatorname{Im}\omega_{T}\rvert$ found in the WKB analysis. This comparison concerns the original coordinate frequency $\omega_{T}$. A comparison at fixed physical mass must employ the normalized time $t_{\infty}$ and the frequency conversion in Eq.~\eqref{freqnorm}. The $\ell=0$ waveform is not dominated by an extended oscillatory stage and passes more rapidly from the prompt response to a nonoscillatory decay, in agreement with the limited accuracy of the WKB approximation in the monopole sector.

\begin{figure}[t]
\centering
\begin{minipage}{0.48\linewidth}
\centering
\includegraphics[width=\linewidth]{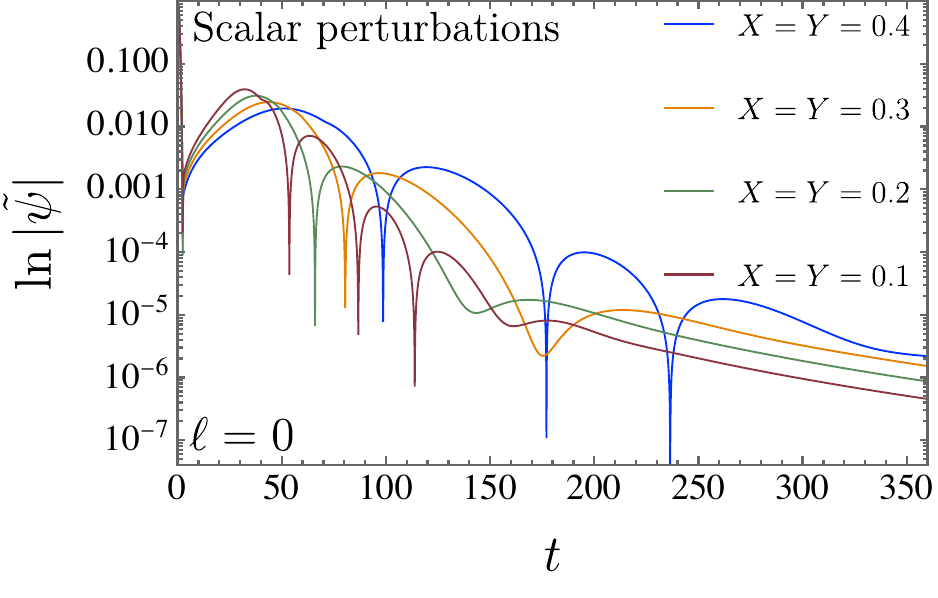}\\[-0.5ex]
{\small (a) $\ell=0$}
\end{minipage}
\hfill
\begin{minipage}{0.48\linewidth}
\centering
\includegraphics[width=\linewidth]{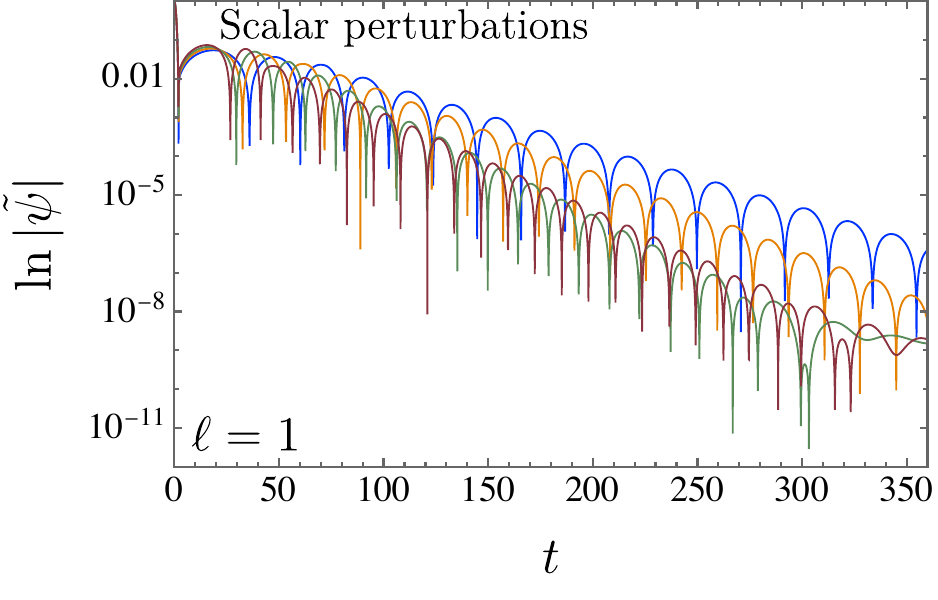}\\[-0.5ex]
{\small (b) $\ell=1$}
\end{minipage}

\vspace{0.8em}

\begin{minipage}{0.48\linewidth}
\centering
\includegraphics[width=\linewidth]{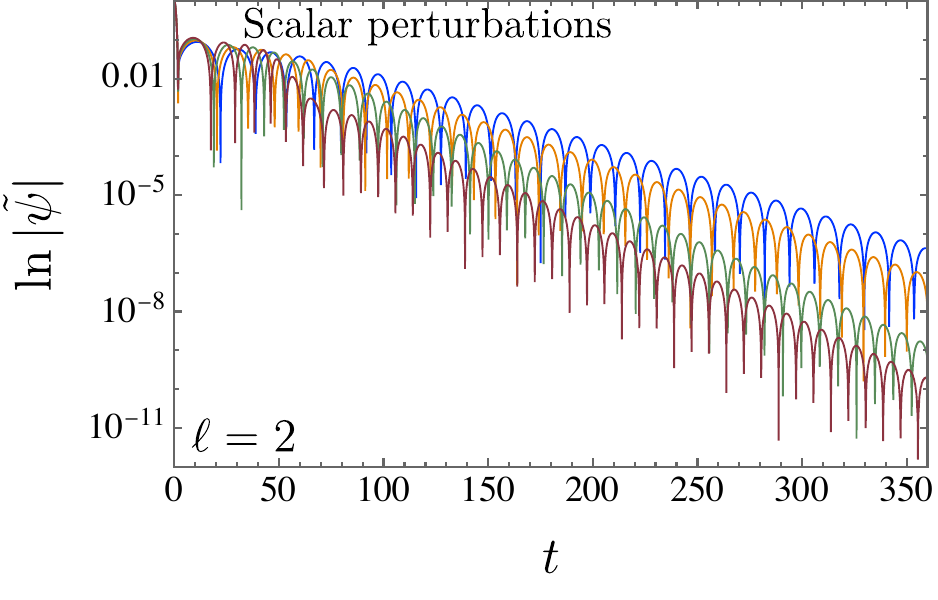}\\[-0.5ex]
{\small (c) $\ell=2$}
\end{minipage}
\caption{Absolute scalar amplitude $\lvert\widetilde{\psi}\rvert$ as a function of the original time coordinate $T$ on a logarithmic vertical scale for $M=1$ and the displayed values of $X=Y$. Panels (a), (b), and (c) correspond to $\ell=0$, $\ell=1$, and $\ell=2$, respectively.}
\label{psilntimedomain_l0}
\end{figure}

Fig.~\ref{psilntimedomain_l0} presents the absolute amplitude on a logarithmic vertical scale. For $\ell=1$ and $\ell=2$, the upper envelopes are approximately linear during the interval dominated by quasinormal ringing, as expected for a signal proportional to $e^{-\lvert\operatorname{Im}\omega\rvert T}$. The narrow downward features arise when the oscillatory waveform crosses zero and should not be interpreted as sudden enhancements of the damping rate. The monopole profile does not contain a comparably extended linear interval, which again indicates that a single oscillatory quasinormal mode does not control its evolution for a sufficiently long time.

\begin{figure}[t]
\centering
\begin{minipage}{0.48\linewidth}
\centering
\includegraphics[width=\linewidth]{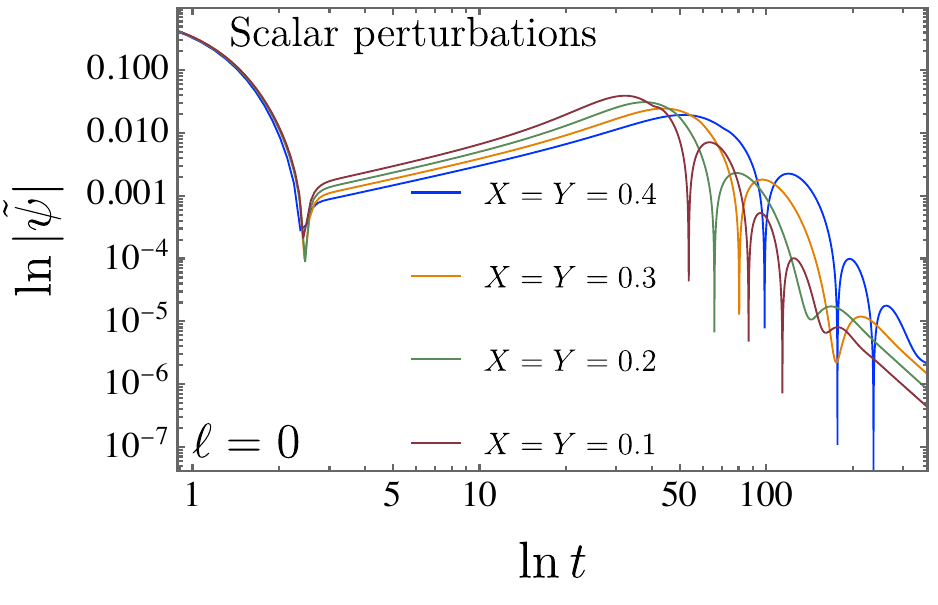}\\[-0.5ex]
{\small (a) $\ell=0$}
\end{minipage}
\hfill
\begin{minipage}{0.48\linewidth}
\centering
\includegraphics[width=\linewidth]{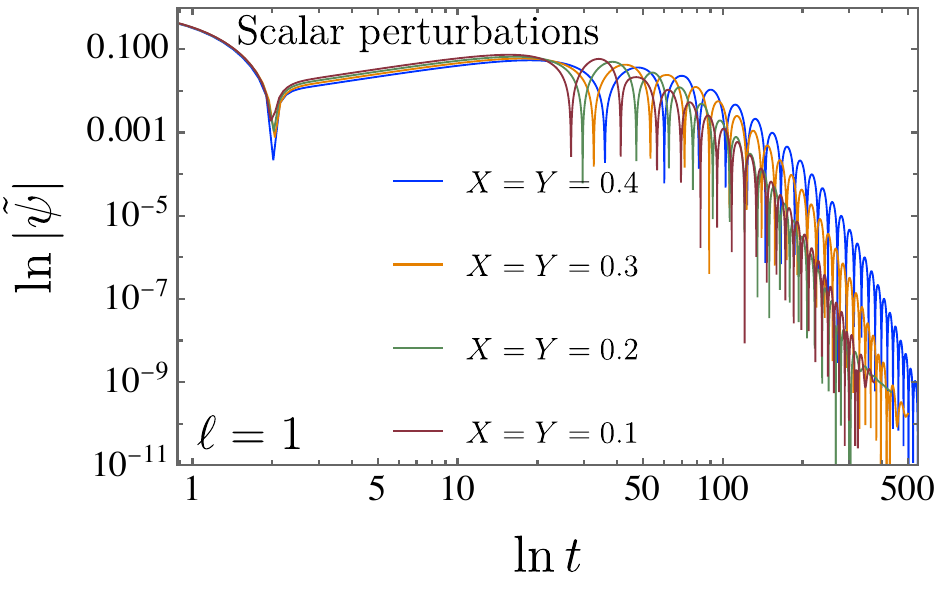}\\[-0.5ex]
{\small (b) $\ell=1$}
\end{minipage}

\vspace{0.8em}

\begin{minipage}{0.48\linewidth}
\centering
\includegraphics[width=\linewidth]{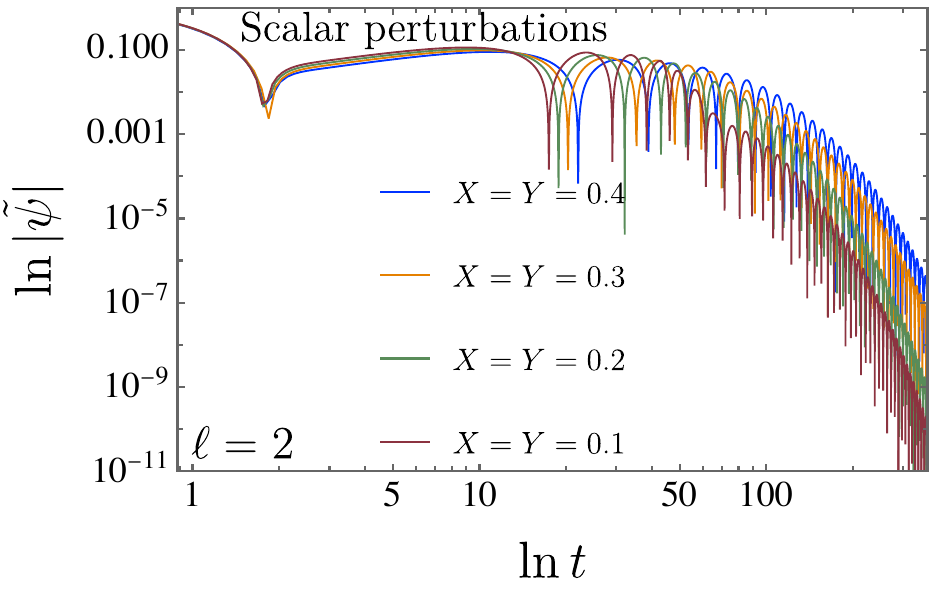}\\[-0.5ex]
{\small (c) $\ell=2$}
\end{minipage}
\caption{Late-time scalar amplitude $\lvert\widetilde{\psi}\rvert$ as a function of the original time coordinate $T$ on logarithmic axes for $M=1$ and the displayed values of $X=Y$. Panels (a), (b), and (c) correspond to $\ell=0$, $\ell=1$, and $\ell=2$, respectively.}
\label{psilnlntimedomain_l0}
\end{figure}

The same signals are represented on logarithmic axes in Fig.~\ref{psilnlntimedomain_l0}. This representation separates the late time behavior from the preceding exponential stage and makes deviations from the quasinormal envelope easier to identify. A power law tail would appear as an approximately linear envelope in this representation. The figures alone, however, do not support a reliable determination of the corresponding exponent because the raw time series, the extraction radius dependence, and evolutions at different grid spacings are required to distinguish a physical tail from discretization errors and the numerical noise floor. For the same reason, no independent Prony frequency or measured convergence order is assigned to the displayed profiles.

\section{Solar System bounds on the invariant deformation }
\label{sec:solar}

The comparison with Solar System observations must be carried out using the areal radius and the unit normalized Killing time introduced in Eq.~\eqref{normalizedmetric}. In these variables, the physical geometry is
\begin{equation}
\mathrm{d}s^{2}=-F(R)\mathrm{d}t_{\infty}^{2}+\frac{D}{F(R)}\mathrm{d}R^{2}+R^{2}\mathrm{d}\Omega^{2},\qquad F(R)=1-\frac{2\mathcal M}{R},
\label{solar_normalized_metric}
\end{equation}
and the entire Lorentz--violating dependence of minimally coupled massive and massless probes is contained in the single scalar combination
\begin{equation}
\varepsilon_{\rm LV}(X,Y)\equiv D(X,Y)-1 = \frac{1+3X/4}{(1-X/4)\left[1-Y\sqrt{(1-X/4)/(1+3X/4)}\right]}-1.
\label{epsilonLVdef}
\end{equation}
For small coefficients, this quantity becomes
\begin{equation}
\varepsilon_{\rm LV}=X+Y+\frac{X^{2}}{4}+\frac{XY}{2}+Y^{2}+\mathcal O\!\left(X^{3},X^{2}Y,XY^{2},Y^{3}\right).
\label{epsilonLVexpansion}
\end{equation}
In this case, it is more appropriate to constrain $\varepsilon_{\rm LV}$ than to assign separate limits to $X$ and $Y$. This parametrization also avoids attributing physical meaning to the coordinate displacement of the horizon in the original variables. At fixed $\mathcal M$, the horizon remains at $R_{h}=2\mathcal M$, whereas the radial normalization carries the measurable departure from the Schwarzschild geometry, as argued in the previous sections.

The asymptotic region of Eq.~\eqref{solar_normalized_metric} is not Minkowskian when $D\neq1$. Introducing the proper radial coordinate $\rho=\sqrt{D}\,R$ gives
\begin{equation}
\mathrm{d}s^{2}\simeq-\mathrm{d}t_{\infty}^{2}+\mathrm{d}\rho^{2}+\frac{\rho^{2}}{D}\mathrm{d}\Omega^{2},\qquad \Delta\Omega=4\pi\left(1-\frac{1}{D}\right)=4\pi\varepsilon_{\rm LV}+\mathcal O\!\left(\varepsilon_{\rm LV}^{2}\right),
\label{conical_asymptotics_solar}
\end{equation}
where $\Delta\Omega$ denotes the solid angle deficit, with a negative value corresponding to a solid angle surplus. Standard asymptotically flat PPN formulae cannot be transferred directly to this geometry because radial ranging, angular measurements, and the construction of the barycentric reference system are simultaneously affected. The leading effects are consequently derived from the geodesic equations and compared with the precision reached by the corresponding experiments. Similar procedures have been applied to the Schwarzschild--like bumblebee solution and to other Lorentz--violating or non--Schwarzschild geometries \cite{Casana:2017jkc,Yang:2023wtu,Wang:2024solar,Fathi:2025solar}. The numerical limits obtained below must nevertheless be interpreted as attainable sensitivities until Eq.~\eqref{solar_normalized_metric} is implemented in a global ephemeris adjustment \cite{FiengaMinazzoli:2024,Huang:2024solar}.

For geodesic motion on the equatorial plane, the conserved energy and angular momentum per unit mass are $E=F\dot t_{\infty}$ and $L=R^{2}\dot\phi$. The normalization $g_{\mu\nu}\dot x^{\mu}\dot x^{\nu}=-\kappa$, with $\kappa=1$ for massive particles and $\kappa=0$ for photons, gives the common radial first integral
\begin{equation}
D\dot R^{2}=E^{2}-F(R)\left(\kappa+\frac{L^{2}}{R^{2}}\right).
\label{solar_radial_first_integral}
\end{equation}
This equation provides the classical tests considered below without introducing a PPN identification at the level of the metric.


\subsection{Perihelion advance }

For a timelike orbit, the substitution $u=1/R$ in Eq.~\eqref{solar_radial_first_integral} gives
\begin{equation}
D\frac{\mathrm{d}^{2}u}{\mathrm{d}\phi^{2}}+u=\frac{\mathcal M}{L^{2}}+3\mathcal M u^{2}.
\label{solar_timelike_orbit}
\end{equation}
The redefinition $\bar\phi=\phi/\sqrt D$ transforms Eq.~\eqref{solar_timelike_orbit} into the Schwarzschild orbit equation. Writing $p=a_{\rm orb}(1-e^{2})$, the radial period in the original azimuthal coordinate is
\begin{equation}
\Phi_{r}=\frac{2\pi\sqrt D}{1-3\mathcal M/p}+\mathcal O\!\left(\frac{\mathcal M^{2}}{p^{2}}\right),
\label{solar_radial_period}
\end{equation}
so that the perihelion advance per revolution becomes
\begin{equation}
\Delta\varpi=2\pi\left(\sqrt D-1\right)+\frac{6\pi\sqrt D\,\mathcal M}{p}+\mathcal O\!\left(\frac{\mathcal M^{2}}{p^{2}}\right).
\label{solar_perihelion_exactD}
\end{equation}
Expanding around $D=1$ separates the Schwarzschild contribution from the leading Lorentz--violating correction,
\begin{equation}
\Delta\varpi=\frac{6\pi\mathcal M}{p}+\pi\varepsilon_{\rm LV}+\frac{3\pi\mathcal M}{p}\varepsilon_{\rm LV}+\mathcal O\!\left(\varepsilon_{\rm LV}^{2},\frac{\mathcal M^{2}}{p^{2}}\right).
\label{solar_perihelion_expanded}
\end{equation}
The mixed term is negligible for planetary orbits, while the conical contribution accumulates once per revolution. The INPOP10a supplementary precession for Mercury is $0.4\pm0.6\,\mathrm{mas\,century^{-1}}$ \cite{Fienga:2011}. Mercury completes approximately $N_{\rm Mer}=415.2$ revolutions per century, and requiring $|\pi\varepsilon_{\rm LV}|$ to remain below the quoted $1\sigma$ uncertainty per revolution gives
\begin{equation}
|\varepsilon_{\rm LV}|\lesssim\frac{0.6\,\mathrm{mas}}{\pi N_{\rm Mer}}\left(\frac{\pi}{180\times3600\times10^{3}}\right)=2.3\times10^{-12}.
\label{solar_perihelion_bound}
\end{equation}
This estimate improves the original Casana et al. sensitivity because it uses the supplementary precession extracted from a modern planetary ephemeris \cite{Casana:2017jkc}. It is not a statistically independent confidence interval, since the published residual was obtained without including $D$ among the adjusted parameters.


\subsection{Light deflection }

For null motion, Eq.~\eqref{solar_radial_first_integral} yields
\begin{equation}
D\frac{\mathrm{d}^{2}u}{\mathrm{d}\phi^{2}}+u=3\mathcal M u^{2},\qquad b\equiv\frac{L}{E}.
\label{solar_null_orbit}
\end{equation}
The total azimuthal variation between the two asymptotic portions of the trajectory is
\begin{equation}
\Delta\phi=\sqrt D\left[\pi+\frac{4\mathcal M}{b}+\mathcal O\!\left(\frac{\mathcal M^{2}}{b^{2}}\right)\right].
\label{solar_total_azimuth}
\end{equation}
Relative to the Euclidean $\pi$ convention, the bending angle is consequently
\begin{equation}
\alpha_{\rm light}=\pi\left(\sqrt D-1\right)+\frac{4\sqrt D\,\mathcal M}{b}+\mathcal O\!\left(\frac{\mathcal M^{2}}{b^{2}}\right),
\label{solar_light_deflection}
\end{equation}
which reduces at leading order to
\begin{equation}
\alpha_{\rm light}=\frac{4\mathcal M}{b}+\frac{\pi}{2}\varepsilon_{\rm LV}+\frac{2\mathcal M}{b}\varepsilon_{\rm LV}+\mathcal O\!\left(\varepsilon_{\rm LV}^{2},\frac{\mathcal M^{2}}{b^{2}}\right).
\label{solar_light_expanded}
\end{equation}
The mass independent term is the angular manifestation of the conical asymptotics and is analogous to the contribution found for a global monopole \cite{Barriola:1989hx,Ono:2018ybw}. The improved VLBI analysis of Lambert and Le Poncin-Lafitte found $\gamma=0.99992\pm0.00012$ \cite{Lambert:2011gamma}. For a ray grazing the solar limb, $\alpha_{\rm GR}=1.7516687''$, and the corresponding $1\sigma$ uncertainty in the deflection is $\sigma_{\alpha}=(\sigma_{\gamma}/2)\alpha_{\rm GR}=1.05\times10^{-4}$. Imposing $|\pi\varepsilon_{\rm LV}|/2\leq\sigma_{\alpha}$ gives
\begin{equation}
|\varepsilon_{\rm LV}|\lesssim\frac{2\sigma_{\alpha}}{\pi}=3.2\times10^{-10},
\label{solar_light_bound}
\end{equation}
where $\sigma_{\alpha}$ is understood in radians in the last equality. If the $\mathcal M=0$ conical geometry is adopted as the reference background, the mass independent term in Eq.~\eqref{solar_light_deflection} is subtracted and the coordinate mass dependent bending becomes $4\sqrt D\,\mathcal M/b$. Matching only this contribution to the PPN expression gives $\gamma_{\rm eff}=2\sqrt D-1$ and the much weaker sensitivity $|\varepsilon_{\rm LV}|\lesssim1.2\times10^{-4}$. It is important to mention that the difference between these estimates is not an algebraic ambiguity. Instead, it reflects whether the global angular defect is included in the observational reference geometry.


\subsection{Shapiro time delay }

Let $R_{0}$ denote the distance of closest approach of a photon. From Eq.~\eqref{solar_radial_first_integral}, $b^{2}=R_{0}^{2}/F(R_{0})$, and the coordinate propagation time satisfies
\begin{equation}
\frac{\mathrm{d}t_{\infty}}{\mathrm{d}R}=\frac{\sqrt D}{F(R)}\left[1-\frac{F(R)}{F(R_{0})}\frac{R_{0}^{2}}{R^{2}}\right]^{-1/2}.
\label{solar_time_integrand}
\end{equation}
Expansion to first order in $\mathcal M/R$ gives
\begin{equation}
t(R,R_{0})=\sqrt D\left[\sqrt{R^{2}-R_{0}^{2}}+\mathcal M\sqrt{\frac{R-R_{0}}{R+R_{0}}}+2\mathcal M\ln\!\left(\frac{R+\sqrt{R^{2}-R_{0}^{2}}}{R_{0}}\right)\right]+\mathcal O(\mathcal M^{2}).
\label{solar_oneway_time}
\end{equation}
For a round trip between an emitter at $R_{E}$ and a receiver at $R_{R}$, with $R_{0}\ll R_{E},R_{R}$, Eq.~\eqref{solar_oneway_time} becomes
\begin{equation}
T_{D}=\sqrt D\left\{T_{0}+4\mathcal M\left[1+\ln\!\left(\frac{4R_{E}R_{R}}{R_{0}^{2}}\right)\right]\right\},\qquad T_{0}=2(R_{E}+R_{R}).
\label{solar_roundtrip_time}
\end{equation}
Two comparisons must be distinguished. Relative to the massless conical background, whose travel time is $T_{0}^{(D)}=\sqrt D\,T_{0}$, the gravitational delay is
\begin{equation}
\Delta T_{\rm grav}^{(D)}=4\sqrt D\,\mathcal M\left[1+\ln\!\left(\frac{4R_{E}R_{R}}{R_{0}^{2}}\right)\right].
\label{solar_shapiro_conical}
\end{equation}
The impact parameter dependent logarithmic term can be compared with the Cassini determination $\gamma-1=(2.1\pm2.3)\times10^{-5}$ \cite{Bertotti:2003rm}. Since its coefficient corresponds to $\gamma_{\rm eff}=2\sqrt D-1$, the Cassini result gives the conservative estimate
\begin{equation}
|\varepsilon_{\rm LV}|\lesssim2.3\times10^{-5}.
\label{solar_shapiro_conservative_bound}
\end{equation}
Alternatively, following the prescription used in the original Schwarzschild--like bumblebee analysis \cite{Casana:2017jkc}, one may compare Eq.~\eqref{solar_roundtrip_time} directly with a Minkowski travel time $T_{0}$. The additional contribution is then
\begin{equation}
\delta T_{\rm LV}=\left(\sqrt D-1\right)\left(T_{0}+\Delta T_{\rm GR}\right)\simeq\frac{\varepsilon_{\rm LV}}{2}T_{0}.
\label{solar_shapiro_fulltime_correction}
\end{equation}
Taking $R_{E}=1\,\mathrm{AU}$, $R_{R}=8.43\,\mathrm{AU}$, and $R_{0}=1.6R_{\odot}$ gives the formal sensitivity
\begin{equation}
|\varepsilon_{\rm LV}|\lesssim6.4\times10^{-13}.
\label{solar_shapiro_formal_bound}
\end{equation}
This number is useful for comparison with Casana et al., but it should not be quoted as a direct Cassini confidence bound. The Cassini observable was a Doppler signature generated as the impact parameter varied during solar conjunction, whereas the term proportional to $T_{0}$ is strongly correlated with the ranging scale, orbital initial conditions, and the definition of the reference coordinates. Establishing a bound at the level of Eq.~\eqref{solar_shapiro_formal_bound} requires a complete reanalysis of the tracking data in the conical geometry.


\subsection{Geodetic precession }

An independent constraint follows from the parallel transport of a gyroscope along a circular orbit. The orbital angular velocity remains Schwarzschild--like, $\Omega^{2}=\mathcal M/R^{3}$, but the spin rotation accumulated during one orbital period depends on the radial normalization. Direct parallel transport gives
\begin{equation}
\Delta\Psi_{D}=2\pi\left[1-\sqrt{\frac{1-3\mathcal M/R}{D}}\right]=\frac{3\pi\mathcal M}{R}+\pi\varepsilon_{\rm LV}+\mathcal O\!\left(\varepsilon_{\rm LV}^{2},\frac{\mathcal M}{R}\varepsilon_{\rm LV},\frac{\mathcal M^{2}}{R^{2}}\right).
\label{solar_geodetic_precession}
\end{equation}
Gravity Probe B measured a geodetic drift of $-6601.8\pm18.3\,\mathrm{mas\,yr^{-1}}$, while the general relativistic prediction is $-6606.1\,\mathrm{mas\,yr^{-1}}$ \cite{Everitt:2011hp}. Using the $642\,\mathrm{km}$ orbital altitude and requiring the additional contribution in Eq.~\eqref{solar_geodetic_precession} to remain below the experimental uncertainty gives
\begin{equation}
|\varepsilon_{\rm LV}|\lesssim\frac{18.3}{6606.1}\frac{3\mathcal M_{\oplus}}{R_{\rm GPB}}=5.3\times10^{-12}.
\label{solar_gpb_bound}
\end{equation}
In addition, the geodetic channel is insensitive to the frame dragging contribution at the order retained here because Eq.~\eqref{solar_normalized_metric} is static. 

\subsection{Observables without a leading bound}

The normalized temporal sector of Eq.~\eqref{solar_normalized_metric} is exactly Schwarzschild--like. The frequency ratio measured by two static observers and the angular velocity of a circular geodesic are
\begin{equation}
\frac{\nu_{O}}{\nu_{E}}=\sqrt{\frac{F(R_{E})}{F(R_{O})}},\qquad \Omega^{2}=\frac{F'(R)}{2R}=\frac{\mathcal M}{R^{3}}.
\label{solar_null_tests}
\end{equation}
Gravitational redshift and the circular Kepler law receive no correction at fixed normalized mass coefficient $\mathcal M$. They cannot constrain $D$, $X$, or $Y$ within the vacuum geometry considered here. On the other hand, a nonvanishing effect in either observable would require a direct coupling of matter to the bumblebee or to the independent connection, or a modification of the temporal potential beyond Eq.~\eqref{solar_normalized_metric}.


\subsection{Bounds on $X$ and $Y$ }

The exact translation from a bound on $\varepsilon_{\rm LV}$ to the original coefficients is most transparent after defining $a_{X}=1-X/4$ and $c_{X}=1+3X/4$. Solving the relation $D=1+\varepsilon_{\rm LV}$ for $Y$ gives
\begin{equation}
Y(D;X)=\sqrt{\frac{c_{X}}{a_{X}}}\left(1-\frac{c_{X}}{a_{X}D}\right).
\label{solar_exact_Y_of_D}
\end{equation}
Accordingly, an observational sensitivity $|\varepsilon_{\rm LV}|\leq\delta$ defines the exact allowed strip
\begin{equation}
Y(1-\delta;X)\leq Y\leq Y(1+\delta;X),
\label{solar_exact_allowed_strip}
\end{equation}
subject to the signature and reality conditions established previously. If one coefficient is set to zero, Eq.~\eqref{solar_exact_allowed_strip} reduces to
\begin{equation}
Y=0:\quad -\frac{\delta}{1-\delta/4}\leq X\leq\frac{\delta}{1+\delta/4},\qquad X=0:\quad -\frac{\delta}{1-\delta}\leq Y\leq\frac{\delta}{1+\delta}.
\label{solar_one_parameter_bounds}
\end{equation}
The purely metric tests cannot distinguish points along the exact degeneracy curve
\begin{equation}
Y_{0}(X)=-\frac{X}{a_{X}}\sqrt{\frac{c_{X}}{a_{X}}}=-X-\frac{3X^{2}}{4}+\mathcal O(X^{3}),
\label{solar_degeneracy_curve}
\end{equation}
for which $D=1$ and the normalized physical metric is exactly Schwarzschild. Using the Mercury sensitivity in Eq.~\eqref{solar_perihelion_bound}, the principal result may be written as
\begin{equation}
|\varepsilon_{\rm LV}|\lesssim2.3\times10^{-12},\qquad |X+Y|\lesssim2.3\times10^{-12}.
\label{solar_principal_bound}
\end{equation}
If $Y=0$ or $X=0$, the corresponding one-parameter limit is $|X|\lesssim2.3\times10^{-12}$ or $|Y|\lesssim2.3\times10^{-12}$. Along the illustrative slice $X=Y$, Eq.~\eqref{solar_principal_bound} gives
\begin{equation}
|X|=|Y|\lesssim1.2\times10^{-12}.
\label{solar_equal_parameter_bound}
\end{equation}
The results are collected in Table~\ref{tab:solar_bounds}. The most stringent number is the formal full time Cassini sensitivity, while the Mercury estimate is the appropriate principal bound to quote without claiming a reanalysis of spacecraft tracking data. Separating $X$ from $Y$ requires a nonmetric observable that probes the temporal and norm sectors independently, such as direct bumblebee interactions or \textit{metric--affine} matter couplings \cite{kostelecky2011data,Delhom:2021bumblebee,Heidari:2024scattering,AraujoFilho:2026gwma}.

\begin{table*}[t]
\caption{Indicative Solar System sensitivities to the invariant deformation $\varepsilon_{\rm LV}=D-1$. The last column uses $X=Y$ and the leading relation $\varepsilon_{\rm LV}\simeq2X$. }
\label{tab:solar_bounds}
\begin{ruledtabular}
\begin{tabular}{lcccc}
Observable & Leading anomalous contribution & Experimental input & Bound on $|\varepsilon_{\rm LV}|$ & Bound for $X=Y$ \\
Mercury perihelion advance & $\pi\varepsilon_{\rm LV}$ per orbit & $0.6\,\mathrm{mas\,century^{-1}}$ & $2.3\times10^{-12}$ & $1.2\times10^{-12}$ \\
Gravity Probe B geodetic drift & $\pi\varepsilon_{\rm LV}$ per orbit & $18.3\,\mathrm{mas\,yr^{-1}}$ & $5.3\times10^{-12}$ & $2.7\times10^{-12}$ \\
Solar light deflection, conical term & $\pi\varepsilon_{\rm LV}/2$ & $\sigma_{\gamma}=1.2\times10^{-4}$ & $3.2\times10^{-10}$ & $1.6\times10^{-10}$ \\
Cassini, conical-background delay & $\gamma_{\rm eff}-1\simeq\varepsilon_{\rm LV}$ & $\sigma_{\gamma}=2.3\times10^{-5}$ & $2.3\times10^{-5}$ & $1.2\times10^{-5}$ \\
Cassini, full time prescription & $\varepsilon_{\rm LV}T_{0}/2$ & Cassini conjunction geometry & $6.4\times10^{-13}$ & $3.2\times10^{-13}$ \\
Gravitational redshift & $0$ & Schwarzschild temporal sector & no bound & no bound \\
\end{tabular}
\end{ruledtabular}
\end{table*}


\section{Conclusion }
\label{sec:conclusion}

In this work, we obtained a general static and spherically symmetric black hole supported by a bumblebee vacuum in traceless \textit{metric--affine} gravity. Starting from $b_{\mu}=(b_{t}(r),b_{r}(r),0,0)$, the field equations required $b_{t}'(r)=0$, so that the temporal component reduced to a constant and the Maxwell--like bumblebee field strength vanished. The auxiliary geometry consequently satisfied the vacuum Einstein equations, whereas Lorentz symmetry breaking entered the physical spacetime through the exact inverse deformation map. This restriction selected the vacuum configuration investigated here and did not exclude sourced static solutions with $b_{t}'(r)\neq0$.

In the original parametrization, the physical metric depended separately on $X=\xi b^{2}$ and $Y=\xi b_{0}^{2}$, while its coordinate horizon also varied with these quantities. The invariant content became transparent after introducing the areal radius and normalizing the timelike Killing coordinate at spatial infinity. The resulting line element assumed the form $\mathrm{d}s^{2}=-F(R)\mathrm{d}t_{\infty}^{2}+D F^{-1}(R)\mathrm{d}R^{2}+R^{2}\mathrm{d}\Omega_{2}^{2}$, with $F(R)=1-2\mathcal M/R$. At fixed normalized lapse mass coefficient $\mathcal M$, every observable constructed solely from the physical metric depended on the Lorentz--violating background through the single combination $D(X,Y)$. The horizon remained at $R_{h}=2\mathcal M$, whereas $D\neq1$ generated an asymptotically conical spatial geometry. The curvature invariants remained finite at the Killing horizon and diverged only at $R=0$, while the maximal extension preserved the causal structure of the Schwarzschild spacetime.

The thermodynamic analysis showed that the unit-normalized Hawking temperature was $T_{H}^{(\infty)}=1/(8\pi\mathcal M\sqrt{D})$. Positive values of the illustrative deformation $X=Y$ increased $D$ and reduced the temperature at fixed $\mathcal M$, but no finite extremal configuration or zero temperature remnant appeared within the admissible black hole sector. The horizon area remained $A_{H}=4\pi R_{h}^{2}=16\pi\mathcal M^{2}$ and contained no explicit dependence on $D$ when expressed in terms of the normalized mass coefficient. Since the gravitational action was first order and contained a nonminimal bumblebee--curvature coupling, $A_{H}/4$ was interpreted only as a geometric area law quantity. The corresponding geometric response satisfied $C_{A}=-2S_{A}$ and remained negative throughout the allowed domain. The temperature based thermodynamic vector field possessed no finite zero, so its total topological charge vanished and no thermodynamic critical point occurred for fixed $X$ and $Y$.

The radial free fall analysis separated genuine Lorentz--violating effects from those produced by the original coordinate normalization. At fixed $\mathcal M$, $R$, and release radius, the velocity measured by a static observer retained the Schwarzschild form and approached the speed of light as an exterior limit at the horizon. The deformation nevertheless modified proper fall times, radial photon propagation intervals, and proper radial distances through the factor $\sqrt{D}$. The frequency exchange exhibited the usual asymmetry near the horizon: an outgoing signal emitted by the freely falling observer became infinitely redshifted when received by a static observer, whereas the frequency ratio for an ingoing signal received by the infaller approached $1/2$. These limiting ratios were independent of $D$, since the same radial normalization entered the timelike and null trajectories.

The tidal field retained a direct invariant dependence on the Lorentz--violating geometry. In the freely falling frame, the radial and transverse eigenvalues became $\lambda_{r}=2\mathcal M/(D R^{3})$ and $\lambda_{\perp}=-\mathcal M/(D R^{3})$, respectively. The deformation therefore rescaled the magnitude of the tidal forces without changing their character: neighboring geodesics were stretched radially and compressed along both angular directions. The relation $\lambda_{r}=-2\lambda_{\perp}$ remained unchanged, showing that the Schwarzschild ratio between radial stretching and transverse compression survived the \textit{metric--affine} deformation.

For massless scalar perturbations, the normalized effective potential reduced to $V_{\mathrm{eff}}=F(R)\left[\ell(\ell+1)/R^{2}+2\mathcal M/(D R^{3})\right]$. It remained positive outside the event horizon, vanished at both asymptotic ends of the tortoise coordinate, and formed a single barrier without an additional trapping region. Along the illustrative choice $X=Y$, the quasinormal frequencies calculated in the original coordinates showed smaller oscillation and damping magnitudes as the deformation increased. After normalizing the asymptotic time, the variation of the dimensionless oscillation frequency became milder, while the reduction of the damping magnitude remained present. The characteristic evolution exhibited decaying scalar signals throughout the sampled parameter range and reproduced the qualitative behavior inferred from the WKB spectrum.

The Solar System analysis was formulated directly in terms of the normalized asymptotically conical geometry, avoiding an unqualified identification with the standard asymptotically flat PPN metric. Perihelion advance, light deflection, Shapiro delay, and geodetic precession constrained the invariant deformation $\epsilon_{\mathrm{LV}}=D-1$, while gravitational redshift and the circular Kepler law remained unchanged at fixed $\mathcal M$. The Mercury perihelion estimate supplied the principal sensitivity $|\epsilon_{\mathrm{LV}}|\lesssim2.3\times10^{-12}$, which became $|X+Y|\lesssim2.3\times10^{-12}$ at leading order and $|X|=|Y|\lesssim1.2\times10^{-12}$ along the illustrative choice $X=Y$. Purely metric observations could not separate $X$ from $Y$ and remained insensitive along the exact degeneracy curve $D(X,Y)=1$, for which the normalized physical metric coincided with the Schwarzschild geometry.

As a further perspective, the analysis of particle and photon geodesics, gravitational lensing in the weak-- and strong--deflection regimes, matter accretion, neutrino oscillations, and gravitational radiation emitted by periodic orbits could provide complementary probes of the invariant deformation introduced by the \textit{metric--affine} bumblebee background.


\section*{Acknowledgments}
\hspace{0.5cm} A. A. Araújo Filho is supported by Conselho Nacional de Desenvolvimento Cient\'{\i}fico e Tecnol\'{o}gico (CNPq) -- [150223/2025-0]. 
N. H. is supported by Conselho Nacional de Desenvolvimento Cient\'{\i}fico e Tecnol\'{o}gico — CNPq, project number 152891/2025-0. Also, N. H. is grateful for the support provided by three COST Actions: CA21106 (COSMIC WISPers in the Dark Universe: Theory, Astrophysics and Experiments), CA21136 (Addressing Observational Tensions in Cosmology with Systematics and Fundamental Physics, also known as CosmoVerse), and CA23130 (Bridging High and Low Energies in Search of Quantum Gravity, or BridgeQG). V. B. Bezerra is partially supported by Conselho Nacional de Desenvolvimento Cient\'{\i}fico e Tecnol\'{o}gico (CNPq) - Brazil, Grant number 311847/2026-9. 
FSNL acknowledges support from the Funda\c{c}\~{a}o para a Ci\^{e}ncia e a Tecnologia (FCT) Scientific Employment Stimulus contract with reference CEECINST/00032/2018, and funding through the research grant UID/04434/2025.

\section*{Data Availability Statement}

Data associated with this study consist of the analytical expressions and numerical figures presented in the manuscript. No additional data set is required to reproduce the analytical results.

\bibliographystyle{apsrev4-2}
\bibliography{main}

\end{document}